\documentclass[epj]{svjour}
\usepackage{graphics}
\usepackage{appendix}
\usepackage[dvipsnames]{xcolor}
\usepackage[T1]{fontenc}
\usepackage{graphicx}
\usepackage{amsmath}
\usepackage{amssymb}
\usepackage{xspace}
\usepackage[numbers,sort&compress]{natbib}
\usepackage{subfig}
\usepackage[shortlabels]{enumitem}
\usepackage{float}
\usepackage{pstricks}
\usepackage{graphicx}
\usepackage{xspace}
\usepackage{hyperref}
\usepackage{lipsum}
\usepackage{multirow}
\usepackage{citehack}
\usepackage{placeins}
\usepackage{braket}
\usepackage[compat=1.1.0]{tikz-feynman}
\usepackage{contour}
\usepackage{tikz}
\usepackage{placeins}

\newcommand{\phiphantK}{ \varphi_{i_{k}^{\phantom{\prime}}}}
\newcommand{\phiphantJ}{ \varphi_{i_{j}^{\phantom{\prime}}}}

\newcommand{\phiphantJK}{ \varphi_{i_{j/k}^{\phantom{\prime}}}}

\newcommand{\LO}{\text{LO}\xspace}

\newcommand{\EW}{\ensuremath{\text{EW}}\xspace}

\newcommand{\NVI}{\ensuremath{\text{NLO}_{\text{VI}}~\EW}\xspace}
\newcommand{\NLL}{\ensuremath{\text{NLL}_{\text{VI}}~\EW}\xspace}
\newcommand{\NLLp}{\ensuremath{\text{NLL'}_{\text{VI}}~\EW}\xspace}
\newcommand{\NLLpzero}{\ensuremath{\text{NLL'}_{\text{no Im, VI}}~\EW}\xspace}
\newcommand{\NLLpMREXT}{\ensuremath{\text{NLL'}_{\text{V}_{\text{MR}}}\EW \hspace{0.1 cm} \text{ext-only}}\xspace}
\newcommand{\NLLMR}{\ensuremath{\text{NLL}_{\text{V}_{\text{MR}}}\EW}\xspace}
\newcommand{\NLLpMR}{\ensuremath{\text{NLL'}_{\text{V}_{\text{MR}}}\EW}\xspace}
\newcommand{\LSC}{\ensuremath{\text{LSC}}\xspace}
\newcommand{\SSC}{\ensuremath{\text{SSC}}\xspace}
\newcommand{\SSSC}{\ensuremath{\text{S-SSC}}\xspace}

\newcommand{\COLL}{\ensuremath{\text{C}}\xspace}
\newcommand{\PR}{\ensuremath{\text{PR}}\xspace}
\newcommand{\WFRC}{\ensuremath{\text{WFRC}}\xspace}
\newcommand{\YUK}{\ensuremath{\text{YUK}}\xspace}

\newcommand{\dLSC}{\ensuremath{\hat\delta^{\text{LSC}}}\xspace}
\newcommand{\dSSC}{\ensuremath{\hat\delta^{\text{SSC}}}\xspace}
\newcommand{\dSSSC}{\ensuremath{\hat\delta^{\text{S-SSC}}}\xspace}

\newcommand{\dC}{\ensuremath{\hat\delta^{\text{C}}}\xspace}
\newcommand{\dYUK}{\ensuremath{\hat\delta^{\YUK}}\xspace}
\newcommand{\dWFRC}{\ensuremath{\hat\delta^{\WFRC}}\xspace}

\newcommand{\Sherpa}{{\rmfamily\tt Sherpa}\xspace}

\newcommand{\OpenLoops}{{\rmfamily\tt OpenLoops}\xspace}

\newcommand{\Rivet}{{\rmfamily\tt Rivet}\xspace}

\newcommand{\rs}{\mathrm{s}}

\newcommand{\alphaS}{\alpha_{\rs}}
\newcommand{\sqrtS}{\ensuremath{\sqrt{s}}\xspace}
\newcommand{\mureg}{\ensuremath{\mu_{\rm D}}\xspace}
\newcommand{\muew}{\ensuremath{\mu_{\rm PR}}\xspace}
\newcommand{\muren}{\ensuremath{\mu_{\rm R}}\xspace}
\newcommand{\ord}{\mathcal{O}}

\newcommand{\beqar}{\begin{eqnarray}}
\newcommand{\eeqar}{\end{eqnarray}}
\newcommand{\beq}{\begin{equation}}
\newcommand{\eeq}{\end{equation}}
\newcommand{\bit}{\begin{itemize}}
\newcommand{\eit}{\end{itemize}}

\def\reffi#1{\mbox{Fig.~\ref{#1}}}

\def\citere#1{\mbox{Ref.~\cite{#1}}}

\def\max{\mathrm{max}}

\def\setrelwidth{0.47}

\newcommand{\DP}{{\sc DP}\xspace}
\newcommand{\DPA}{\DP algorithm\xspace}

\usepackage[left,mathlines]{lineno} 
\usepackage{soul} 
\begin{document}
\title{Logarithmic EW corrections at one-loop}
\author{J. M. Lindert, L. Mai 
}                     
%
%
\institute{Department of Physics and Astronomy, University of Sussex, Brighton BN1 9QH, UK
} 
\date{Received: date / Revised version: date}
%
\abstract{
We present a fully automated implementation of next-to-leading order electroweak (NLO EW) corrections in the logarithmic approximation in OpenLoops. For energies above the electroweak scale NLO EW corrections are logarithmically enhanced and in tails of kinematic distributions of crucial LHC processes yield correction factors of several tens of percent. The implementation of the logarithmic Sudakov EW approximation in the amplitude generator OpenLoops is fully general, largely model independent, it supports the computation of EW corrections to resonant processes, and it is suitable for extensions to the two-loop NNLO EW level. The implementation is based on an efficient representation of the logarithmic approximation in terms of an effective vertex approach. Investigating a set of representative LHC processes we find excellent agreement between the logarithmic approximation and full one-loop results in observables where the assumptions of the EW Sudakov approximation are fulfilled.
%
\PACS{
      {PACS-key}{discribing text of that key}   \and
      {PACS-key}{discribing text of that key}
     } 
} 
\maketitle
%

\section{Introduction} \label{intro}
Analyses performed by the experimental collaborations at the Large
Hadron Collider (LHC) are stress-testing the Standard Model (SM)
of particle physics at unprecedented levels of energies and precision.
Without a clear sign of new physics in terms of new resonances, these
analyses are required to look for small percent-level deviations in the tails
of kinematic distributions measured in complex final states. In order to
match the required precision in theory simulations higher-order QCD and
electroweak (EW) predictions are required for a large number of processes.
Nowadays, one-loop QCD and EW calculations can be carried out with a number
of automated and widely applicable tools \cite{Buccioni:2019sur, Actis:2016mpe, GoSam:2014iqq,
Frederix:2018nkq, Denner:2016kdg, Peraro:2014cba}, and are part of automated multi-purpose NLO Monte
Carlo generators \cite{Gleisberg:2008ta,Alioli:2010xd,Bevilacqua:2011xh,Alwall:2014hca,Bellm:2015jjp,Bredt:2022zpn}.

Above the EW scale higher-order EW corrections of relative $\ord(\alpha)$ are strongly
enhanced, and in this regime they can easily reach a similar size as QCD
higher-order corrections. Thus, a very good control of these effects and
in particular of remaining theoretical uncertainties is critical for a success
of many physics endeavours at the LHC. To be more precise, in the Sudakov regime,
where all kinematic invariants 
are of the same order $s$ and much larger than the
electroweak scale, 
EW corrections are dominated by
double/leading logarithms (LL) and single/next-to-leading logarithms (NLL) of
the ratio of the energy scale to the weak gauge-boson masses~\cite{Ciafaloni:1998xg,Beccaria:1998qe,Beccaria:1998qe,Kuhn:1999de,Fadin:1999bq,Beccaria:2000jz,Denner:2000jv,Ciafaloni:2000df,Hori:2000tm,Melles:2001ye}. These EW Sudakov
logarithms originate from virtual gauge bosons that couple to one or two on-shell
(or nearly on-shell) external particles in the soft and/or collinear limits, as well as from the running of physical parameters from the EW scale to the hard scale. 

For processes that are not mass-suppressed in the high-energy limit,
EW Sudakov corrections are universal and general factorisation formulas for
the LLs and NLLs applicable to any Standard Model process at one loop are
available~\cite{Denner:2000jv,Denner:2001gw,Pozzorini:2001rs}. 
This factorisation relies on the soft/ collinear approximation of scattering
amplitudes. Based on this factorisation as part of
Refs.~\cite{Denner:2000jv,Denner:2001gw,Pozzorini:2001rs}
an algorithm as been formulated, which allows for the
determination of EW Sudakov corrections in a process-independent way.
In addition, this factorisation allows for the resummation
of EW Sudakov corrections~\cite{Fadin:1999bq,Kuhn:1999nn,Melles:2000gw,Melles:2000ia,Melles:2001dh,Ciafaloni:2005fm}, also employing soft-collinear effective theory~\cite{Chiu:2007yn,Chiu:2007dg,Chiu:2008vv,Chiu:2009mg,Chiu:2009ft,Fuhrer:2010eu,Bauer:2017bnh}. 
The algorithm of Refs.~\cite{Denner:2000jv,Denner:2001gw,Pozzorini:2001rs} by Denner and Pozzorini has been applied for the computation of EW corrections to a
multitude of important LHC and LEP scattering processes~\cite{Denner:2000jv,Layssac:2001ur,Kuhn:2004em,Kuhn:2005gv,Kuhn:2007qc,Accomando:2006hq,Mishra:2013una,Granata:2017iod}.
More recently the algorithm has been implemented in three automated
tools~\cite{Chiesa:2013yma,Bothmann:2020sxm,Pagani:2021vyk}.
In particular the implementations in {\tt Sherpa}~\cite{Bothmann:2020sxm} and
{\tt MadGraph5\_aMC@NLO}~\cite{Pagani:2021vyk,Pagani:2023wgc} are fully general and applicable to any
hard SM process where all kinematic invariants are much larger than the EW scale, e.g. to processes without internal resonances.

In this paper we present a novel fully automated implementation for the computation of one-loop EW  corrections in logarithmic approximation within the amplitude generator \OpenLoops. The implementation is based on a formulation of the
algorithm by Denner and Pozzorini in terms of EW Sudakov pseudo-counterterms, which are process independent and will
allow for a streamlined extension to the two-loop NNLO EW level~\cite{Hori:2000tm,Kuhn:2001hz,Beenakker:2001kf,Denner:2003wi,Feucht:2003yx,Pozzorini:2004rm,Penin:2004kw,Feucht:2004rp,Jantzen:2005xi,Jantzen:2005az,Jantzen:2006jv,Denner:2006jr,Denner:2008yn}, for the computation of mass-suppressed corrections, and to theories beyond the SM, which can similarly also be achieved within the implementations presented in~\cite{Bothmann:2020sxm,Pagani:2021vyk}. 
The computation of NLL single logarithms originating from the UV
employs the standard EW one-loop renormalisation as implemented in \OpenLoops supplemented
with an appropriate logarithmic approximation of the contributing self-energy integrals.
In our implementation infrared singular photon/QED contributions can be regularised in mass or dimensional regularisation.  
We investigate the scope of our implementation for a set of representative LHC processes, including for on-shell production of $V$+jets, $VV$+jets, $VVV$, and $t\bar t+X$ processes. 

The standard implementation of logarithmically enhanced one-loop EW corrections is only applicable to hard scattering processes, where all scales are of the same order and much larger than the EW scale. Based on this standard implementation we introduce an algorithm which allows to apply the obtained one-loop EW corrections in logarithmic approximation to resonant processes involving the decay of heavy particles. This algorithm  correctly describes the transition between on-shell and off-shell configurations. We showcase this implementation for off-shell resonant processes considering hadronic $l^+l^-+$jet and $\ell^+\ell^-\ell'^+\ell'^-$ production.

The structure of the article is organised as follows: in Section \ref{sec2} we
revisit the structure of one-loop EW Sudakov logarithms following Refs.~\cite{Denner:2000jv,Denner:2001gw,Pozzorini:2001rs}. In Section \ref{sec3} we
introduce the implementation of the algorithm for the evaluation of one-loop EW corrections in logarithmic approximation in \OpenLoops focussing on the versatility of the employed pseudo-counterterm approach. 
In Section~\ref{sec4}  we present numerical results of our implementation. 
We conclude in Section~\ref{sec5}.
\\

\section{EW Sudakov corrections at one-loop} \label{sec2}

In this section we discuss the structure of one-loop EW corrections in the high-energy limit where the algorithm of Ref.~\cite{Denner:2000jv,Denner:2001gw,Pozzorini:2001rs} allows to evaluate such corrections in logarithmic approximation (LA).  In the following Section~\ref{sec3} we will discuss the implementation of these corrections in \OpenLoops.

\subsection{Notation, conventions and Logarithmic Approximation}

For the discussion of the structure of the EW corrections in logarithmic approximation we use the same conventions as in Refs.~\cite{Denner:2000jv}, and consider a $n \to 0$  process
\begin{equation} \label{process_convention}
            \varphi_{i_{1}}(p_1) \ldots \varphi_{i_{n}}(p_n) \to 0\,,
        \end{equation}
where $\varphi_i$ stands for the $i$ component of a multiplet $\varphi$ which can be any field in the Standard Model and the momenta $p_1,...,p_n$ are all incoming; the corresponding predictions for the physical process 
 \begin{equation}
    \varphi_{i_{1}}(p_1^{\text{in}}) \ldots \varphi_{i_{m}}(p_m^{\text{in}}) \to \varphi_{j_{1}}(p_{1}^{\text{out}}) \ldots \varphi_{j_{n-m}}(p_{n-m}^{\text{out}})
\end{equation}
can be obtained by means of crossing symmetries which are automatically applied in the \OpenLoops evaluation. 

We assume that the Born-level amplitude $\mathcal{M}_0^{\varphi_{i_1}...\varphi_{i_n}}$  for the process \eqref{process_convention} is not mass-suppressed. That means for large energies 
 $E=\sqrt{s} \gg m_{W}$ we have
\begin{align}
\mathcal{M}_0^{\varphi_{i_1}...\varphi_{i_n}} \sim E^d\,,
\end{align}
with $d$ the mass dimension of the Born matrix element. 
For such processes and in the kinematic region where all external momenta are on-shell,  $p_i^2=m_{\varphi_i}^2$, and all other invariants $r_{jk}$ are much larger than the gauge boson masses, i.e. 
\begin{equation}\label{la}
|r_{jk}|=|(p_j+p_k)^2| \approx 2|p_jp_k| \sim s\gg m_{W}^2,    \hspace{1 cm} j \neq k,
\end{equation}
the one-loop amplitude for the relative $\mathcal{O}(\alpha)$ EW corrections to the    process \eqref{process_convention} 
are of the form~\cite{Kuhn:1999de,Fadin:1999bq,Denner:2000jv,Pozzorini:2001rs}
 \begin{equation}\label{DSLcorrections}
 \begin{aligned}
 \mathcal{M}_1^{\varphi_{i_1}...\varphi_{i_n}} = (\hat\delta^{\rm{DL}} + \hat\delta^{\rm{SL}}) \mathcal{M}_0^{\varphi_{i_1}...\varphi_{i_n}}   \sim \frac{{\alpha}}{4\pi}  (L + l) E^d\,,
\end{aligned}
\end{equation}
where  $\hat\delta^{\rm{DL}}$ and $\hat\delta^{\rm{SL}}$ are the double- and single-logarithmic correction operators, which we will discuss in detail in the following sections \ref{section:dl} and \ref{section:sl} respectively. The factors $L$ and $l$ represent explicit double and single logarithms of the form

\begin{equation}\label{DSLogs}
L(|r_{jk}|,m_i^2):= \log^2{\frac{|r_{jk}|}{m_i^2}}, \hspace{1 cm} l(|r_{jk}|,m_i^2):= \log{\frac{|r_{jk}|}{m_i^2}}.
\end{equation}

In the high-energy limit \eqref{la}, logarithmic corrections of the form Eq.~\eqref{DSLcorrections} are the dominant part of one-loop EW corrections. These logarithmic corrections are universal and factorise with respect to the corresponding Born or SU(2)-flipped Born amplitudes. That means they can be implemented in a process-independent way. 
The form of the EW corrections in Eq.~\eqref{DSLcorrections} defines the LA. 
Not covered by the LA are mass-suppressed logarithmic contributions to the $\mathcal{O}(\alpha)$ EW corrections which are of the form $\alpha m^n_{W} E^{d-n}L$ and $\alpha m^n_{W} E^{d-n}l$  with $n > 0$, as well as any corrections of the form $\alpha E^d$, i.e. terms that are constant relative to the Born matrix element. A necessary condition to apply the LA to a given process is that at least one helicity configuration of the matrix element is not mass suppressed. 

For the evaluation of one-loop EW corrections to longitudinally polarised gauge-bosons $Z_L$  and $W_L^{\pm}$ the algorithm of Denner and Pozzorini employs the Goldstone-boson equivalence theorem (GBET)~\cite{Cornwall:1974km,Gounaris:1986cr,Yao:1988aj,Bagger:1989fc,He:1992nga,He:1993yd,Espriu:1994ep}. That means corrections to longitudinal gauge boson modes are instead computed for the respective Goldstone bosons. Details of our implementation of the GBET will be discussed in Section \ref{sec:goldstoneTH}.

Our notation and conventions for the discussion of the explicit SL and DL terms follow those in~\citere{Denner:2000jv}: chiral fermions and antifermions are denoted respectively by $f_{j,\sigma}^{\kappa}$ and $\bar{f}_{j,\sigma}^{\kappa}$, where $f = Q,L$ corresponds to quarks and leptons, $\kappa=L,R$ specifies the chirality, the index $\sigma=\pm$ stands for the weak isospin and $j=1,2,3$ is the generation index. Gauge bosons are denoted by $V=A, Z, W^{\pm}$ and can be transversely (T) or longitudinally (L) polarised. Employing the GBET longitudinal vector-boson modes are replaced by their respective  
Goldstone bosons $\chi$ and $\phi^{\pm}$. The Goldstone bosons are part of the scalar
doublet $\Phi$ which also contains the physical Higgs boson $H$. 

In the following, we will extensively use the short-hand notations
\begin{equation}\label{ournotation}
\begin{aligned} 
&l^{\text{LSC}, X} \equiv L\left(s, m_X^2\right), \quad  
l^{\text{SSC}} \equiv l\left(|r_{jk}|, s\right), \quad \\
&l^{\text{S-SSC}} \equiv L\left(|r_{jk}|, s\right), \quad
l^{\text{SL}, X} \equiv l\left(s, m_X^2\right), \quad\\
&l^{\text{SL}, A} \equiv l\left(s, \lambda^2 \right), \quad
l^{\text{phase}, X} \equiv l\left(|r_{jk}|, m_X^2\right), \quad \\
&L^{\text{Reg,}\varphi}\equiv\log ^2\left(\frac{m_{\mathrm{\varphi}}^{2}}{\lambda^{2}}\right),\quad
l^{\text{Reg,}\varphi}\equiv\log \left(\frac{m_{\mathrm{\varphi}}^{2}}{\lambda^{2}}\right),\quad\\
&L^{\varphi, X}\equiv\log ^2\left(\frac{m_{\mathrm{\varphi}}^{2}}{m_X^{2}}\right),\quad
\end{aligned}
\end{equation}
with $\varphi$ any charged SM field and $X$ any massive particle $X= Z, W^{\pm}, H, t$~\footnote{By default we only consider the top-quark as massive fermion. Alternative flavour schemes with e.g. $m_b > 0 $ are also supported by our implementation.}.
These logarithms 
depend on the invariants $r_{jk}$ and mass scales $m_X$. 
$\lambda$ is a fictitious photon mass for the regularisation of IR divergencies in the QED logarithms $l^{\text{SL}, A}$ and  $l^{\text{Reg,}\varphi}$. The latter also depends on the mass of a charged fermion $m_{\mathrm{\varphi}}$, which can be a light fermion $f \neq t$. In the following we summarise explicit correction factors for the DL and SL terms employing mass regularisation (MR), i.e. in terms of $\lambda$ and $m_f$. Alternatively to mass regularisation we also consider the translation to conventional dimensional regularisation (DR) for these contributions. 

\subsection{Double Logarithms}
\label{DLtheory}
\label{section:dl}

The double logarithmic corrections (DL) are induced by loop diagrams where a virtual gauge boson $V$ connects two on-shell/external legs, i.e via diagrams of the type
 \begin{equation} \label{DL_diagrams}
     \vcenter{\hbox{\begin{tikzpicture}
    \begin{feynman}
      \vertex (a) at ( 1, 0);
      \vertex (c) at ( -1.5, 1.5);
      \vertex (t) at ( 2, 1);
      \vertex (r) at ( 2, -1);
      \vertex[dot] (k) at ( -0.75, 0.75) {\contour{black}{}};
      \vertex[dot] (z) at ( -0.75, -0.75) {\contour{black}{}};
      \vertex (d) at ( -1.5, -1.5);
        \vertex[draw,circle,minimum size=0.75cm] (q) at ( 0, 0) {\contour{black}{}};
      \diagram* {
    (q)--[edge label=\(\varphi_{{i_k^{\prime}}}\)] (z),
       (z) -- [ edge label=\(\phiphantK\)] (d),
        (q) -- [edge label'=\(\varphi_{{i_j^{\prime}}}\)] (k),
       (k) -- [edge label'=\(\phiphantJ\)]  (c),
       (z) -- [photon, edge label=\(V\)]  (k)
         };
\end{feynman}
  \end{tikzpicture}}}
\end{equation}
where  $\phiphantJ, \phiphantK, \varphi_{i_j^{\prime}} , \varphi_{i^{\prime}_k}$ are SM fields according to the EW Feynman rules. For $Z,\gamma$ exchange 
$\phiphantJK=\varphi_{i_{j/k}^{\prime}}$, while for $W^{\pm}$ exchange $\phiphantJK$ and $\varphi_{i_{j/k}^{\prime}}$ are the respective SU(2) partners.
This one-loop amplitude develops a logarithmic enhancement when integrated over the regime where the virtual gauge boson $V$ becomes soft or collinear to one of the external states.
  
The DL one-loop correction to a LO amplitude $\mathcal{M}_0$ due to the exchange of a vector boson $V$ can be evaluated in the eikonal approximation and is given by~\cite{Denner:2000jv}
\begin{equation} \label{DLiniz}
\begin{aligned}
\hat\delta^{\text{DL}, V}_{\phiphantJ \phiphantK}\mathcal{M}^{\varphi_{i_{1}^{\phantom{\prime}}} \ldots \varphi_{i_{n}^{\phantom{\prime}}}}_0 \overset{\text{LA}}{=} 
\frac{1}{16 \pi^2}
\sum_{\varphi_{i_{j}^{\prime}}, \varphi_{i_{k}^{\prime}}} & 4p_j p_k I_{\varphi_{i_{j}^{\prime}} \phiphantJ}^{V} I_{\varphi_{i_{k}^{\prime}} \phiphantK}^{{V}} C_0^{\text{LA}} \\
&\times \mathcal{M}_{0}^{\varphi_{i_{1}^{\phantom{\prime}}}\ldots \varphi_{i_{j}^{\prime}} \ldots \varphi_{i_{k}^{\prime}} \ldots \varphi_{i_{n}^{\phantom{\prime}}}} \,.
\end{aligned}
\end{equation}
Here,  the $I_{\varphi_{i_{j}^{\prime}} \phiphantJ}^{V}$  
 coefficients denote the scalar part of the tree-level EW vertices $\phiphantJ-\varphi_{i_{j}^{\prime}}-V$, whose expression in LA reads
\begin{equation}
\label{eq:threepointvertices}
 \vcenter{\hbox{
 \begin{tikzpicture}
     \begin{feynman}
      \vertex (a) at ( 1, 0);
      \vertex  (e) at ( 0, 0);
      \vertex (c) at ( -2, 0);
      \vertex (k) at ( -1, 0);
      \vertex (u) at (-1, 0.75);
       
      \diagram* {
             (e) -- [plain, edge label=\(\varphi_{i_{j}^{\prime}}\)]
         (k),
          (k) -- [plain, edge label=\(\phiphantJ\)]
         (c),
                  (k) -- [photon, edge label=\(V\)]
         (u)
         };
    \end{feynman}
  \end{tikzpicture}
  }}  \overset{\text{LA}}{=} 2p^{\mu}_j
I_{\varphi_{i_{j}^{\prime}} \phiphantJ}^{V},
\end{equation}
implicitly encoding a factor $e$, as well as sign factors in case the soft boson is a charged $W$-boson. In the case  of a soft $W$-boson exchange 
 the matrix element on the rhs of Eq.~\eqref{DLiniz} differs from the corresponding Born-level matrix element: the external states $\phiphantJ$ and $\phiphantK$ are replaced by the corresponding SU(2)-flipped states  $\varphi_{i_{j}^{\prime}}$ and $\varphi_{i_{k}^{\prime}}$.
The scalar three-point function  $C_0^{\text{LA}}$ in \eqref{DLiniz} is evaluated in the high energy limit where it explicitly reads~\cite{Roth:1996pd}, 
\begin{equation} \label{C0imaginary} 
\begin{aligned}
C_0^{\text{LA}}&\equiv C_0^{\text{LA}} (p_j ,p_k,m_{V},m_{\varphi_{i_{j}^{\prime}}},m_{\varphi_{i_{k}^{\prime}}}) \\ 
&=\frac{1}{2 r_{jk}}
  \left[ \log ^{2}\frac{|r_{jk}|}{m_{V}^{2}} - 2 i\pi \Theta(r_{jk})\log \frac{|r_{jk}|}{m_{V}^2}   \right],
\end{aligned}
\end{equation}
with  the Heaviside step function denoted as $\Theta$.        

The imaginary term in Eq. \eqref{C0imaginary} arises via analytical continuation and its contribution in the context of EW corrections in the LA has first been pointed out in Ref.~\cite{Pagani:2021vyk}; it was  omitted in Ref.~\cite{Denner:2000jv}. This imaginary term only contributes to tree$\times$loop interference when also the Born amplitude already develops an imaginary part, wich can only occur for $2 \to n$ processes with $n > 2$. We include this imaginary term throughout.

The total DL contribution sums over all possible $V$ exchanges between all possible external legs and reads
\begin{equation} \label{DLtotal}
\hat\delta^{\text{DL}}\mathcal{M}_0^{\varphi_{i_{1}} \ldots \varphi_{i_{n}}}\overset{\text{LA}}{=}
\sum_{j=1}^{n} \sum_{k<j} \sum_{V=A,Z,W^{\pm}} 
\hat\delta_{\varphi_{i_{j}} \varphi_{i_{k}}}^{\mathrm{DL}, V}
\mathcal{M}_0^{\varphi_{i_{1}} \ldots \varphi_{i_{n}}}\,.
\end{equation}

In Section~\ref{sec3} we will present an efficient implementation of this triple sum in \OpenLoops based on two-point correlators of external legs via dedicated Feynman rules for EW Sudakov pseudo-counterterms.
However, before proceeding we want to note that in order to define the DL in strict LA the log-squared term in \eqref{C0imaginary} should be rewritten as

\begin{equation} \label{log}
\begin{aligned}
   & \log ^{2}\left(\frac{\left|r_{jk}\right|}{m_V^{2}}\right) =\\
   & \underbrace{\log ^{2}\left(\frac{s}{m_V^{2}}\right)}_{\propto \quad l^{\text{{LSC}}, V} }+2 \underbrace{\log\left(\frac{s}{m_V^{2}}\right)}_{\propto \quad  l^{\text{{SL}}, V}} \hspace{0.1 cm}\underbrace{\log \left(\frac{\left|r_{jk}\right|}{s}\right)}_{\propto \quad l^\text{{SSC}}}
    +\underbrace{\log ^{2}\left(\frac{\left|r_{jk}\right|}{s}\right)}_{\propto \quad l^\text{{S-SSC}}}\,,
\end{aligned}
\end{equation}
where we employ the short-hand notations from \eqref{ournotation}.

In this form we can isolate the angular-independent leading soft-collinear (LSC) contribution $\propto l^{\text{LSC}, V}$ to the DL correction, and the angular-dependent sub-leading soft-collinear (SSC) contribution $\propto l^{\text{SL}, V} l^\text{{SSC}}$. The last term $\propto l^\text{{S-SSC}}$, which we denote as sub-sub-leading soft-collinear contribution, does not contribute in the strict limit \eqref{la} of the LA and has therefore been omitted in~\cite{Denner:2000jv}. In fact, in the LA there is no control of such angular-dependent terms, which can arise for example also in the high-energy limit of the scalar four-point function~\cite{Roth:1996pd}. 
However, as has been noted and investigated in Ref.~\cite{Pagani:2021vyk}, these \SSSC terms can become numerically relevant for $r_{jk} \gg r_{j'k'}$, i.e. in phase-space regions where not all kinematic invariant are of the order of the center-of-mass energy $s$. Our implementation in \OpenLoops allows to compute the LSC, SSC and S-SSC terms individually in a fully automated and efficient way as discussed in the following, or their sum via direct application of Eq.~\eqref{DLtotal}. In our numerical analysis we investigate if 
the magnitude of the \SSSC terms can be used to estimate the size of $\mathcal{O}(\alpha)$ logarithmically-enhanced corrections beyond the LA. Following Eq. \eqref{log} we define the splitting of the total DL correction as
\begin{equation}\label{DLsplit}
\hat\delta^{\mathrm{DL}} =  \hat\delta^{\mathrm{LSC}}+ \hat\delta^{\mathrm{SSC}}+ \hat\delta^{\mathrm{S-SSC} }.
\end {equation}
In the following we discuss the structure of the individual correction operators $\dLSC$, $\dSSC$, $\dSSSC$.

\subsubsection*{LSC: leading soft-collinear logarithms}

Neglecting mass suppressed contributions the angular - independent LSC corrections can be recasted  in the compact form  
\begin{equation}\label{eq:lsc}
    \hat\delta^{\LSC} \mathcal{M}_0^{\varphi_{i_{1}} \ldots \varphi_{i n}}=\sum_{j=1}^{n} \sum_{\varphi_{i_{j}^{\prime}}} \delta_{\varphi_{i_{j}^{\prime}} \phiphantJ}^{\mathrm{LSC}} \mathcal{M}_{0}^{\varphi_{i_{1}^{\phantom{\prime}}} \ldots \varphi_{i_{j}^{\prime}} \ldots \varphi_{i_{n}^{\phantom{\prime}}}}
\end{equation}
which explicitly indicates the factorisation of LSC correction factors with respect to 
underlying Born matrix elements as a single sum over external legs. The LSC correction factors  $ \delta_{\varphi_{i_{j}^{\prime}}\phiphantJ}^{\mathrm{LSC}}$
are diagonal for all external states $i_{j}$ but for neutral gauge bosons and respective Goldstone bosons. For diagonal correction factors the LSC correction factorises with respect to the original Born matrix element $\mathcal{M}_0^{\varphi_{i_{1}} \ldots \varphi_{i n}}$. For non-diagonal correction factors there is an additional contribution that factorises with respect to the SU(2)-associated Born matrix element $\mathcal{M}_{0}^{\varphi_{i_{1}^{\phantom{\prime}}} \ldots \varphi_{i_{j}^{\prime}} \ldots \varphi_{i_{n}^{\phantom{\prime}}}}$.
The LSC correction factors $ \delta_{\varphi_{i_{j}^{\prime}}\phiphantJ}^{\mathrm{LSC}}$read

 \begin{equation} \label{LSC1}
 \begin{aligned}
     \delta_{\varphi_{i_{j}^{\prime}}\phiphantJ}^{\mathrm{LSC}}=-\frac{1}{32 \pi^2} &\sum_{V=A,Z,W^{\pm}} \sum_{\varphi_{i_{j}''}
     } I^{V}_{\phiphantJ \varphi_{i_{j}''}} I^{V}_{\varphi_{i_{j}''} \varphi_{i_{j}'}} \\  
     &\times\left[l^{\text{LSC}, V} -  \delta_{V,A}  L^{\text{Reg,}\varphi}   \right] 
\end{aligned}
 \end{equation}
where the second term $\sim L^{\text{Reg,}\varphi}$
originates from virtual photon exchange in the regime where these become collinear to external states.

In Eq.~\eqref{LSC1} we refrain from rephrasing the sums in $V$ over the coupling factors $I^{\bar{V}}_{\phiphantJ\varphi_{i_{j}'}}$ in terms of the electroweak Casimir operators as done in Ref.~\cite{Denner:2000jv}. Instead, in our implementation in \OpenLoops we perform explicit sums over the soft vector bosons $V={A,Z,W^{\pm}}$, as discussed in Section~\ref{sec3}. Therefore, in Eq.~\eqref{LSC1} also additional logarithms of the form $\log{m_Z^2/m_W^2}$ are absent. Lastly, the sum over $\varphi_{i_{j}''}$ has to be understood as a sum over all possible internal fields allowed by the EW Feynamn rules, i.e. pictographically this sum can be understood as,

 \begin{equation} 
     \vcenter{\hbox{\begin{tikzpicture}
    \begin{feynman}
      \vertex (a) at ( 0, 0);
      \vertex (o1) at ( -1, 0);
      \vertex (o2) at ( 2, 0);
      \vertex(b) at ( 1, 0);
      \vertex (d) at ( -1.5, -1.5);
       \vertex[draw,circle,minimum size=0.75cm] (q) at ( 2.35, 0) {\contour{black}{}};
      \diagram* {	
       (a) -- [photon, half left, edge label=\(V\)]   (b),
        (a) -- [plain , edge label'=\(\varphi_{{i_j^{\prime \prime}}}\)]   (b),
       (a) -- [plain, edge label=\(\varphi_{{i_j}}\)]   (o1),
       (b) -- [plain, edge label'=\(\varphi_{{i_j^{\prime}}}\)]   (o2)
         };
\end{feynman}
  \end{tikzpicture}}}
\end{equation}
where we have highlighted the $j$th leg, i.e. the blob corresponds to the remaining part of the born matrix element. For external fermions and with a diagonal CKM matrix, as we use throughout, there is only one term in this sum over $\varphi_{i_{j}''}$, however for external $W^{\pm}$ bosons the two neutral states $\gamma, Z$ contribute.

\subsubsection*{QED contributions in mass and dimensional regularisation}\label{IRsingDRMR}

The total QED contribution to Eq.~\eqref{LSC1} can  be identified as
\begin{equation}\label{LSC2}
L^{\mathrm{QED, MR}}_{\text{tot}} \equiv L^{\mathrm{QED, MR}}_{\text{tot}}(s, \lambda^2, m_{\varphi}^2)= l^{\text{LSC}, A} - L^{\text{Reg,}\varphi} \,,
\end{equation}
and  can be either kept in MR as in Eq. \eqref{LSC2}
, or as discussed in the following it can be expressed in dimensional regularisation (DR) as employed in all modern NLO Monte Carlo generators.
Having the coefficients of the DR $1/\epsilon$ poles in LA allows for pole cancellation checks with the universal pole structure~\cite{Catani:1996vz,Catani:2002hc} in the Sudakov limit.
The IR divergencies regularised via the fictitious photon mass $\lambda$ and a light-fermion mass $m_f$ appearing in Eq. \eqref{LSC2} can directly be be translated into the corresponding $1/\epsilon^n$ ($n=1,2$) DR poles~\cite{Basso:2015gca}. The procedure to perform such identification depends on the mass of the emitter particle:

\begin{itemize}
\item massive emitter, $m_{\varphi} \neq 0$: in this case there are only single poles ($1/\epsilon$ in DR and $\ln \left(\lambda^2\right)$ in mass regularisation) and it is sufficient to apply the substitution
\begin{equation}
\ln \left(\lambda^2\right) \rightarrow \frac{C_\epsilon \mu_{\text{IR}}^{2 \epsilon}}{\epsilon}+\mathcal{O}(\epsilon),
\end{equation}
 where $ \mu_{\text{IR}}$ is the dimensional regularisation scale, and
 \begin{equation}
C_\epsilon=(4 \pi)^\epsilon \Gamma(1+\epsilon)=\frac{(4 \pi)^\epsilon}{\Gamma(1-\epsilon)}+\frac{\pi^2}{6} \epsilon^2+\mathcal{O}\left(\epsilon^3\right),
\end{equation}
\item  massless emitter, $m_{\varphi} =0$: in this case mass regularisation will lead to single and double IR divergent logarithms, which are all properly contained in the generic function
 \begin{equation}\label{Lmassreg}
 \begin{aligned}
\mathcal{L}\left(P^2, m_{\varphi} ^2\right) &=\ln \left(\frac{m_{\varphi} ^2}{P^2}\right) \ln \left(\frac{\lambda^2}{P^2}\right)+\ln \left(\frac{\lambda^2}{P^2}\right)\\ 
&-\frac{1}{2} \ln ^2\left(\frac{m_{\varphi} ^2}{P^2}\right)+\frac{1}{2} \ln \left(\frac{m_{\varphi} ^2}{P^2}\right)\,,
\end{aligned}
\end{equation}
where $P^2$ is an appropriate kinematical invariant. To identify the correspondence with DR, one can match these logarithms with the poles appearing in the expression of $\mathcal{L}\left(P^2, m_{\varphi} ^2\right)$ obtained in DR for $m_{\varphi} =0$, which is
 \begin{equation}
 \begin{aligned}
 \mathcal{L}\left(P^2, 0\right)&=\left(\frac{\mu_{\text{IR}}^2}{P^2}\right)^\epsilon C_\epsilon\left(\frac{1}{\epsilon^2}+\frac{3}{2 \epsilon}\right)+2 \\
 &= C_\epsilon \left[\frac{1}{\epsilon^2} + \frac{3/2 + \log \left( \mu_{\text{IR}}^2/P^2\right)}{\epsilon} \right. \\
 &\left.+\frac{3}{2} \log\left( \frac{\mu_{\text{IR}}^2}{P^2}\right)+ \frac{1}{2}\log^2 \left(\frac{\mu_{\text{IR}}^2}{P^2} \right)\right].
 \end{aligned}
\end{equation}
\end{itemize}
We can then define the total QED contribution $L^{\mathrm{QED}}_{\text{tot}} $ as 
 \begin{equation} 
L^{\mathrm{QED}}_{\text{tot}} \equiv L^{\mathrm{QED}}_{\text{tot}} (s,\lambda^2,m_{\varphi}^2) = \begin{cases} 
L^{\mathrm{QED, MR}}_{\text{tot}} 
 & \text { in MR } \\
L^{\mathrm{QED, DR}}_{\text{tot}}   & \text { in DR }\,,
 \end{cases}
\end{equation}
where
\begin{equation} 
 \begin{aligned}
L^{\mathrm{QED, DR}}_{\text{tot}}
&= l^{\text{LSC}, \mu_{\text{IR}}}  \\ 
&+ C_{\varepsilon}\begin{cases} 
 -\frac{1}{\varepsilon ^2}  & \text { for } \quad m_{\varphi }= 0 \\
 L^{\varphi, \mu_{\text{IR}}} - \frac{1}{\varepsilon} \log\left( \frac{m_{\varphi}^2}{s}\right) & \text { for } \quad m_{\varphi} \ne 0 
 \end{cases}
  \end{aligned}
\end{equation}
and $L^{\mathrm{QED, MR}}_{\text{tot}}$ is as defined in Eq. \eqref{LSC2}. Here, a natural choice is to use $\mu_{\text{IR}}=m_W$ to recast the $l^{\text{LSC}, A}$ term in a similar form to $l^{\text{LSC}, W/Z}$ in Eq. \eqref{LSC1}, which effectively corresponds to integrating out QED radiation up to the scale $m_W$.
The conversion from MR to DR for QED contributions has for the finite contributions also been performed in Ref.~\cite{Pagani:2021vyk}
which is equivalent to our formulation with the choice $\lambda=Q=\mu_{\text{IR}}$. Eventually, 
any explicit dependence on $\mu_{\text{IR}}$ will cancel when the virtual one-loop EW amplitude is combined with the corresponding (integrated) real radiation - for example as part of a standard NLO QED subtraction scheme.

We want to note that in our construction of one-loop EW corrections any mass-suppressed contributions are neglected. These mass-suppressed contributions lead to additional IR divergent logarithms suppressed e.g. by $m_{\varphi} ^2/s$. This means that while in the case of massless emitters we perfectly recover the correct IR divergent structure, in the case of massive emitters we only reconstruct the correct IR structure up to $\mathcal{O} (m_{\varphi} ^2/s)$.

\subsubsection*{SSC: subleading soft-collinear logarithms}

The angular dependent SSC contribution coming from Eq. \eqref{log} remains a sum over pairs of external legs,

\begin{equation} \label{SSC1}
\begin{aligned}
  \hat  \delta^{\mathrm{SSC}} \mathcal{M}_0^{\varphi_{i_{1}} \ldots \varphi_{i_{n}}} =&\sum_{j=1}^{n} \sum_{k<j} 
    \sum_{\varphi_{i_{j}^{\prime}}, \varphi_{i_{k}^{\prime}}} \delta_{\varphi_{i_{j}^{\prime}}  \phiphantJ   \varphi_{i_{k}^{\prime}} \phiphantK}^{\mathrm{SSC}} \\ 
    &\times \mathcal{M}_{0}^{\varphi_{i_{1}^{\phantom{\prime}}} \ldots \varphi_{i_{j}^{\prime}} \ldots \varphi_{i_{k}^{\prime}} \ldots \varphi_{i_{n}^{\phantom{\prime}}}}\,,
\end{aligned}
\end{equation}
with
\begin{equation} \label{SSCteo}
\begin{aligned}
\delta_{\varphi_{i^{\prime}_j} \phiphantJ \varphi_{i^{\prime}_{k}}\phiphantK} ^{\mathrm{SSC}}=&\frac{1}{8 \pi^2}\sum_{V=A,Z,W^{\pm}}   I_{\varphi_{i_{j}^{\prime}} \phiphantJ}^{V} I_{\varphi_{i_{k}^{\prime}} \phiphantK}^{V} \\ 
&\times \left[l^{\text{SL}, V}l^{\text{SSC}}  - i\pi \Theta(r_{jk}) l^{\text{phase}, V}\right]\,.
\end{aligned}
\end{equation}
Here the photon couplings are always diagonal, $I_{\varphi_{i^{\prime}} \varphi_{i}}^{A}=\delta_{\varphi_{i^{\prime}} \varphi_{i}} I_{\varphi_{i}}^{A}$, as are the $Z$-couplings except in the neutral scalar sector; on contrary, owing to the non-diagonal matrices $I^{\pm}$, flipped contributions of the SU(2)-transformed Born matrix elements appear on the r.h.s. of \eqref{SSC1} in the case of $V=W^{\pm}$ exchange.
As for the LSC term, we can define a total photon contribution  $l^{\text{QED}}_{\text{tot}}$ which is fixed by the regularisation scheme as
  \begin{align} 
l^{\mathrm{QED}}_{\text{tot}} &\equiv l^{\mathrm{QED}}_{\text{tot}} (r_{jk}, \lambda^2, m_{\varphi}^2)=\\
&= \begin{cases} 
l^{\mathrm{QED, MR}}_{\text{tot}}=  l^{\text{SL}, A}l^{\text{SSC}} & \text { in MR } \\
l^{\mathrm{QED, DR}}_{\text{tot}}   & \text { in DR } \,, 
 \end{cases}
\end{align}
with
\begin{equation}
\begin{aligned}
 l^{\text{QED, DR}}_{\text{tot}} (r_{jk},\lambda^2,m_{\varphi}^2) &= l^{\text{SSC}}  l^{\text{SL}, \mu_{\text{IR}}} \\ 
 &+ C_{\varepsilon} \begin{cases} \frac{1}{\varepsilon } \log \left(\frac{\mu^2}{|r_{jk}|}\right) & \text { for } \quad m_{\varphi }= 0 \\ 
 \frac{1}{\varepsilon} \log\left( \frac{s}{|r_{jk}|}\right) & \text { for } \quad m_{\varphi} \ne 0 \,.
\end{cases}
\end{aligned}
\end{equation}

\subsubsection*{S-SSC: sub-subleading soft-collinear logarithms} \label{SSSCth}

Similarly to the SSC contribution in Eq. \eqref{SSC1}, also the angular dependent S-SSC correction involves a sum over pairs of external legs

\begin{equation} \label{SSSCteo}
\begin{aligned}
\hat    \delta^{\mathrm{S-SSC}} \mathcal{M}_0^{\varphi_{i_{1}} \ldots \varphi_{i_{n}}}=&\sum_{j=1}^{n} \sum_{k<j} 
    \sum_{\varphi_{i_{j}^{\prime}}, \varphi_{i_{k}^{\prime}}} \delta_{\varphi_{i_{j}^{\prime}}  \phiphantJ  \varphi_{i_{k}^{\prime}} \phiphantK}^{\mathrm{S-SSC}}\\ 
    &\times \mathcal{M}_{0}^{\varphi_{i_{1}^{\phantom{\prime}}} \ldots \varphi_{i_{j}^{\prime}} \ldots \varphi_{i_{k}^{\prime}} \ldots \varphi_{i_{n}^{\phantom{\prime}}}},
\end{aligned}
\end{equation}
where the relative correction factors can be obtained from Eq. \eqref{SSCteo} performing the replacement
\begin{equation}
l^{\text{SL}, V}l^{\text{SSC}} \to l^{\text{S-SSC}}\,,
\end{equation}
where we notice that the S-SSC contribution is IR finite.



\subsection{Single Logarithms}
\label{section:sl}

Single-logarithmic (SL) corrections have a triple origin, they arise from 
\begin{itemize}
	\item the UV regime via the renormalisation constants of dimensionless parameters (PR),
	\item soft/collinear regions contained in the wavefunction renormalisation constants (WFRCs),
	\item collinear emission of virtual gauge bosons from external states (COLL).
\end{itemize}
The logarithms associated with the collinear emission\linebreak (COLL) of virtual gauge bosons and  the ones originating in the wavefunction renormalisation constants (WFRCs) factorise with respect to the Born amplitude and together yield a gauge invariant SL correction associated to external states (C). 
In the following subsections we discuss these correction factors individually.
The construction of the PR correction is based on a standard on-shell renormalisation at NLO~EW, which in our case exploits the implementation in \OpenLoops~\cite{Buccioni:2019sur} restricted to the LA. The construction in Ref.~\cite{Buccioni:2019sur} involves the computation of WFRCs as part of the UV counterterms, such that renormalised one-loop amplitudes correspond directly to S-matrix elements and no additional LSZ factors are required. In our implementation we allow for the computation of the WFRCs either as part of the standard on-shell renormalisation procedure (denoted as $\delta^{\mathrm{\overline{PR}}}$) or alternatively via dedicated LSZ factors as part of $\dC=\dWFRC+\delta^{\mathrm{COLL}}$. Both alternatives are numerically identical. Therefore, the total SL correction can be separated in the following three equivalent ways
\begin{equation}
\label{slsplitting}
\hat\delta^{\mathrm{SL}}= \hat\delta^{\mathrm{PR}}+ \hat\delta^{\mathrm{COLL}}+  \hat\delta^{\mathrm{WFRC}}  =  \hat\delta^{\mathrm{\overline{PR}}} + \hat\delta^{\mathrm{COLL}} = \hat\delta^{\mathrm{PR}} +  \hat\delta^{\mathrm{C}}  \,.
\end {equation} 
In our numerical results presented in Section~\ref{sec4} we include the WFRC into $\dC$, i.e. we choose the separation $\hat\delta^{\mathrm{SL}}=\hat\delta^{\mathrm{PR}} +  \hat\delta^{\mathrm{C}}$.

\subsubsection{Parameter Renormalisation}\label{PRtheory}

The UV regime induces single-logarithmic corrections due to the running of renormalised couplings from the renormalisation scale (effectively the EW scale for an on-shell renormalisation in the EW sector) to the hard scale $\sqrtS$. These effects arise via a combination of $\sim\alpha\log\left(s/\mureg^2\right)$ terms from regularised one-loop diagrams with \linebreak $\sim\alpha\log\left(\mureg^2/\muren^2\right)$ terms from the UV counterterms (CTs), where $\mureg$ is the dimensional regularisation scale and $\muren$ the renormalisation scale. In the sum of these terms any \mureg dependence cancels and an overall $\sim\alpha \log \left(s/\muren ^2 \right)$ dependence remains. In order to reconstruct this overall dependence without explicit computation of one-loop diagrams we can identify
\begin{equation}
\label{eq:muregshat}
\mureg \to \muew = \sqrtS
\end{equation}
in the computation of the UV counterterms for any dimensionless coupling parameters.  
We base the computation of these UV counterterms on the EW renormalisation procedure implemented in \OpenLoops~\cite{Buccioni:2019sur} restricting the contributing two-point self-energies to their values in LA. In the following, we revisit the EW renormalisation
in \OpenLoops for the case of real-valued masses starting from
 \begin{equation}\label{RenormalisationRelations}
 \begin{aligned}
m_{i, 0}^2 & =m_i^2+\delta m_i^2\,, \\
g_{i, 0} & =g_i+\delta g_i=\left(1+\delta Z_{g_i}\right) g_i\,,
\end{aligned}
 \end{equation}
where $m_{i, 0}^2, g_{i, 0} $ are the bare masses and couplings, and $\delta m_i^2, \delta Z_{g_i}$ the relative UV CTs. 

\paragraph{Mass counterterms}

For real masses \OpenLoops employs a conventional on-shell renormalisation~\cite{Denner:1991kt}, i.e.
\begin{equation} \label{realmassCT}
\begin{aligned}
\delta m_H^2&=\widetilde \Sigma^H\left(m_H^2\right)\,, \\
\delta m_V^2&=\widetilde\Sigma_{\mathrm{T}}^V\left(m_V^2\right)\,, \\
\delta m_f&=\frac{m_f}{2} \widetilde \Sigma_{\mathrm{LRS}}^f\left(m_f^2\right)\,,
\end{aligned}
\end{equation}
where $\Sigma^H_{\rm{T}}$ is the Higgs-boson self-energy at one-loop, $\Sigma_{\rm{T}}$ denotes the transverse part of the gauge-boson propagator at one-loop, and $\Sigma_{\mathrm{LRS}}$ is the following combination of left-handed (L), right-handed (R) and scalar (S) self-energy contributions for fermions 
\begin{equation}
\Sigma_{\mathrm{LRS}}^f\left(p^2\right)=\Sigma_{\mathrm{L}}^f\left(p^2\right)+\Sigma_{\mathrm{R}}^f\left(p^2\right)+2 \Sigma_{\mathrm{S}}^f\left(p^2\right)\,.
\end{equation}
The $\widetilde{\phantom{x}}$ in Eq. \eqref{realmassCT} indicates that these self-energies are evaluated in LA as detailed in Appendix~\ref{appendixLA}, together with the replacement \ref{eq:muregshat}. 
In the EW SM Yukawa masses and physical fermion masses are related. For the respective
counterterms this implies
\begin{equation}
\frac{\delta \lambda_f}{\lambda_f}=\frac{\delta m_f}{m_f}\, .
\end{equation}

The renormalisation of masses in propagators or in couplings with mass dimension, through LA of two-point functions as in Appendix~\ref{appendixLA},  yields only mass-suppressed corrections which can be then safely neglected for not mass-suppressed processes; therefore, mass CTs contribute only indirectly via Yukawa and gauge coupling CTs. For these reasons, in our implementation we consistently set all two-point mass CTs to zero and we limit to the renormalisation of EW dimensionless couplings.


\paragraph{EW coupling counterterms}

The renormalisation of the EW gauge couplings \eqref{RenormalisationRelations} is implemented through CTs for the QED coupling $e$ and the weak mixing angle $\theta_{\mathrm{w}}$. 
In \OpenLoops the QED coupling $e$ can be defined according to different input schemes, which correspond to different renormalisation conditions and the form of the related counterterm $\delta Z_e$ depends on it. By default in \OpenLoops we use the $G_{\mu}$ input and renormalisation scheme, in which the QED coupling is defined through the Fermi constant by the relation
\begin{equation}
\left.\alpha\right|_{G_\mu}=\frac{\sqrt{2}}{\pi} G_\mu\left|\mu_W^2 \sin ^2 \theta_{\mathrm{w}}\right|.
\end{equation}
The corresponding CT is given by
\begin{equation}
\delta Z_e\vert_{G_\mu} = \delta Z_e\vert_{\alpha(0)} - \frac{1}{2}\operatorname{Re} \Delta r\,,
\end{equation}
where $\Delta r$ represents the radiative corrections to the muon decay~\cite{Denner:1991kt}
and $\delta  Z_e\vert_{\alpha(0)}$ is the counterterm in the $\alpha(0)$ scheme (in this scheme the QED coupling is defined in the Thomson limit with momentum transfer $Q^2\to 0$), which is given by
\begin{equation}
\delta Z_e\vert_{\alpha(0)} = -\frac{1}{2} \operatorname{Re} \left(\delta Z_{AA} + \frac{\sin \theta_{\mathrm{w}}}{\cos \theta_{\mathrm{w}}} \delta Z_{ZA} \right)\,.
\end{equation}
Here $\delta Z_{AA}, \delta Z_{ZA}$ are the WFRCs in the neutral gauge boson sector, as detailed below in Section~\ref{sec:wfrcs}.
For the other available input and renormalisation schemes, which can also be employed for the computation of NLO EW corrections in LA see Ref.~\cite{Buccioni:2019sur}. 
In the computation of these renormalisation constants we retain finite contributions beyond the LA.

The weak mixing angle $\theta_{\mathrm{w}}$ is defined in terms of the weak-boson masses by imposing
\begin{equation}
\cos ^2 \theta_{\mathrm{w}}=\frac{m_W^2}{m_Z^2}
\end{equation}
 to all orders. The resulting mixing angle counterterm reads
\begin{equation}
\frac{\delta \cos ^2 \theta_{\mathrm{w}}}{\cos ^2 \theta_{\mathrm{w}}}=-\frac{\delta \sin ^2 \theta_{\mathrm{w}}}{\cos ^2 \theta_{\mathrm{w}}}=\frac{\delta m_W^2}{m_W^2}-\frac{\delta m_Z^2}{m_Z^2},
\end{equation}
where the mass counterterms are computed as in Eq. \eqref{realmassCT}.

Finally, the renormalisation of the scalar self-couplings receives a contribution from the tadpole renormalisation $\delta t$, which we compute as in \citere{Degrassi:1992ff} consistent with \citere{Buccioni:2019sur} applying the LA according to Eq.~\eqref{eq:azero}. This tadpole renormalisation agrees with the one in Ref.~\cite{Denner:2000jv}.

\subsubsection{Wave-function renormalisation constants}
\label{sec:wfrcs}

In the NLO EW renormalisation procedure in \OpenLoops~\cite{Buccioni:2019sur}, wave-function renormalisation constants $\delta Z_{\varphi_i \varphi_j}$ are defined in a way that one-loop propagators do not mix, and their residues are normalised to one, i.e.
 \begin{equation}\label{RenormalisationRelationsWFRC}
 \begin{aligned}
\varphi_{i, 0} & =\left(1+\frac{1}{2} \delta Z_{\varphi_i \varphi_j}\right) \varphi_j \,,
\end{aligned}
 \end{equation}
where $\varphi_{i, 0}$ are the bare fields. 
For vector and scalar bosons we have for the diagonal WFRCs
\begin{equation}\begin{aligned}
& \delta Z_{A A}=-\operatorname{Re}  \widetilde  \Sigma_T^{\prime A}(0)\,, \quad \quad \quad \quad \delta Z_{Z Z}=-\operatorname{Re} \widetilde  \Sigma_T^{\prime Z Z}\left(m_Z^2\right)\,, \\
& \delta Z_{W W}=-\operatorname{Re} \widetilde  \Sigma_T^{\prime W}\left(m_W^2\right), \quad \quad \delta Z_H=-\operatorname{Re} \widetilde  \Sigma^{\prime H}\left(m_H^2\right)\,, \\
\end{aligned}
\end{equation}
and for the non-diagonal WFRCs associated with $\gamma-Z$ mixing\,,
\begin{equation}\begin{aligned}
&\delta Z_{Z A}  =2 \frac{{\operatorname{Re}} \widetilde \Sigma_{\mathrm{T}}^{A Z}(0)}{m_Z^2}\,, \\
&\delta Z_{A Z} =-2 \frac{{\operatorname{Re}}  \widetilde \Sigma_{\mathrm{T}}^{A Z}\left(m_Z^2\right)}{m_Z^2}\,,
\end{aligned}
\end{equation}
where $\Sigma_{\rm{T}}'$ is the derivative of the transverse part of the self-energy $\Sigma_{\rm{T}}$ evaluated in LA, as summarised in Appendix~\ref{appendixLA}, where we identify $\mureg\to \mu_{\rm EW}=\sqrtS$ as in \eqref{eq:muregshat}.
In case of fermionic fields the explicit form of the corresponding WFRCs depends on their mass, since in the massive case there is a mixture of left- and right-handed helicity states $\lambda=L,R$ which is absent in the massless scenario:
\begin{equation}
\delta Z_{f_\lambda}= \begin{cases}-\operatorname{Re} \widetilde \Sigma_\lambda^{\prime}(0), & \text {for}\,  m_f=0 \\
 -{\operatorname{Re}}\widetilde  \Sigma_\lambda^f\left(m_f^2\right)-m_f^2 {\operatorname{Re}} \widetilde \Sigma_{\mathrm{LRS}}^{\prime f}\left(m_f^2\right) & \text {for}\,  m_f>0\,.\end{cases}
\end{equation}
Based on the explicit expressions of these WFRCs in LA (see Appendix~\ref{appendixLA}) the single-logarithmic correction $\dWFRC$ can be constructed as an LSZ factor, i.e. as a single sum over the external legs

\begin{equation}\label{wfrccorrections}
     \dWFRC \mathcal{M}_0^{\varphi_{i_{1}} \ldots \varphi_{i_{n}}}=\sum_{j=1}^{n} \sum_{\varphi_{i_{j}^{\prime}}}   \delta^{\WFRC}_{\varphi_{i_{j}'} \phiphantJ}  \mathcal{M}_{0}^{\varphi_{i_{1}^{\phantom{\prime}}} \ldots \varphi_{i_{j}^{\prime}} \ldots \varphi_{i_{n}^{\phantom{\prime}}}},
\end{equation}
with the relative WFRC correction factors 
\begin{equation}
  \delta^{\WFRC}_{\varphi_{i_{j}'} \phiphantJ} = \frac{1}{2}\delta Z_{\varphi_{i_{j}'} \phiphantJ}\,,
\end{equation}	
with off-diagonal contributions only for the neutral gauge bosons.

\subsubsection*{Yukawa logarithms} 
In the spirit of Refs.~\cite{Denner:2000jv,Bothmann:2020sxm} in the WFRC
corrections, terms $\sim m_f/m_V$ or $\sim m_f/m_H$ can be separated, where $m_f$ is the mass of a heavy fermion (in particular the top-quark). Such terms appear for external heavy quarks or scalar fields (Higgs or Goldstone bosons), and are denoted as Yukawa (YUK) corrections. 
As for the original $\dWFRC$ corrections these $\dYUK$ corrections can be constructed as LSZ factors via

\begin{equation}\label{yukawacorrections}
     \dYUK \mathcal{M}_0^{\varphi_{i_{1}} \ldots \varphi_{i_{n}}}=\sum_{j=1}^{n} \sum_{\varphi_{i_{j}^{\prime}}}   \delta^{\YUK}_{\varphi_{i_{j}'} \phiphantJ}  \mathcal{M}_{0}^{\varphi_{i_{1}^{\phantom{\prime}}} \ldots \varphi_{i_{j}^{\prime}} \ldots \varphi_{i_{n}^{\phantom{\prime}}}}\,.
\end{equation}

\subsubsection{Collinear logarithms}

A further source of single logarithmic corrections at NLO~EW arises from collinear emission of gauge bosons from external states~\cite{Kinoshita:1962ur}. In Ref.~\cite{Denner:2000jv,Denner:2001gw} such contributions have been extracted in LA via Ward identities in the collinear limit, and they have been found to factorise as collinear correction factors with respect to the Born amplitude in a single sum over the external legs, i.e. they have the same form as an LSZ factor and read 
\begin{equation}
          \hat\delta^{\mathrm{COLL}} \mathcal{M}_0^{\varphi_{i_{1}} \ldots \varphi_{i^{\phantom{\prime}}} }=\sum_{j=1}^{n} \sum_{\varphi_{i_{j}'}}\delta^{\text{COLL}}_{\varphi_{i_{j}'}\phiphantJ} \mathcal{M}_{0}^{\varphi_{i_{1}^{\phantom{\prime}}} \ldots \varphi_{i_{j}^{\prime}} \ldots \varphi_{i_{n}^{\phantom{\prime}}}}
\end{equation}
where  $\delta^{\text{COLL}}_{\varphi_{i_{j}'}\phiphantJ}$ strictly depends on the external states and polarisations and has the general form~\cite{Denner:2001gw}
\begin{equation}
\begin{aligned}
\delta^{\text{COLL}}_{\varphi_{i_{j}'}\phiphantJ} =  \quad  \frac{K_{\varphi_i}}{16\pi^2}
\sum_{V=A,Z,W^{\pm}}\sum_{\varphi_{i_{j}''}} &I^{V}_{\phiphantJ \varphi_{i_{j}''}} I^{{V}}_{\varphi_{i_{j}''} \varphi_{i_{j}'}} \\
&\times\left[ l^{\text{SL}, V} + \delta_{V,A}    l^{\rm{Reg},\varphi} \right],
\end{aligned}
\end{equation}
where $K_{\varphi_i}={1,2}$ respectively for bosons and fermions and the sum over $\varphi_{i_{j}''}$ can be understood as discussed in the context of Eq. \eqref{LSC1}. The $l^{\rm{Reg},\varphi} $ contribution originates from collinear photon radiation regularised via the mass $m_{\varphi}$.

\paragraph{Combined collinear plus WFRC LSZ factors} 

Both the collinear corrections $\hat\delta^{\mathrm{COLL}}$ and the WFRC corrections $\hat\delta^{\mathrm{WFRC}}$ factorise as LSZ factors. The combination of these two corrections is gauge invariant and can be constructed as combined SL correction $\hat\delta^{\mathrm{C}}$ associated to any external leg,
         \begin{equation}\label{contribC}
   \delta^{\mathrm{C}} \mathcal{M}_0^{\varphi_{i_{1}} \ldots \varphi_{i_{n}}}=\sum_{j=1}^{n} \sum_{\varphi_{i_{j}'}}
    \delta^{\text{C}}_{\varphi_{i_{j}'}\phiphantJ}  
\mathcal{M}_{0}^{\varphi_{i_{1}^{\phantom{\prime}}} \ldots \varphi_{i_{j}^{\prime}} \ldots \varphi_{i_{n}^{\phantom{\prime}}}}\,,
       \end{equation}
        where
        \begin{equation}
    \delta^{\text{C}}_{\varphi_{i_{j}'}\phiphantJ} = \left.\left(\delta^{\text{COLL}}_{\varphi_{i_{j}'}\phiphantJ} + \delta^{\text{WFRC}}_{\varphi_{i_{j}'}\phiphantJ} \right)  \right|_{\mureg^2=s} . 
        \end{equation}   
As in Section \ref{IRsingDRMR} the total QED contribution $ l^{\text{QED}}_{\text{C, tot}}$ associated to 
 the $\hat\delta^{\mathrm{C}}$ corrections can be identified in MR and re-expressed in DR as
  \begin{equation} 
  \begin{aligned}
l^{\mathrm{QED}}_{\text{C, tot}} &\equiv l^{\mathrm{QED}}_{\text{C, tot}} (m_{\varphi}^2)\\ 
&=
\begin{cases} 
l^{\mathrm{QED, MR}}_{\text{C, tot}}= l^{\text{SL}, A}  + l^{\rm{Reg},\varphi}  
& \text { in MR }\\
l^{\mathrm{QED, DR}}_{\text{C, tot}}   & \text { in DR }  \,,
 \end{cases}
 \end{aligned}
\end{equation}
where 
\begin{equation} 
l^{\text{QED, DR}}_{\text{C, tot}}= l^{\text{SL}, \mu_{\text{IR}}} +  C_{\varepsilon}\begin{cases} 
-\frac{3}{2\varepsilon} & \text { for } \quad m_{\varphi }= 0 \\
- \frac{1}{\varepsilon}  & \text { for } \quad m_{\varphi} \ne 0\,. 
 \end{cases}
\end{equation}

\section{Implementation in \OpenLoops} \label{sec3}


The standard implementation of one-loop EW corrections in LA in \OpenLoops directly follows the structure presented in Section~\ref{sec2}. This standard implementation will be discussed in the following Section \ref{sec:implementation} and is applicable for hard scatting processes, with $|r_{jk}|\gg m_W^2$. 
In Section \ref{sec:InternalinsTheory} we present an extension of this implementation 
which can be applied to processes with internal resonances and/or processes subject to a separation of scales, e.g. for the computation of processes with VBF/VBS topologies. Representative results based on our implementation will be presented in Section~\ref{Onshell} for hard on-shell processes and in Section~\ref{Offshell} for processes with internal resonances. In Appendix~\ref{app:parameters} we list all relevant \OpenLoops parameters for the evaluation of one-loop EW corrections in LA.

\subsection{Standard implementation}
\label{sec:implementation}

The logarithmic approximation allows to compute one-loop EW corrections for any process which is not mass suppressed, i.e. a process which features at least one helicity configuration that is not mass suppressed for a given phase-space region, in a process-independent way evaluating only tree-level matrix elements. Corrections in this approximation are split into DL contributions \eqref{DLsplit} and SL collinear contributions \eqref{contribC}, which both factorise as as double- or single-sum(s) over the external particles with respect to underlying Born matrix elements, which might differ with respect to the LO Born matrix element by $SU(2)$ flips. The implementation of these corrections in  \OpenLoops will be discussed in the following. Here we also discuss the additional SL contributions of UV origin which can be captured via an evaluation of the $\ord(\alpha)$ parameter renormalisation in logarithmic approximation. Subsequently, in Section~\ref{sec:goldstoneTH} we discuss details on the implementation of the GBET which is used for the evaluation of processes involving longitudinal gauge bosons. We present a summary of the steps we performed to validate our implementation in Section~\ref{sec:validation}.

\subsubsection*{DL and collinear logarithms} 

The implementation of the general soft-collinear DL contribution, Eq. \eqref{DLtotal}, 
and also of the expanded \SSC and \SSSC contributions, Eqs.~\eqref{SSCteo} and~\eqref{SSSCteo} respectively, 
require the construction of the double-sum over all external legs, while summing also over all soft EW vector bosons $V={A,Z,W^\pm}$. These sums involve the evaluation of underlying Born matrix elements of the LO process, as well as of $SU(2)$-flipped tree amplitudes in the case of $W^\pm$-boson exchange. Similarly, the construction of the expanded \LSC contribution, Eq.~\eqref{LSC1}, and of the SL \COLL contribution (collinear plus WFRC), Eq.~\eqref{contribC}, involve single-sums over the external legs with also in general  $SU(2)$-flipped underlying Born matrix elements. We implement these double- and single-sums over external legs including the correct EW couplings via effective EW Sudakov external-leg pseudo-counterterm contributions, which can be automatically generated for any process in \OpenLoops. 

We illustrate this procedure in the following considering the Drell-Yann process $q(k) \bar q(k') \to \ell^+(j)  \ell^-(j')$ and its $\ord(\alpha)$ one-loop correction. One of the topologies which yields DL contributions at high-energies is given by the following vertex correction\footnote{Dirac fermion arrows are implicitly understood.}

 \begin{equation} \label{exampleDrell}
    \vcenter{\hbox{\begin{tikzpicture}
     \begin{feynman}
      \vertex (a) at ( 1, 0);
      \vertex  (e) at ( 0, 0);
       \vertex (k) at ( -0.5, 0.5);
      \vertex (z) at ( -0.5, -0.5);
      \vertex (c) at ( -1, 1);
      \vertex (d) at ( -1, -1);
      \vertex (f) at (2, 1);
      \vertex (g) at (2, -1);
      \diagram* {
        (a) -- [photon,edge label'=\( V^\prime\)]
         (e),
         (e) -- [plain, edge label=\(\bar q^{\prime}\)]
         (z),
          (z) -- [plain, edge label=\(\bar q\)]
         (d),
             (e) -- [plain, edge label' =\(q^{\prime}\)]
         (k),
          (k) -- [plain, edge label'    =\(q\)]
         (c),
         (a) -- [plain, edge label=\( \ell^+\)]
         (f),
             (a) -- [plain, edge label'=\( \ell^-\)]
         (g),
         (z) -- [photon,edge label=\(V\)] 
         (k),
         };
    \end{feynman}
  \end{tikzpicture}}} 
\end{equation}
In the case of a soft boson $V=W^{\pm}$ the internal particles $q^{\prime}$ and $\bar q^{\prime}$ are the $SU(2)$ partners of $q$ and $\bar q$.  
In logarithmic approximation the virtual soft $V$ propagator can be cut and 
together with the internal fermion propagators it can be replaced by the 
correlator between two external-leg pseudo-counterterms:
\begin{equation}
    \vcenter{\hbox{\begin{tikzpicture}
     \begin{feynman}
      \vertex (a) at ( 1, 0);
      \vertex  (e) at ( 0, 0);
       \vertex (k) at ( -0.5, 0.5);
       \vertex (o) at ( -0.7, 0.1);
       \vertex (i) at ( -0.7, -0.1);
      \vertex (z) at ( -0.5, -0.5);
      \vertex (c) at ( -1, 1);
      \vertex (d) at ( -1, -1);
      \vertex (f) at (2, 1);
      \vertex (g) at (2, -1);
      \diagram* {
        (a) -- [photon,edge label'=\( V^\prime\)]
         (e),
         (e) -- [plain, edge label=\(\bar q^{\prime}\)]
         (z),
          (z) -- [plain, edge label=\(\bar q\)]
         (d),
             (e) -- [plain, edge label' =\(q^{\prime}\)]
         (k),
          (k) -- [plain, edge label' =\(q\)]
         (c),
         (a) -- [plain, edge label=\( \ell^+\)]
         (f),
             (a) -- [plain, edge label'=\( \ell^-\)]
         (g),
         (z) -- [photon,edge label=\(V\)] 
         (i),
             (k) -- [photon,edge label'=\(V\)] 
         (o),
         };
    \end{feynman}
  \end{tikzpicture}}} 
     \hspace{0.1 cm} \overset{\text{CT}}{\longrightarrow} 
     \vcenter{\hbox{\begin{tikzpicture}
     \begin{feynman}
      \vertex (a) at ( 1, 0);
      \vertex  (e) at ( 0, 0);
      \vertex (c) at ( -1, 1);
      \vertex (t) at ( 2, 1);
      \vertex (r) at ( 2, -1);
      \vertex[dot] (k) at ( -0.5, 0.5) {\contour{black}{}};
      \vertex[dot] (z) at ( -0.5, -0.5) {\contour{black}{}};
     \vertex (u) at (-0.8, -0.5) {\(V\)}; 
       \vertex (u) at (-0.8, 0.45) {\(V\)};  
      \vertex (d) at ( -1, -1);
      \diagram* {
        (a) -- [photon,edge label'=\( V ^\prime\)]
         (e),
           (e) -- [plain,  insertion={[size=2 pt, style=thick]0.5}, edge label=\(\bar q^{\prime}\)]
         (z),
          (z) -- [plain, edge label=\(\bar q\)]
         (d),
             (e) -- [plain,  insertion={[size=2 pt, style=thick]0.5 }, edge label' =\(q^{\prime}\)]
         (k),
          (k) -- [plain, edge label'=\(q\)]
         (c),
         (a) -- [plain,  edge label=\( \ell^+\)]
         (t),
         (a) -- [plain,  edge label'=\( \ell^-\)]
         (r)
         };
    \end{feynman}
  \end{tikzpicture}}} 
   \end{equation}
Here, the cross on the fermion line indicates that the corresponding propagator is removed.    
In this way it is possible to translate any soft-collinear NLO topology to a corresponding
Born topology with two pseudo-counterterm insertions on the external legs, which remain on-shell. In this construction
effects due to mass differences $m_q \neq m_q'$ are mass suppressed.
In order to automatise this procedure in \OpenLoops, we define suitable effective helicity-dependent two-point vertex rules as 
\begin{equation} \label{eq:effectiveCTrule_fermions}
 \vcenter{\hbox{\begin{tikzpicture}
     \begin{feynman}
      \vertex (a) at ( 1, 0);
      \vertex  (e) at ( 0, 0);
      \vertex (c) at ( -2, 0);
      \vertex (u) at (-1, 0.75);
      \vertex (k) at ( -1, 0);
      
      \diagram* {
             (e) -- [plain, edge label=\(f^{\prime \kappa}\)]
         (k),
          (k) -- [plain, edge label=\(f^{\kappa\phantom{\prime}}\)]
         (c),
         (k) -- [photon, edge label=\(V\)]
         (u)
         };
    \end{feynman} $\hspace{0.5 cm} \longrightarrow \hspace{0.5 cm}$
  \end{tikzpicture}
 $\hspace{0.5 cm} \hspace{0.8 cm}$ 
 \begin{tikzpicture}
     \begin{feynman}
      \vertex (a) at ( 1, 0);
      \vertex  (e) at ( 0, 0);
      \vertex (c) at ( -2, 0);
      \vertex[dot] (k) at ( -1, 0) {\contour{black}{}};
       \vertex (u) at (-1, 0.3) {\(V\)};
       
      \diagram* {
             (e) -- [plain,  insertion={[size=2 pt, style=thick]0.5}, edge label=\(f^{\prime \kappa}\)]
         (k),
          (k) -- [plain, edge label=\(f^{\kappa\phantom{\prime}}\)]
         (c),
         };
    \end{feynman}
  \end{tikzpicture}
  }}
    = 
  ie \mathcal{I}_{f^{\kappa}}^{V} \,,
\end{equation}
where in the case of external fermions these effective counterterms project onto the different chiralities of the external states. 
The $\mathcal{I}_{f^{\kappa}}^{V}$ factors are derived from the corresponding constant terms of the three-point vertices, eq.~\eqref{eq:threepointvertices}, 
as listed in Appendix \ref{CTEffectiverules} for all relevant effective EW Sudakov vertices in the SM. This construction based on \linebreak helicity-dependent effective two-point counterterms allows for a streamlined power-counting and generation of all underlying $SU(2)$-flipped Born matrix elements in \OpenLoops including appropriate projections on the helicities of the external states. 

We employ the same double counterterm insertions to construct the single-sums over external legs required for the computation of the \LSC contributions $\mathcal{M}_1^{{\rm CT, \LSC}}$, Eq.~\eqref{eq:lsc}, and of the collinear SL contribution $\mathcal{M}_1^{{\rm CT, SL}}$, Eq.~\eqref{contribC}. In these cases the two pseudo-counterterm vertices are inserted on the same external line. The chosen logarithmic structure which multiplies a given double-insertion can be selected via the \OpenLoops parameter \texttt{nll\_ewswitch}, as documented in Appendix~\ref{app:parameters}.

Considering again the example of $q \bar q \to \ell^+ \ell^-$ as above, the resulting counterterm amplitude $\mathcal{M}^{\rm CT,DL,q\bar q}_1$ of the considered DL contribution of relative $\ord(\alpha)$ is given by the underlying Born matrix elements times a factor containing the proper logarithmic structure as in Eqs.~\eqref{DLiniz}-\eqref{C0imaginary},
\begin{align}\label{exampleCTs} \nonumber
\mathcal{M}_1^{{\rm CT, DL, q\bar q}} 
&=\frac{\alpha}{4 \pi}  \sum_{V=A,Z,W^{\pm}} \mathcal{I}^V_{q} \mathcal{I}^V_{\bar q } \mathcal{M}_0^{q'\bar q' \to  \ell^+  \ell^-}\\ 
&\quad\times\left[ \log ^{2}\frac{|r_{kk'}|}{m_{V}^{2}} - 2 i\pi \Theta(r_{kk'})\log \frac{|r_{kk'}|}{m_{V}^2}   \right] . 
\end{align}
 For the construction of the entire DL contribution we sum over all counterterm diagrams with two insertions on the external legs yielding $\mathcal{M}_1^{{\rm CT, DL}}$. 
%

This construction of the EW correction factors in LA based on pseudo-counterterms is well suited to support the process-independent implementation of corresponding EW corrections in LA up to NNLO, based on Refs.~\cite{Denner:2003wi,Pozzorini:2004rm,Jantzen:2005xi,Jantzen:2005az,Denner:2006jr,Denner:2008yn}. It allows for a streamlined bookkeeping of the required SU(2)-correlated Born matrix elements with up to four soft vector boson  insertions.

\subsubsection*{PR logarithms}

For the evaluation of PR logarithms we employ the standard \OpenLoops UV counterterm machinery with single counterterm insertions at $\ord(\alpha)$, as detailed in Section~\ref{PRtheory}. We discard any WFRC contributions and mass counterterms. For the evaluation of self-energies in the renormalisation procedure we employ the logarithmic approximation, as detailed in Appendix~\ref{appendixLA}. All $\ord(\alpha)$ renormalisation schemes available for the computation of NLO EW amplitudes in \OpenLoops can also be used in logarithmic approximation and can be selected via the parameter $\texttt{ew\_scheme}$, see Tab.~1 in Ref.~\cite{Buccioni:2019sur}.         

%

\subsubsection*{Tree-loop interference}

The one-loop EW scattering probability in logarithmic approximation is finally given by the interference of the complete counterterm amplitude with two insertions on all external legs, $\mathcal{M}_1^{{\rm CT, DL+\COLL}}$, together with the standard single PR vertex-insertions $\mathcal{M}_1^{{\rm CT, PR}}$ with the Born amplitude,
\begin{align}
\nonumber
\mathcal{W}_{01}^{\rm LA} &= 2 \braket{\mathcal{M}_0|\mathcal{M}_1^{\rm CT, DL+\COLL}+\mathcal{M}_1^{\rm CT, PR}} \\
&= \frac{1}{N_{\rm hcs}} \sum_{\rm hel}\sum_{\rm col} 2\, {\rm Re} \left[\mathcal{M}_0^* \left(\mathcal{M}_1^{\rm CT, DL+\COLL}+\mathcal{M}_1^{\rm CT, PR}\right)\right],
\end{align}
where we include the explicit sums over helicities and colour, and the relevant average factor $N_{\rm hcs}$, see Eq.~(2.2) in Ref.~\cite{Buccioni:2019sur}.

\subsection{Longitudinally polarised gauge bosons} \label{sec:goldstoneTH}

For the evaluation of one-loop EW corrections to longitudinally polarised gauge-bosons $Z_L$  and $W_L^{\pm}$ we employ the GBET. This means for any process with external massive gauge bosons $Z,W^{\pm}$ we restrict the evaluation of the construction described in Section~\ref{sec:implementation} to the transverse modes and automatically add additional processes where each individual massive vector boson is replaced by its goldstone boson, i.e. $W^{\pm} \to \phi^{\pm}$ and $Z \to \chi$ in all combinations. In turn, for these additional processes we apply the construction of the one-loop EW corrections in LA based on single and double counterterm insertions, as described above. 


The GBET strictly only holds for sufficiently large energies $s \gg m_W^2$ whereas it fails in the low energy regime. Nevertheless, we employ the GBET for the evaluation of one-loop EW corrections in LA to longitudinal gauge bosons in the entire phase-space. Therefore, to ensure the appropriate computation of the tree-loop interference amplitude $\mathcal{W}_{01}^{\rm LA}$ we apply the following reweighting
\begin{align}\label{eq:gbetreweight}
	\mathcal{W}_{01}^{\rm LA} = \frac{\mathcal{W}_{00}}{\mathcal{W}_{00}^{\rm GBET}}\, \mathcal{W}_{01}^{\rm LA, GBET} \,,
\end{align}
where $\mathcal{W}_{00}$ is the standard LO scattering probability, while $\mathcal{W}_{00}^{\rm GBET}$
is the LO scattering probability, where the longitudinal modes are evaluated via the GBET. For energies far above the EW scale where the GBET is strictly valid and where EW corrections in LA become phenomenologically relevant this reweighting factor tends to one. 

%

\subsection{Validation}\label{sec:validation}

In order to validate our implementation of the one-loop EW corrections in logarithmic approximation in \OpenLoops we compared predictions at the amplitude-squared level against various analytical results available in the literature~\cite{Denner:2000jv,Denner:2001gw,Pozzorini:2001rs}, and to additional original analytical computations. For these checks we compared the numerical coefficients in front of the logarithms for the individual one-loop EW contributions: $\LSC, \SSC, \COLL, \YUK, \PR$. In the \OpenLoops implementation these coefficients are obtained via a linear fit in the argument of the respective logarithms. We performed such detailed checks for the following processes
\begin{itemize}
\item four-fermion neutral current: $e^+ e^- \to \mu^+ \mu^-$, $e^+ e^- \to \bar{t} t$, $e^+ e^- \to \bar{b}b$
\item $W$-boson pair production: $e^+ e^- \to W^+ W^-$, $\bar{q} q \to W^+ W^-$
\item neutral gauge bosons pair production: $e^+ e^- \to N_{\rm{1}} N_{\rm{2}}$, with $N_{\rm{1,2}}=A,Z$
\item $W^+ N$ production in $\bar{d} u$ annihilations, with $N=A,Z$
\item $\bar{u} u \to H W^+W^-$ 
\end{itemize}
and found excellent agreement for the  coefficients of all contributions.
This comparison has been performed for all possible non mass-suppressed helicity configurations, exploiting the GBET for external gauge bosons with longitudinal polarisation, as discussed in the previous section.
Furthermore, we validated the splitting of the DL correction into \LSC, \SSC and \SSSC terms with respect to the un-expanded construction for all processes above and any further processes studied in Section~\ref{sec4}. We found excellent agreement for all non-mass suppressed helicity configurations of all considered processes.

Finally, we checked our implementation of the EW corrections in logarithmic approximation at the amplitude-squared level against full one-loop results as provided by \OpenLoops. Considering scans in energy and in angles we found no residual logarithmic effects, in phase-space regions where Eq.~\eqref{la} is fulfilled. We document a representative selection of these numerical checks in Appendix~\ref{appendixEnergyScans}.

\subsection{Internal insertions} \label{sec:InternalinsTheory}

As a consequence of the assumption in Eq.~\eqref{la} processes involving 
resonant decays of unstable particles cannot be directly computed based
on the standard implementation described in Section~\ref{sec:implementation}.
In fact, for a process involving the resonant two-body decay of an unstable
particle $X \to i j$ the standard implementation only associates DL and \COLL 
correction factors to the final-state decay products $i,j$. 
However, in the resonant region $s_{ij} \approx m^2_{X}$ the requirement 
Eq.~\eqref{la} is clearly violated. In fact, in this region the particle $X$ 
is nearly on-shell and the EW logarithmic correction factors should be applied
to the hard scattering process with particle $X$ as external state.  This can for
example be achieved in the narrow-width approximation (NWA), where the hard
scattering process for an on-shell particle $X$ is generated first including
EW corrections, and then the decay $X \to i j$ is subsequently added.
Additionally, LO off-shell effects can be retained via a Breit-Wigner smearing, as can 
LO spin correlations. Such a procedure is for instance implemented in \texttt{MadSpin}~\cite{Artoisenet:2012st} 
or via the HDH decay handler in \Sherpa~\cite{Sherpa:2019gpd}. Corresponding simulations
with logarithmic EW corrections in the hard scattering process have for example been presented as part of the recent study Ref.~\cite{Pagani:2023wgc}.   

Here we present a generalisation of the implementation presented
in Section~\ref{sec:implementation}, which allows for a computation of resonant
processes including EW corrections in logarithmic approximation. 
The implementation preserves
non-resonant effects at $\ord(\alpha)$ in the LA, and can
be applied to any process with internal resonances e.g. involving vector-boson 
or top-quark decays. 
The implementation exploits the formulation of the logarithmic approximation in terms of 
pseudo-counter\-term vertices: in addition to the double-insertions of EW pseudo-counterterms 
on all external legs, we also insert such helicity-dependent vertices on all internal lines associated to an unstable particles. Based on kinematic projectors we then select for every phase-space point a certain resonance topology (which corresponds to a combination of internal and external pseudo CT insertions) that 
generates the logarithmic EW correction structure. The kinematic projectors are implemented via
a hit-and-miss strategy in order to preserve unitarity.

Considering the partonic process $q \bar{q} \to \ell^+  \ell^- \ell'^+ \ell'^- $,  in the following we illustrate the structure of such internal insertions for the computation of resonant processes in logarithmic approximation.  At LO a double-resonant topology to the process at hand is given by
\begin{equation}
    \vcenter{\hbox{\begin{tikzpicture}
     \begin{feynman}
      \vertex (a) at ( 0, 0);
      \vertex  (b) at ( 0, -2);
       \vertex (c) at ( -1, 0);
       \vertex (d) at ( -1, -2);
       \vertex (e) at ( 1, -2);
      \vertex (f) at ( 1, 0);
      \vertex (g) at ( 1.5, 0.5);
      \vertex (h) at ( 1.5, -0.5);
      \vertex (i) at ( 1.5, -1.5);
      \vertex (l) at ( 1.5, -2.5);    
       \vertex (lname) at ( 1.7, 0.4) {\(\ell^+\)};   
        \vertex (lbarname) at (1.7, -0.2) {\(\ell^-\)};   
         \vertex (lpname) at ( 1.7, -1.7) {\(\ell'^+\)};   
        \vertex (lpbarname) at ( 1.7, -2.2) {\(\ell'^-\)};   
      \diagram* {
        (a) -- [plain]
         (b),
    (a) -- [plain, edge label'=\(q\)]
         (c),
          (b) -- [plain, edge label=\(\Bar{q}\)]
         (d),
             (b) -- [photon, edge label =\(V'_N\)]
         (e),
          (a) -- [photon, edge label =\(V_N\)]
         (f),
         (f) -- [plain]
         (g),
             (f) -- [plain]
         (h),
         (e) -- [plain]
         (i),
             (e) -- [plain]
         (l),        
 
         };
    \end{feynman}
  \end{tikzpicture}}} \hspace{0.5 cm}
   \end{equation}
where $V_N=\gamma, Z$. In the standard algorithm only the external states generate 
the DL and \COLL corrections, e.g. via the following pseudo CT insertions on the external legs,
\begin{equation}
    \vcenter{\hbox{\begin{tikzpicture}
     \begin{feynman}
      \vertex (a) at ( 0, 0);
      \vertex  (b) at ( 0, -2);
       \vertex (c) at ( -1.5, 0);
       \vertex (d) at ( -1.5, -2);
       \vertex (e) at ( 1, -2);
      \vertex (f) at ( 1, 0);
      \vertex (g) at ( 1.5, 0.5);
      \vertex (h) at ( 1.5, -0.5);
      \vertex (i) at ( 1.5, -1.5);
      \vertex (l) at ( 1.5, -2.5);    
       \vertex (lname) at ( 1.7, 0.4) {\( \ell^+\)};   
        \vertex (lbarname) at (1.7, -0.2) {\( \ell^- \)};   
         \vertex (lpname) at ( 1.7, -1.7) {\( \ell'^+\)};   
        \vertex (lpbarname) at ( 1.7, -2.2) {\( \ell'^- \)};   
            \vertex[dot] (x) at ( -0.75, 0) {\contour{black}{}};
        \vertex[dot] (y) at ( -0.75, -2) {\contour{black}{}};
                     \vertex (qname) at ( -1.2, 0.25) {\(q\)}; 
       \vertex (qbarname) at ( -1.2, -2.25) {\(\Bar{q}\) }; 
      \diagram* {
        (a) -- [plain]
         (b),
    (a) -- [plain, insertion={[size=2 pt, style=thick]-0.75}]
         (c),
          (b) -- [plain, insertion={[size=2 pt, style=thick]-0.75}]
         (d),
             (b) -- [photon, edge label =\(V'_N\)]
         (e),
          (a) -- [photon, edge label =\(V_N\)]
         (f),
         (f) -- [plain]
         (g),
             (f) -- [plain]
         (h),
         (e) -- [plain]
         (i),
             (e) -- [plain]
         (l),        
 
         };
    \end{feynman}
  \end{tikzpicture}}} \hspace{0.5 cm}
  \vcenter{\hbox{\begin{tikzpicture}
     \begin{feynman}
      \vertex (a) at ( 0, 0);
      \vertex  (b) at ( 0, -2);
       \vertex (c) at ( -1.5, 0);
       \vertex (d) at ( -1.5, -2);
       \vertex (e) at ( 1, -2);
      \vertex (f) at ( 1, 0);
      \vertex (g) at ( 1.5, 0.5);
      \vertex (h) at ( 1.5, -0.5);
      \vertex (i) at ( 1.5, -1.5);
      \vertex (l) at ( 1.5, -2.5);    
       \vertex (lname) at ( 1.7, 0.4) {\( \ell^+\)};   
        \vertex (lbarname) at (1.7, -0.2) {\( \ell^- \)};   
         \vertex (lpname) at ( 1.7, -1.7) {\( \ell'^+\)};   
        \vertex (lpbarname) at ( 1.7, -2.2) {\( \ell'^- \)};   
            \vertex[dot] (x) at ( -0.75, 0) {\contour{black}{}};
        \vertex[dot] (y) at ( 1.25, 0.25) {\contour{black}{}};
                     \vertex (qname) at ( -1.2, 0.25) {\(q\)}; 
      \diagram* {
        (a) -- [plain]
         (b),
    (a) -- [plain, insertion={[size=2 pt, style=thick]-0.75}]
         (c),
          (b) -- [plain, edge label=\(\Bar{q}\)]
         (d),
             (b) -- [photon, edge label =\(V'_N\)]
         (e),
          (a) -- [photon, edge label =\(V_N\)]
         (f),
         (f) -- [plain, insertion={[size=2 pt, style=thick]-0.8}]
         (g),
             (f) -- [plain]
         (h),
         (e) -- [plain]
         (i),
             (e) -- [plain]
         (l),        
 
         };
    \end{feynman}
  \end{tikzpicture}}} 
   \end{equation}
Here and in the following for improved legibility the index $V_N$ of the pseudo-CT insertions is understood implicitly.  \\

Allowing for double-insertions also on the internal legs we generate contributions like 
the following,
\begin{equation} \label{diag:internal1}
\vcenter{\hbox{\begin{tikzpicture}
     \begin{feynman}
      \vertex (a) at ( 0, 0);
      \vertex  (b) at ( 0, -2);
       \vertex (c) at ( -1, 0);
       \vertex (d) at ( -1, -2);
       \vertex (e) at ( 1.5, -2);
      \vertex (f) at ( 1.5, 0);
      \vertex (g) at ( 2, 0.5);
      \vertex (h) at ( 2, -0.5);
       \vertex (i) at ( 2, -1.5);
      \vertex (l) at ( 2, -2.5); 
             \vertex (lname) at ( 2.2, 0.4) {\( \ell^+ \)};   
        \vertex (lbarname) at (2.2, -0.2) {\( \ell^- \)};   
         \vertex (lpname) at ( 2.2, -1.7) {\( \ell'^+ \)};   
        \vertex (lpbarname) at ( 2.2, -2.2) {\( \ell'^-\)};   
      \vertex (Zname) at ( 1.2, 0.3) {\(V_N\)};  
      \vertex (Zpname) at ( 1.2, -1.7) {\(V'_N\)}; 
    \vertex[dot] (c1) at ( 0.75, 0) {\contour{black}{}};
        \vertex[dot] (c2) at ( 0.75, -2) {\contour{black}{}};
      \diagram* {
        (a) -- [plain]
         (b),
    (a) -- [plain, edge label'=\(q\)]
         (c),
          (b) -- [plain,  edge label=\(\Bar{q}\)]
         (d),
             (b) -- [photon, insertion={[size=2 pt, style=thick]0.25}]
         (e),
          (a) -- [photon, insertion={[size=2 pt, style=thick]0.25}]
         (f),
         (f) -- [plain]
         (g),
             (f) -- [plain]
         (h),
                  (e) -- [plain]
         (i),
             (e) -- [plain]
         (l),        
         };
    \end{feynman}
  \end{tikzpicture}}} 
     \hspace{0.5 cm}  
    \vcenter{\hbox{\begin{tikzpicture}
     \begin{feynman}
      \vertex (a) at ( 0, 0);
      \vertex  (b) at ( 0, -2);
       \vertex (c) at ( -1, 0);
       \vertex (d) at ( -1, -2);
       \vertex (e) at ( 1.5, -2);
      \vertex (f) at ( 1.5, 0);
      \vertex (g) at ( 2, 0.5);
      \vertex (h) at ( 2,  -0.5);
    \vertex[dot] (x) at ( 0.75, 0) {\contour{black}{}};
        \vertex[dot] (y) at ( -0.5, 0) {\contour{black}{}};
         \vertex (mid) at ( -0.5, 0);
      \vertex (i) at ( 2, -1.5);
      \vertex (l) at ( 2, -2.5);
             \vertex (lname) at ( 2.2, 0.4) {\( \ell^+ \)};   
        \vertex (lbarname) at (2.2, -0.2) {\( \ell^-\)};   
         \vertex (lpname) at ( 2.2, -1.7) {\( \ell'^+ \)};   
        \vertex (lpbarname) at ( 2.2, -2.2) {\( \ell'^-\)};   
            \vertex (Zname) at ( 1.2, 0.3) {\(V_N\)}; 
             \vertex (qname) at ( -0.8, 0.25) {\(q\)}; 
             \vertex (qpname) at ( -0.2, 0.3) {\(q'\)}; 
      \diagram* {
        (a) -- [plain]
         (b),
         (a) -- [plain, insertion={[size=2 pt, style=thick]0.5}]
         (y),
         (y) -- [plain]
         (c),
          (b) -- [plain,  edge label=\(\Bar{q}\)]
         (d),
             (b) -- [photon,,  edge label=\(V'_N\)]
         (e),
          (a) -- [photon, insertion={[size=2 pt, style=thick]0.25}]
         (f),
         (f) -- [plain]
         (g),
             (f) -- [plain]
         (h),
                  (e) -- [plain]
         (i),
             (e) -- [plain]
         (l),        
         };
    \end{feynman}
  \end{tikzpicture}}}
    \end{equation}

\begin{equation}
\label{diag:internal2}
      \vcenter{\hbox{\begin{tikzpicture}
     \begin{feynman}
      \vertex (a) at ( 0, 0);
      \vertex  (b) at ( 0, -2);
       \vertex (c) at ( -1, 0);
       \vertex (d) at ( -1, -2);
       \vertex (e) at ( 1.5, -2);
      \vertex (f) at ( 1.5, 0);
      \vertex (g) at ( 2, 0.5);
      \vertex (h) at ( 2,  -0.5);
             \vertex (lname) at ( 2.2, 0.4) {\( \ell^+ \)};   
        \vertex (lbarname) at (2.2, -0.2) {\( \ell^-\)};   
         \vertex (lpname) at ( 2.2, -1.7) {\( \ell'^+\)};   
        \vertex (lpbarname) at ( 2.2, -2.2) {\( \ell'^-\)};   
    \vertex[dot] (x) at ( 0.75, 0) {\contour{black}{}};
        \vertex[dot] (y) at ( 1.75, -1.75) {\contour{black}{}};
      \vertex (i) at ( 2, -1.5);
      \vertex (l) at ( 2, -2.5);  
            \vertex (Zname) at ( 1.2, 0.3) {\(V_N\)}; 
      \diagram* {
        (a) -- [plain]
         (b),
         (a) -- [plain, edge label'=\(q\)]
         (c),
          (b) -- [plain,  edge label=\(\Bar{q}\)]
         (d),
             (b) -- [photon,  edge label=\(V'_N\)]
         (e),
          (a) -- [photon, insertion={[size=2 pt, style=thick]0.25}]
         (f),
         (f) -- [plain]
         (g),
             (f) -- [plain]
         (h),
            (e) -- [plain, insertion={[size=2 pt, style=thick]0.5}]
         (y),
          (y) -- [plain]
         (i),
             (e) -- [plain]
         (l),
         };
        \end{feynman}
  \end{tikzpicture}}}
     \hspace{0.5 cm} 
     \vcenter{\hbox{\begin{tikzpicture}
     \begin{feynman}
      \vertex (a) at ( 0, 0);
      \vertex  (b) at ( 0, -2);
       \vertex (c) at ( -1, 0);
       \vertex (d) at ( -1, -2);
       \vertex (e) at ( 1.5, -2);
      \vertex (f) at ( 1.5, 0);
      \vertex (g) at ( 2, 0.5);
      \vertex (h) at ( 2,  -0.5);
             \vertex (lname) at ( 2.2, 0.4) {\( \ell^+ \)};   
        \vertex (lbarname) at (2.2, -0.2) {\( \ell^-\)};   
         \vertex (lpname) at ( 2.2, -1.7) {\( \ell'^+\)};   
        \vertex (lpbarname) at ( 2.2, -2.2) {\( \ell'^-\)};   
    \vertex[dot] (x) at ( 0.75, -2) {\contour{black}{}};
        \vertex[dot] (y) at ( 1.75, 0.25) {\contour{black}{}};
      \vertex (i) at ( 2, -1.5);
      \vertex (l) at ( 2, -2.5);  
            \vertex (Zname) at ( 1.2, -1.7) {\(V'_N\)}; 
      \diagram* {
        (a) -- [plain]
         (b),
         (a) -- [plain, edge label'=\(q\)]
         (c),
          (b) -- [plain,  edge label=\(\Bar{q}\)]
         (d),
             (b) -- [photon, insertion={[size=2 pt, style=thick]0.25}]
         (e),
          (a) -- [photon,  edge label=\(V_N\)]
         (f),
         (f) -- [plain, insertion={[size=2 pt, style=thick]0.2}]
         (g),
             (f) -- [plain]
         (h),
            (e) -- [plain]
         (i),
             (e) -- [plain]
         (l),
         };
        \end{feynman}
  \end{tikzpicture}}}
   \end{equation}
As before, the cross denotes, that the corresponding propagator has been removed. 
As for the external ones the internal effective two-point counterterm vertices are helicity-dependent and thus project on the helicity of the combined $\ell^+\ell^-$-current of the external states.
The diagrams in \eqref{diag:internal1} results in \SSC (or \SSSC or full DL) corrections to the hard process $q \bar q \to V_N V_N'$, while the corrections in \eqref{diag:internal2} correspond to \SSC (or \SSSC or full DL) corrections to the hard processes $q \bar q \to V_N \ell'^+ \ell'^- $ and $q \bar q \to  \ell^+  \ell^- V_N'$ respectively. 
 
We map onto the different resonance topologies based on kinematic projectors $P_{X_i}$ for diagrams with any internal insertions on a state $X_i$, and $\bar P_i= 1-P_{X_i}$ for diagrams without any insertions on a state $X_i$ considering all possible resonant states $X_i$ in a process.
For these resonance projectors we employ the following kinematic form 
\begin{equation}\label{projector}
\begin{aligned}
P_{X_i}(k_i)&=\left|\frac{\mu_{X_i}^2 - m_{X_i}^2 w_{\text{rescale}}^2\Gamma_{X_i}^2}{(k_i^2-m_{X_i}^2+iw_{\text{rescale}}\Gamma_{X_i} m_{X_i})^2+\mu_{X_i}^2}\right| \\
&= \begin{cases} 1 & \text { if } k_i^2 \to m_{X_i}^2 \\
0 &  \text { if } k_i^2 \to \infty \end{cases}
\end{aligned}
\end{equation}
where $k_i$ is the momentum of the unstable particle $X_i$ and $\mu^2_{X_i} = m_{X_i}^2 -i\Gamma_{X_i} m_{X_i}$ its complex mass. The shape of these projectors corresponds to normalised Breit-Wigner distributions with effective widths $w_{\text{rescale}} \Gamma_{X_i}$, which determines the resonance regions.  The technical parameter $w_{\text{rescale}}$ should be chosen of order 10 such that 
     the resonance region captures the entire resonance enhancement of the off-shell amplitude. This choice of the projectors ensures correct $\ord(\alpha)$ corrections in the on-shell and off-shell regimes with a smooth transition.
     Any variation of the predictions with $w_{\text{rescale}}$ should be seen as a systematic uncertainty of our method.  
    The parameter $w_{\text{rescale}}$ can be varied via \texttt{nllew\textunderscore rescaling}  as described in Tab.~\ref{tab:par}. 

In order to preserve unitary we apply these projectors via a hit-and-miss strategy, such that for a given phase-space point only one kind of double-insertion resonance structure contributes. In the example above this corresponds to either a $q \bar q \to V_N V_N'$, $q \bar q \to V_N \ell'^+ \ell'^- $, $q \bar q \to  \ell^+  \ell^- V_N'$, or $q \bar q \to  \ell^+  \ell^- \ell'^+ \ell'^-$ hard process. 

In the case of same-flavour final states $\ell =\ell'$ the implementation proceeds in the same manner. Any ambiguity is avoided by reconstructing the appropriate resonant topology via comparison between the invariants of the final states and those of the internal unstable particles. Therefore, projections onto the wrong hard processes $q \bar q \to \ell^+ V_N{''}  \ell^- $ are avoided.
The evaluation of such internal resonance insertions is controlled via the parameter \texttt{nll\_resswitch} as described in Tab. \ref{tab:par}. 

\section{Numerical results} \label{sec4}

\def\setrelwidth{0.45}

In this section we present numerical results of our implementation of EW one-loop corrections in logarithmic approximation in \OpenLoops. Considering a variety of representative processes and distributions we compare the resulting approximate predictions with corresponding exact NLO~EW corrections. In order to focus on the effect of the virtual EW corrections we compare results subject to appropriate infrared subtraction, i.e. for both the approximate results in LA and the exact virtual NLO EW results we regularise infrared divergences in dimensional regularisation and remove the corresponding infrared poles by adding Catani-Seymour's QED $\mathbf{I}_{\rm QED}$-operator~\cite{Catani:1996vz,Catani:2002hc,Dittmaier:1999mb,Dittmaier:2008md,Kallweit:2017khh}, as implemented in \OpenLoops (see Section 3.4 in Ref.~\cite{Buccioni:2019sur}), to the virtual amplitudes. In Ref.~\cite{Kallweit:2015dum} it was found that in kinematic regions dominated by large virtual Sudakov corrections, the agreement of this "VI" approximation with respect to the full NLO EW calculation can be at the small percent level, and that this approach allows for a pragmatic incorporation of EW higher-order corrections into precision QCD simulations including parton-shower matching and multi-jet merging at NLO~QCD, as for example provided via \Sherpa's MEPS@NLO approach~\cite{Hoeche:2012yf}~\footnote{In this approach exclusive QED radiation can be recovered via \Sherpa's resonance-aware implementation of YFS QED radiation~\cite{Yennie:1961ad,Schonherr:2008av,Kallweit:2017khh,Gutschow:2020cug,Flower:2022iew}}~\footnote{Corresponding predictions at MEPS@NLO QCD+EW$_{\rm{VI}}$ level have been investigated for several LHC processes, e.g. in ~\cite{Gutschow:2018tuk,Brauer:2020kfv,Bothmann:2021led}.}.

In the following we denote the results based on the LA in \OpenLoops as NLL~EW and NLL'~EW, where at the DL level the former only includes $\hat\delta^{\mathrm{LSC}}$ and $\hat\delta^{\mathrm{SSC}}$ corrections, while the latter additionally includes $\hat\delta^{\mathrm{S-SSC}}$ corrections, i.e.
\begin{align}
\NLL &= (1+ \hat\delta^{\mathrm{\LSC}} + \hat\delta^{\mathrm{\SSC}} + \hat\delta^{\mathrm{\PR}} +  \hat\delta^{\mathrm{\COLL}}	+ \mathbf{I}_{\rm QED})\, \LO \,,\\
\NLLp &= \NLL + \hat\delta^{\mathrm{\SSSC}}\, \LO \,.
\end{align} 
Here the $\mathbf{I}_{\rm QED}$-operator is exact, i.e.  not in LA. Thus, formally there is a mismatch between the $\mathbf{I}_{\rm QED}$-operator and the bare virtual amplitudes in LA, and the IR poles do not cancel. However, for practical purposes mis-cancellation effects in IR poles or the IR scale $\mu_{\rm IR}$ remain at or below the permil-level for all process and distributions considered in the following. 
Exact $\ord(\alpha)$  EW virtual one-loop corrections are denoted as
\begin{equation}
\NVI=(1+\text{NLO}_{\text{V}} ~ \text{EW} + \mathbf{I}_{\rm QED}) \text{ LO }.
\end{equation}
Differences between the NLL~EW and NLL'~EW predictions might indicate the relevance of angular-dependent sub-sub-leading soft-collinear contributions, which cannot be reliably controlled in LA. 
In fact, in a situation where \SSSC corrections are larger than any differences between the \NLLp and \NVI predictions the \SSSC terms can be seen as reliable approximation of  sub-sub-leading angular dependent contributions beyond the LA. 
We will critically investigate such differences in the following for a set of representative processes.  
On the other hand, in the situation where exact one-loop EW corrections are not known, differences between \NLL and \NLLp might be interpreted as a conservative estimate of uncertainties of the logarithmic approximation.

In Section~\ref{input} we detail our input parameters and numerical setup before presenting results for hard on-shell processes in Section~\ref{Onshell},
and for processes subject to internal resonances in Section~\ref{Offshell}.

\begin{figure*}[tb]
\centering
	\includegraphics[width=\setrelwidth\textwidth]{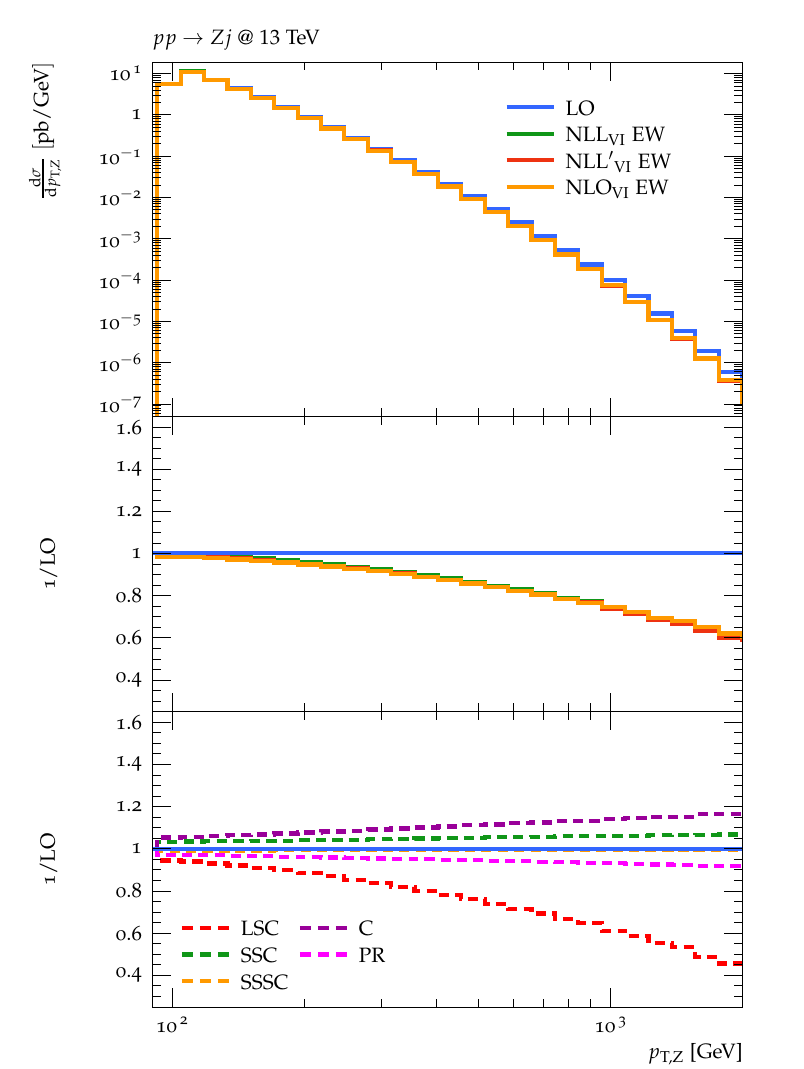}
	\includegraphics[width=\setrelwidth\textwidth]{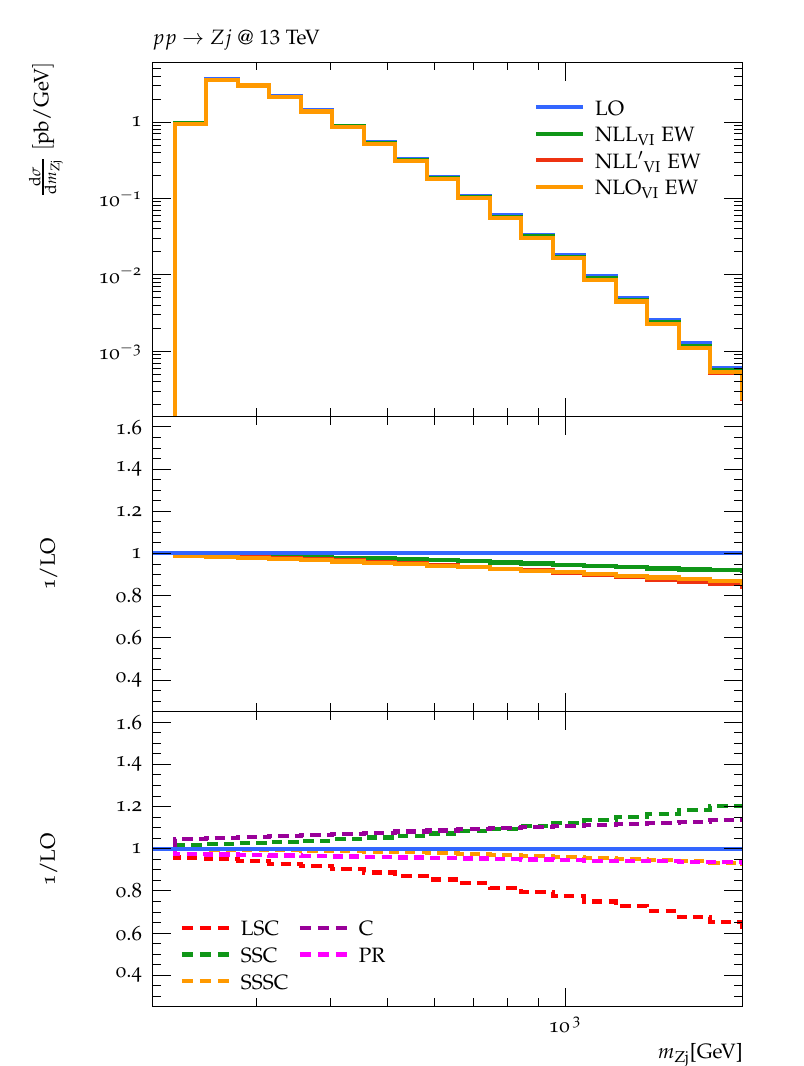}
\caption{Differential distribution in the transverse momentum of the 
$Z$-boson $p_{\rm{T},Z}$ (left) and of the invariant mass of the $Zj$ system $m_{\rm{Zj}}$  (right) 
 in $pp \to Zj$ at $\sqrt{s}=13~\rm{TeV}$. The first panel shows absolute predictions at LO~(blue), \NLL~(green), \NLLp~(red), and \NVI~(orange), where the subscript "VI" indicates that the QED $I$-operator is added to the virtual amplitudes in DR. The second panel shows the relative corrections of the different one-loop EW predictions with respect to LO. The third panel shows the  various contributions at \NLLp in MR, as discussed in Section~\ref{sec2}, normalised to LO:
 \LSC (dashed red),  \SSC (dashed green),  \SSSC (dashed orange), \COLL (dashed purple), \PR (dashed magenta). 
}
\label{fig:zj}
\end{figure*}
\subsection{Input parameters}  \label{input}%

Results for the LHC with $\sqrt{S}=13 \hspace{0.1 cm} \mathrm{TeV}$  presented in the following sections have been obtained in conjunction of \Sherpa~\cite{Sherpa:2019gpd} for the phase-space integration with \OpenLoops as amplitude provider by means of the following input parameters: 
\begin{align}
 m_Z&=91.1876~\mathrm{GeV}\,, \quad m_W=80.399~\mathrm{GeV}\,,  \nonumber\\
 m_H&=125~\mathrm{GeV}, \quad \quad \quad  m_{\mathrm{t}}=173.2~\mathrm{GeV}\,, \nonumber\\
\quad G_\mu &=1.16637 \cdot 10^{-5}  \hspace{0.1 cm}\mathrm{GeV}^{-2}\,. 
\end{align}
All other quarks and leptons are treated as massless. The widths of all particles are set to zero for the evaluation of the on-shell processes in Section~\ref{Onshell}, while for the evaluation of the off-shell processes in Section~\ref{Offshell} we use $\Gamma_Z=2.4952$~GeV and $\Gamma_W=2.085$~GeV. As specified in Eq.~\eqref{eq:muregshat} the scale $\muew$ is set to $\muew = \sqrtS$, where $s$ is the partonic center of mass energy and we use $\mu_{\text{IR}}=100~\mathrm{GeV}$.
We employ the NNPDF3.0 PDF set~\cite{NNPDF:2014otw}~\footnote{
As part of detailed phenomenological analyses including EW higher-order effects 
a PDF set including QED evolution and photon-induced channels should be used.} interfaced via LHAPDF6~\cite{Buckley:2014ana}
and use the provided strong coupling constant $\alpha_S(\mu_{\rm R})$ with $\mu_{\rm R}=\mu_{\rm F}=H_{\rm T}/2$ for the renormalisation scale $\mu_{\rm R}$ and the factorisation scale $\mu_{\rm F}$, where $H_{\rm T}$ is the scalar sum of all final-state transverse momenta. All relative EW corrections shown in the following are independent of this generic scale choice.
Any jets are clustered with the anti-$k_T$ algorithm \cite{Cacciari:2008gp} as implemented in \texttt{FastJet} \cite{Cacciari:2011ma}, with $p_{\rm{T,min}}=30 \hspace{0.1 cm} \text{GeV}$, $R=0.4$, and no rapidity requirement. For any lepton-pair compatible with a $\gamma^*$ we require for the dilepton invariant mass $m_{\ell^+ \ell^-} > 70 \,$GeV. Apart from these jet and lepton requirements if not specified explicitly we consider all processes fully inclusively, i.e. without any phase-space cuts. Partonic events are analysed using \Rivet~\cite{Bierlich:2019rhm}.

\subsection{On-shell processes} \label{Onshell}

In this section we present results including one-loop EW corrections for a selection of important LHC process classes highlighting different features and restrictions of the LA approximation.
Here we consider only on-shell processes, where the standard implementation of the DP algorithm 
is applicable. In particular we consider the following representative process classes in this section: $V$+jets, $VV$+jets, $VVV$, and $t\bar t+X$.

\subsubsection[$V$+jets]{$\mathbf{V}$+jets}

We start our discussion of the numerical results by considering the $V$+jets process category. 
These processes are ubiquitous at the LHC and higher-order EW corrections for the QCD production mode are known with up to three associated jets~\cite{Maina:2004rb,Kuhn:2005gv,Kuhn:2005az,Kuhn:2004em,Kuhn:2005gv,Kuhn:2007qc,Kuhn:2007cv,Denner:2009gj,Denner:2011vu,Denner:2012ts,Denner:2014ina,Kallweit:2014xda,Kallweit:2015dum,Lindert:2017olm}, and for the EW mode with two associated jets~\cite{Lindert:2022ejn}. At large transverse momentum of the vector boson the assumptions of the LA are always fulfilled and the LA approximation is known to reproduce the full NLO EW virtual amplitude at or below the percent level~\cite{Kuhn:2005az,Kuhn:2007qc,Mangano:2016jyj}. We expect the same level of agreement for our process-independent implementation, while larger deviation are expected in the jet transverse momentum distributions for higher-jet multiplicities, and in invariant mass distributions, where at large invariant masses \SSSC terms can become relevant, as \eqref{la} is not necessarily fulfilled.
 
\subsubsection*{$\mathbf{Z+}$jet}

\begin{figure*}[tb]
\centering
	\includegraphics[width=0.4\textwidth]{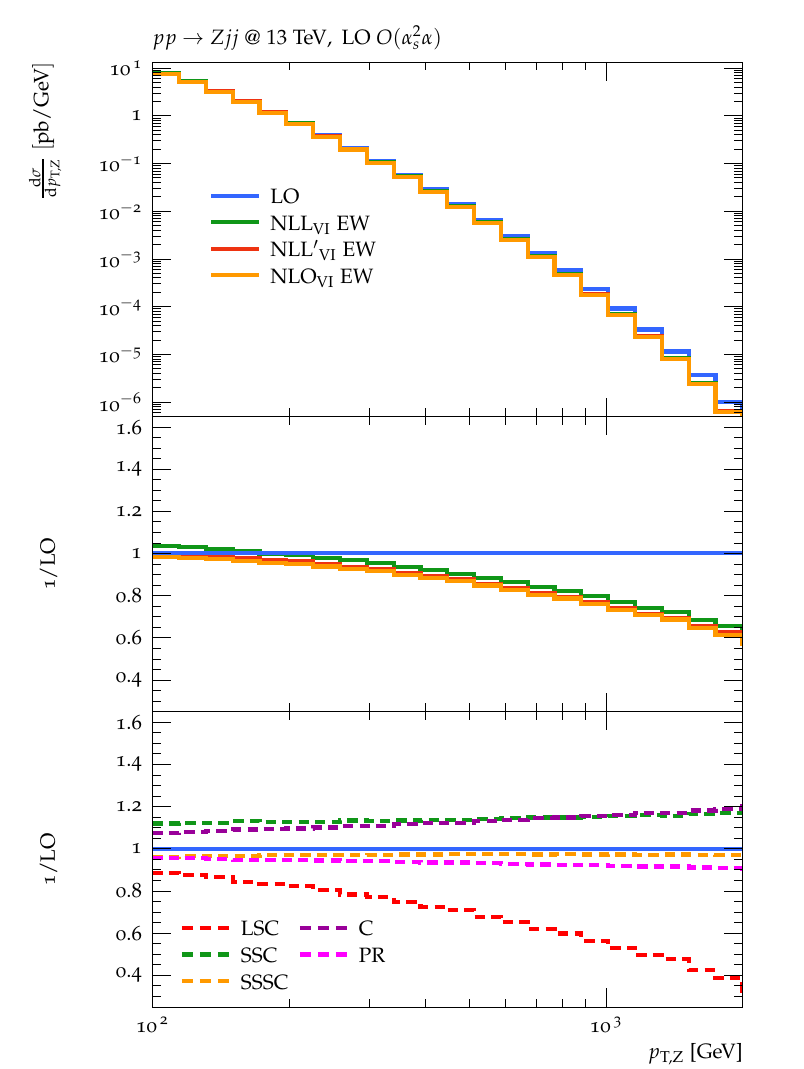}
	\includegraphics[width=0.4\textwidth]{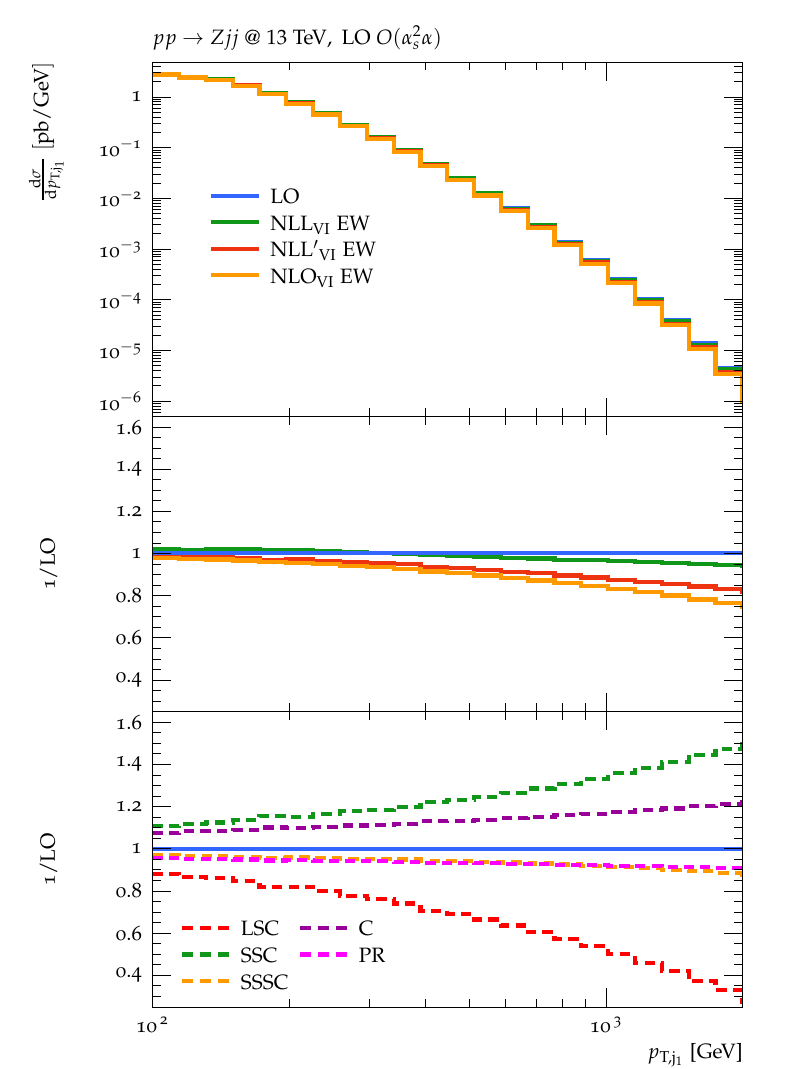}
\caption{Differential distribution in the transverse momentum of the 
$Z$-boson $p_{\rm{T},Z}$ (left) and of the hardest jet $p_{\rm{T},j_1}$  (right) 
 in $pp \to Zjj$ at $\sqrt{s}=13 ~\rm{TeV}$. Curves as in Fig.~\ref{fig:zj}.}
\label{fig:zjj_pT}
\end{figure*}

In Fig.~\ref{fig:zj} we consider differential distributions for $Z$+jet production.
The distribution in the transverse momentum of the $Z$ boson shown on the left is characterised by a clear EW Sudakov-like behaviour: the \LSC correction dominates the EW corrections and the \SSSC corrections is negligible. The \SSC, \PR and \COLL corrections are all of similar size and partly compensate each other. The overall EW corrections reach $-30 \%$ at 
$p_{\rm{T},Z}=1$~TeV with respect to LO, and the \NLL approximation agrees with the full \NVI predictions at the $1\%$ level. Given the negligible size of the \SSSC corrections the \NLL and \NLLp predictions are identical.

In the right plot of Fig.~\ref{fig:zj} we show the invariant mass distribution of the $Zj$ system
in $Z$+jet production. Here the angular-dependent \SSC correction together with the \COLL correction tend to largely compensate the \LSC correction, yielding overall \NLL corrections of only about $-5\%$  at $m_{Zj}=1$~TeV. However, here also the \SSSC correction is relevant indicating a violation of the assumptions of the LA, Eq.~\eqref{la}, yielding itself corrections of about $-5\%$ at $m_{Zj}=1$~TeV.
The resulting combined \NLLp corrections are around $-10\%$ for  $m_{Zj}=1$~TeV and agree at below the percent level with the full \NVI corrections for the entire considered invariant mass range.

\subsubsection*{$\mathbf{Z+2}$~jets}

Next, we consider the EW corrections to hadronic $Z+2$~jets production at $\ord(\alphaS^2 \alpha)$. Differential distributions in $p_{\rm{T},Z}$ and $p_{\rm{T}, j_1}$ are shown in~\reffi{fig:zjj_pT}. In $p_{\rm{T},Z}$ (left) the relative corrections show a very similar picture as for $Z+$~jet production. In fact, the agreement of these relative corrections indicate a large degree of factorisation of QCD and EW effects in this observable. However, in the case of $Z+2$~jets production the \SSSC effects are not completely negligible and yield relative corrections of around $-(3-5)\%$. When these effects are included in the \NLLp prediction there is sub-percent level agreement with the \NVI corrections in the entire considered transverse momentum range.

The \LSC correction to the transverse momentum distribution in the leading jet $p_{\rm{T}, j_1}$ (right) are of the same size as the corresponding corrections in $p_{\rm{T},Z}$. However, these negative \LSC corrections are largely compensated by sizeable positive \SSC and \COLL effects. The negative \SSSC corrections are of the same size as the PR ones and increase up to about $-10\%$ each for $p_{\rm{T}, j_1}=1~$TeV. Summing all these corrections in LA together yields total \NLLp corrections of about $-15\%$ at $p_{\rm{T}, j_1}=1~$TeV differing by about $10\%$ from the \NLL corrections without \SSSC terms. In turn the \NLLp corrections differ from the full \NVI result by about $5\%$ at the TeV scale, indicating angular dependent logarithmic corrections of the same logarithmic order as the \SSSC ones that cannot be controlled in LA. In fact, these differences highlight a violation of the assumption of the LA, i.e. Eq.~\eqref{la}. This violation originates from the fact that at large $p_{\rm{T}, j_1}$ the recoil of the hardest jet can be taken by the second jet, while the $Z$ boson can remain soft. Such configurations yield a separation of scales in violation of Eq.~\eqref{la}.  
Still, the difference between the \NLL and \NLLp predictions overestimates effects beyond the LA.

\begin{figure*}[tb]
\centering
	\includegraphics[width=\setrelwidth\textwidth]{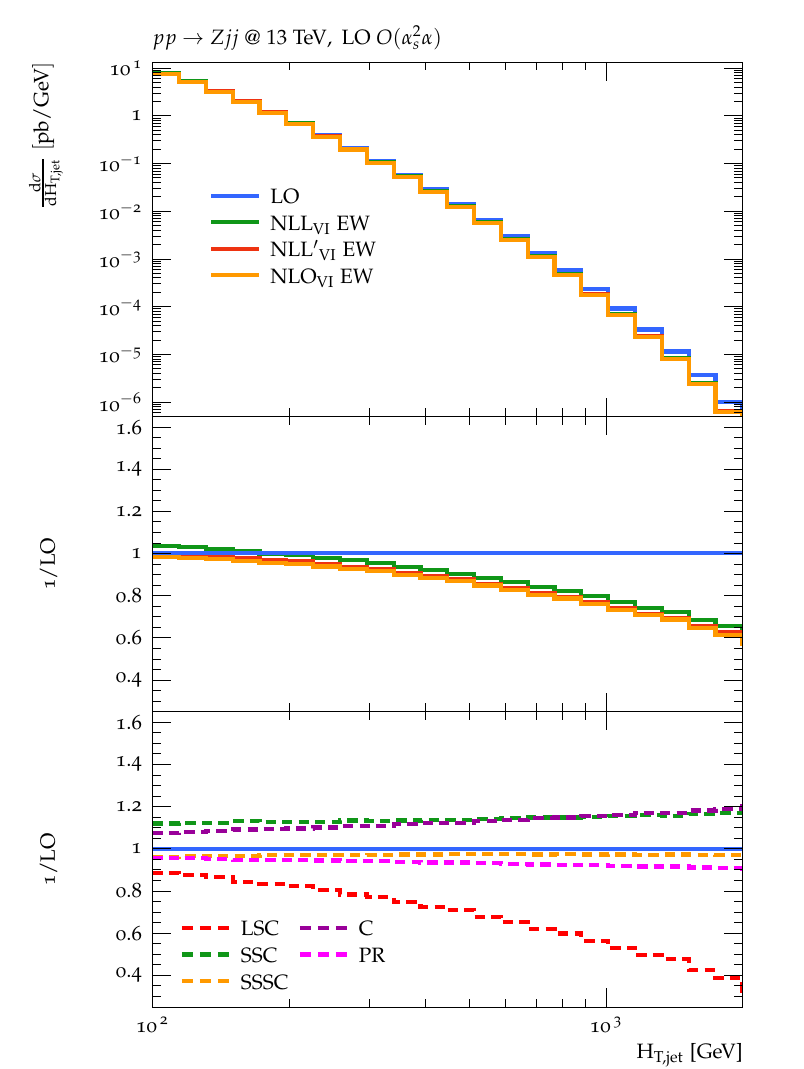}
	\includegraphics[width=\setrelwidth\textwidth]{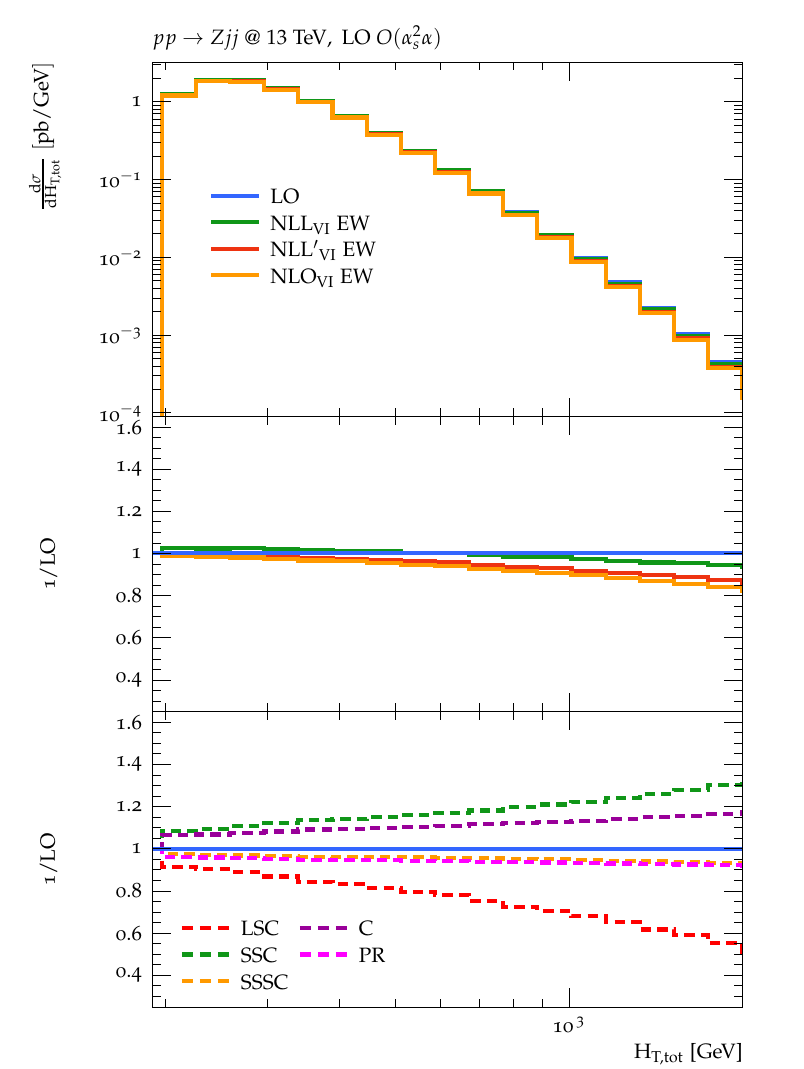}
\caption{Differential $H_{\rm{T}}$ distribution of the 
$jj$ (left) and $Zjj$ systems $H_{\rm{T, jet}}$ and $H_{\rm{T, tot}}$ in $pp \to Zjj$ at $\sqrt{s}=13 ~ \rm{TeV}$. Curves as in Fig.~\ref{fig:zj}.}
\label{fig:zjj_HT}
\end{figure*}

In order to contrast the behaviour of the $p_{\rm{T}, j_1}$ distribution we show differential distributions in $H_{\rm{T, jet}}$, $H_{\rm{T, tot}}$ for $Z+2$~jets production in~\reffi{fig:zjj_HT}, where these observables are defined as
\begin{align}
	H_{\rm{T}, jet}&= p_{\rm{T},j_1} + p_{\rm{T},j_2}\,,\\
H_{\rm{T}, tot}&= H_{\rm{T}, jet}+ E_{\rm{T},Z}\,, \quad \text{with} \quad E_{\rm{T},Z} = \sqrt{p_{\rm{T},Z}^2+M_Z^2}\,.
\end{align}
The $H_{\rm{T}, jet}$ observable is largely correlated to $p_{\rm{T},Z}$ as the entire recoil of the di-jet system is taken by the $Z$-boson. Consequently, relative EW corrections to $H_{\rm{T}, jet}$  are in agreement with the ones observed for $p_{\rm{T},Z}$. On the other hand, the $H_{\rm{T}, tot}$ distribution corresponds to the total sum of all final state momenta and as for the  $p_{\rm{T}, j_1}$ distribution we observe sizeable cancellations between negative \LSC corrections, and positive \SSC and \COLL corrections. For this observable the total \NLLp prediction is in percent-level agreement with the \NVI one with total corrections of about $-10\%$ at $H_{\rm{T}, tot}=1\,$TeV, while here the \NLL prediction only yields around $-5\%$ corrections. In this case considering the difference between 
\NLL and \NLLp as an uncertainty of subleading corrections beyond the LA in a situation where \NVI is not known would overestimate any remaining angular-dependent logarithmic corrections beyond the LA.

\begin{figure*}[tb]
\centering
	\includegraphics[width=\setrelwidth\textwidth]{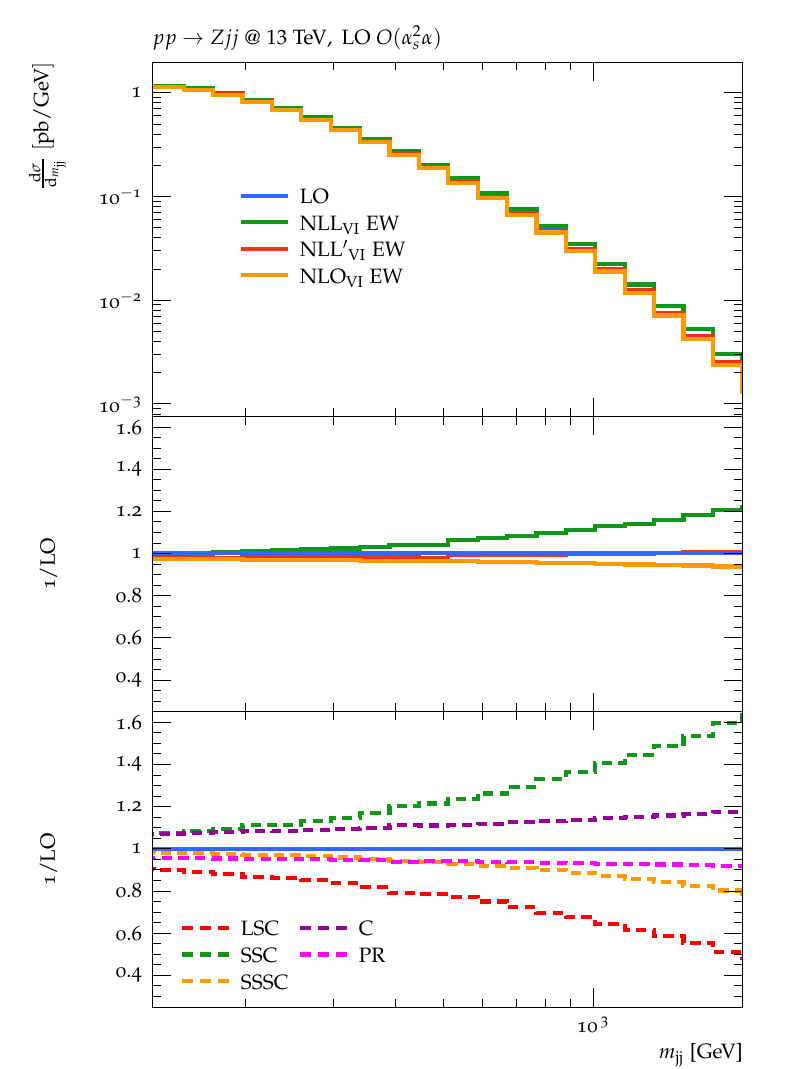}
	\includegraphics[width=\setrelwidth\textwidth]{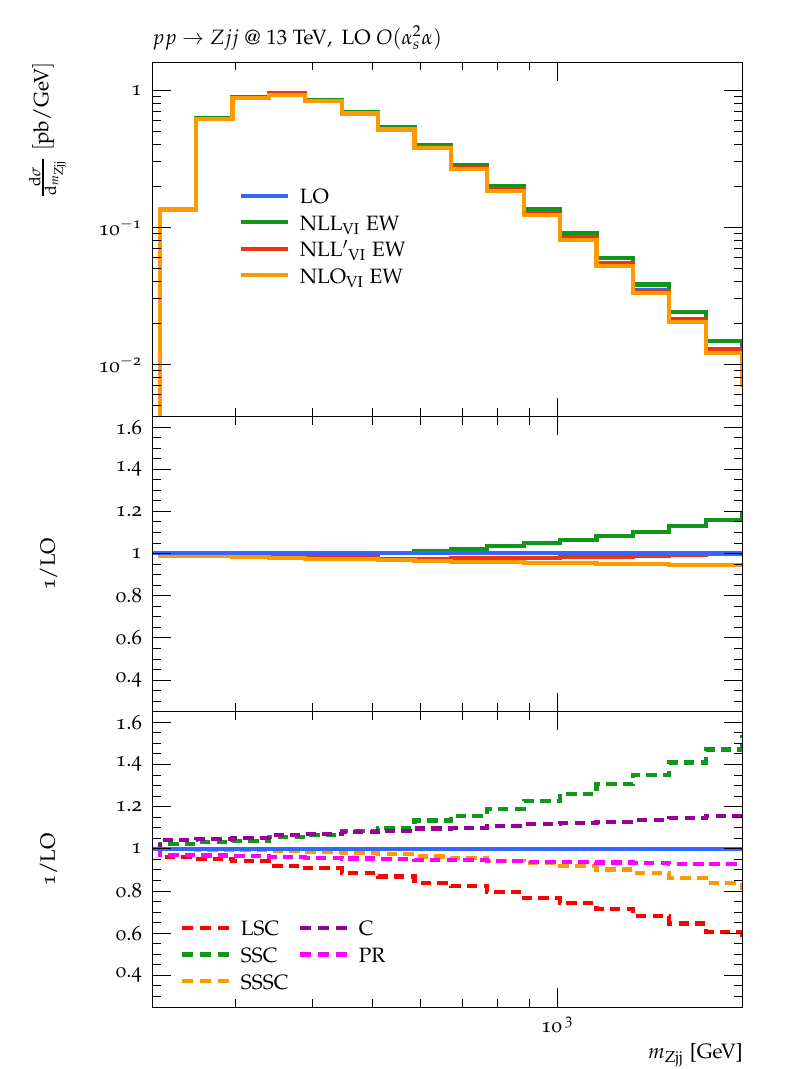}
	\caption{Differential distribution in the invariant mass $m_{\rm{jj}}$ (left) and $m_{\rm{Zjj}}$  
 in $pp \to Zjj$ at $\sqrt{s}=13 ~ \rm{TeV}$ Curves as in Fig.~\ref{fig:zj}.}. 
\label{fig:zjj_mass}
\end{figure*}

In~\reffi{fig:zjj_mass} we turn to invariant mass distributions for $Z+2$~jet production. On the left we show the distribution in the di-jet mass, while on the right we show the invariant mass of the entire $Zj_1j_2$ system. The tails of these distributions indicate a strong violation of the assumptions of the LA. In fact, large invariant mass configurations can be realised via forward configurations in violation of Eq.~\eqref{la}. At large invariant masses the positive \SSC corrections actually dominate over the \LSC corrections resulting in overall positive logarithmically increasing corrections at \NLL reaching $+20\%$ for $m_{jj}=2$~TeV. This picture is significantly altered when considering also the negative \SSSC corrections, which yield overall \NLLp corrections which are close to vanishing for both invariant mass distributions. The exact \NVI corrections also remain small in the entire considered invariant mass ranges only amounting to about $-5\%$ at $m_{jj}=2$~TeV (this can directly be compared with Fig.~5 (left) in Ref.~\cite{Lindert:2022ejn}). Clearly, for invariant mass observables as considered here the inclusion of \SSSC terms is essential for reliable estimates of the one-loop EW corrections. Here, considering the difference between \NLL and \NLLp predictions as an uncertainty in a situation where \NVI is not known would overestimates  differences between \NLLp and \NVI.

\subsubsection[$VV$+jets]{$\mathbf{VV}$+jets} \label{results:ZZjets}

In this subsection we turn to the production of di-boson pairs. To this end we consider on-shell $ZZ$, $W^+W^-$ and $ZZ+$jet production. A discussion of the $ZZ$ production process including off-shell leptonic decays will be presented in Section~\ref{Offshell}. In the case of $W^+W^-$ production a relevant fraction of the cross section originates from longitudinal modes of the $W$-bosons, for which we employ the GBET, as discussed in Section~\ref{sec:goldstoneTH}. 
NLO EW corrections to di-boson processes are known to be large~\cite{Accomando:2004de,Bierweiler:2012kw,Bierweiler:2013dja,Baglio:2013toa,Gieseke:2014gka,Biedermann:2016lvg,Biedermann:2016guo,Biedermann:2017oae,Kallweit:2017khh,Grazzini:2019jkl,Chiesa:2020ttl,Brauer:2020kfv,Bothmann:2021led,Lindert:2022qdd} and their inclusion is mandatory for precision measurements and background predictions at the LHC.
For the case of on-shell $W^+W^-$ production, EW two-loop Sudakov corrections have been
studied up to NNLL accuracy~\cite{Kuhn:2007ca,Kuhn:2011mh}.

\subsubsection*{$\mathbf{ZZ}$}

\begin{figure*}[tb]
\centering
	\includegraphics[width=\setrelwidth\textwidth]{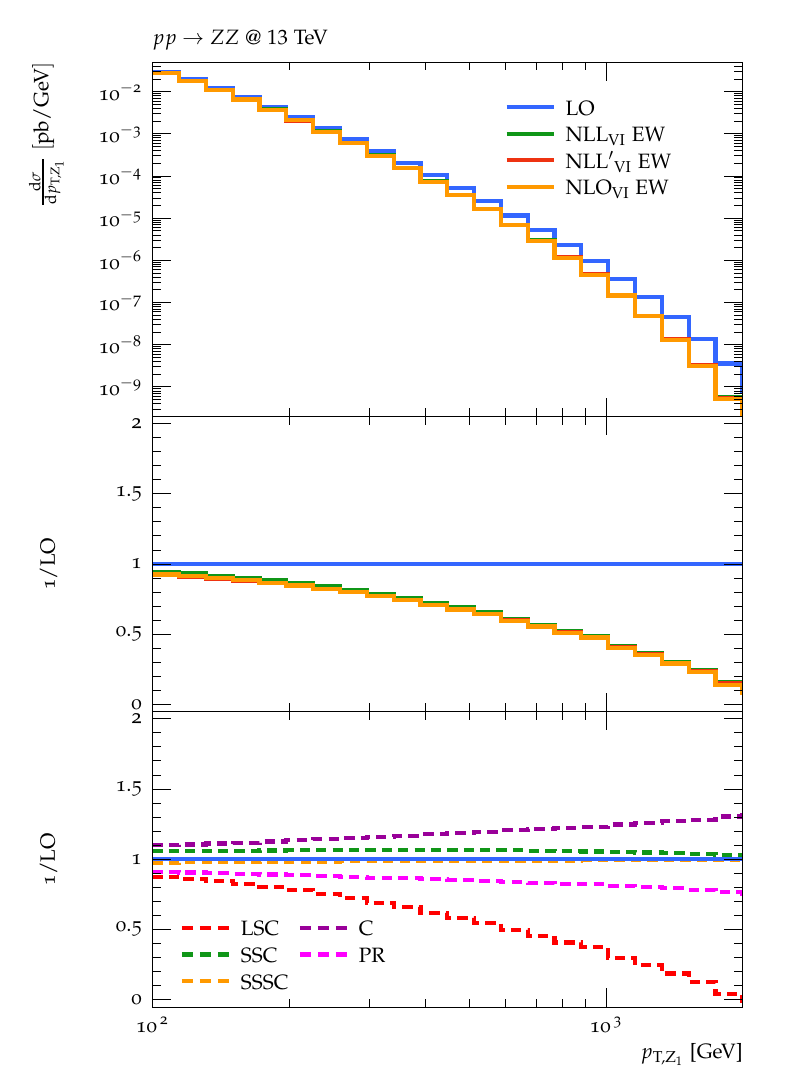}
	   \includegraphics[width=\setrelwidth\textwidth]{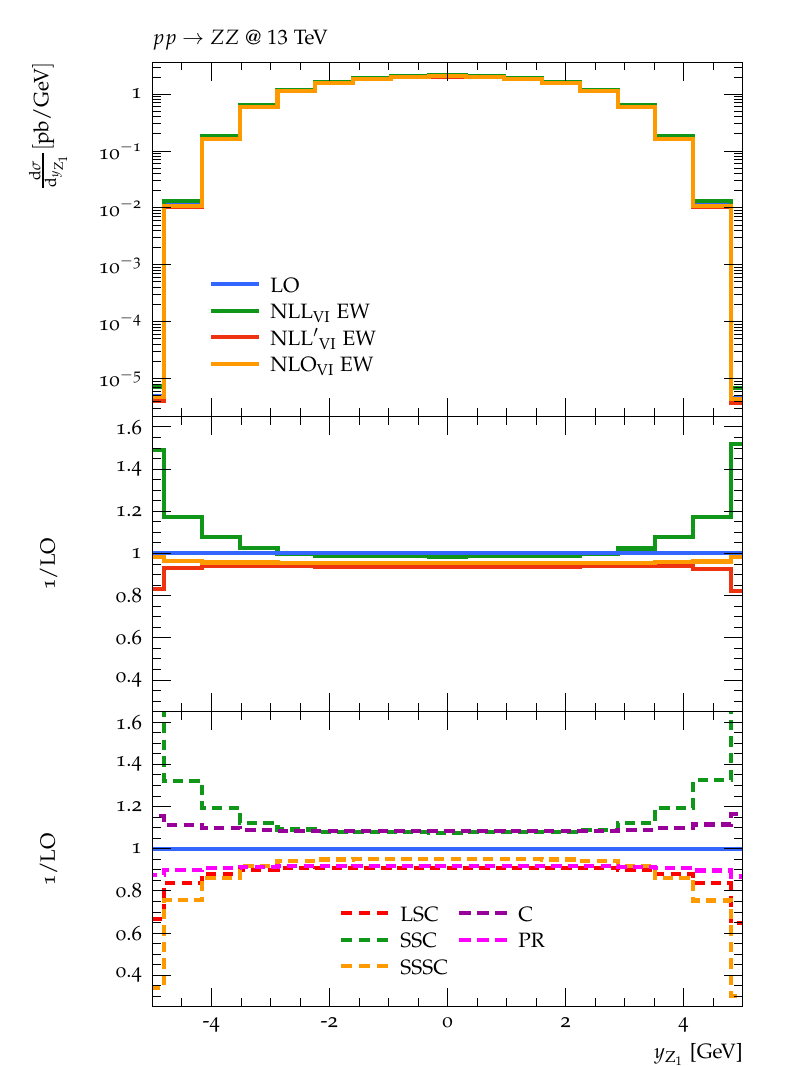}
\caption{Differential distribution in the transverse momentum  $p_{\rm{T},Z_1}$ (left) and of the rapidity $y_{Z_1}$ (right) of the hardest
$Z$-boson in $pp \to ZZ$ at $\sqrt{s}=13 ~ \rm{TeV}$. Curves as in Fig.~\ref{fig:zj}.
 }
\label{fig:zz}
\end{figure*}

In~\reffi{fig:zz} we show the transverse momentum distribution of one of the $Z$-bosons (left), which are back-to-back in this $2\to 2$ scattering configuration,  
and of the $Z$ pseudorapidity distribution (right) for the process $pp \to ZZ$ with on-shell Z-bosons in the fully inclusive phase-space.
At $p_{{\rm T},Z_1}=1$~TeV the total EW corrections yield $-50\%$ effects with all three one-loop predictions agreeing at the one-percent level. 
\LSC effects clearly dominate over all other logarithmic contributions, while \SSC effects are small, \SSSC effects are negligible, and the \COLL and \PR corrections tend to compensate each other. For such large EW corrections as observed here the inclusion of higher-order predictions at NNLO and/or explicit resummation will become necessary. 

When looking at the pseudorapidity shown in the right plot of~\reffi{fig:zz} we observe overall EW corrections in the central region at the level of $5\%$ corresponding to the inclusive EW correction factor. However,  towards large absolute rapidities we see a strong relative increase of the angular-dependent \SSC and \SSSC corrections due to large hierarchies of scales in the forward regions. In this regime eventually the agreement between \NLLp and \NVI starts to deteriorate. In a realistic experimental analysis such phase-space regions are removed due to phase-space cuts on the Z-boson decay products.

\begin{figure*}[tb]
\centering
         \includegraphics[width=\setrelwidth\textwidth]{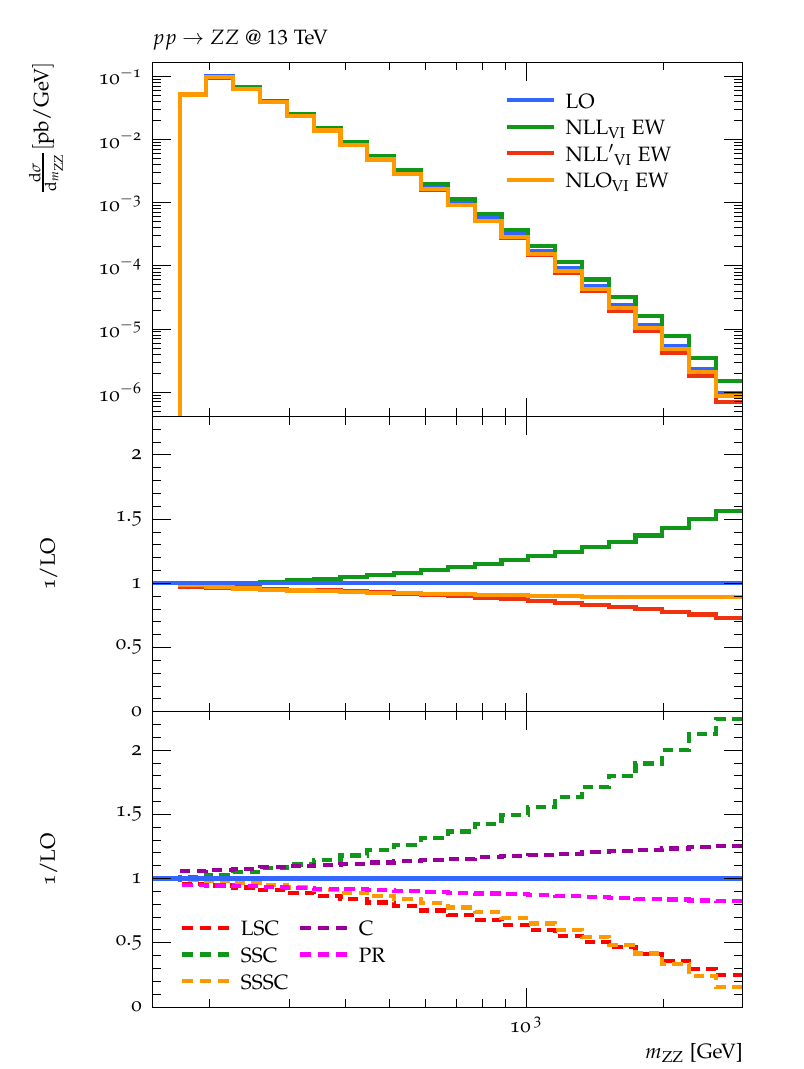}
         \includegraphics[width=\setrelwidth\textwidth]{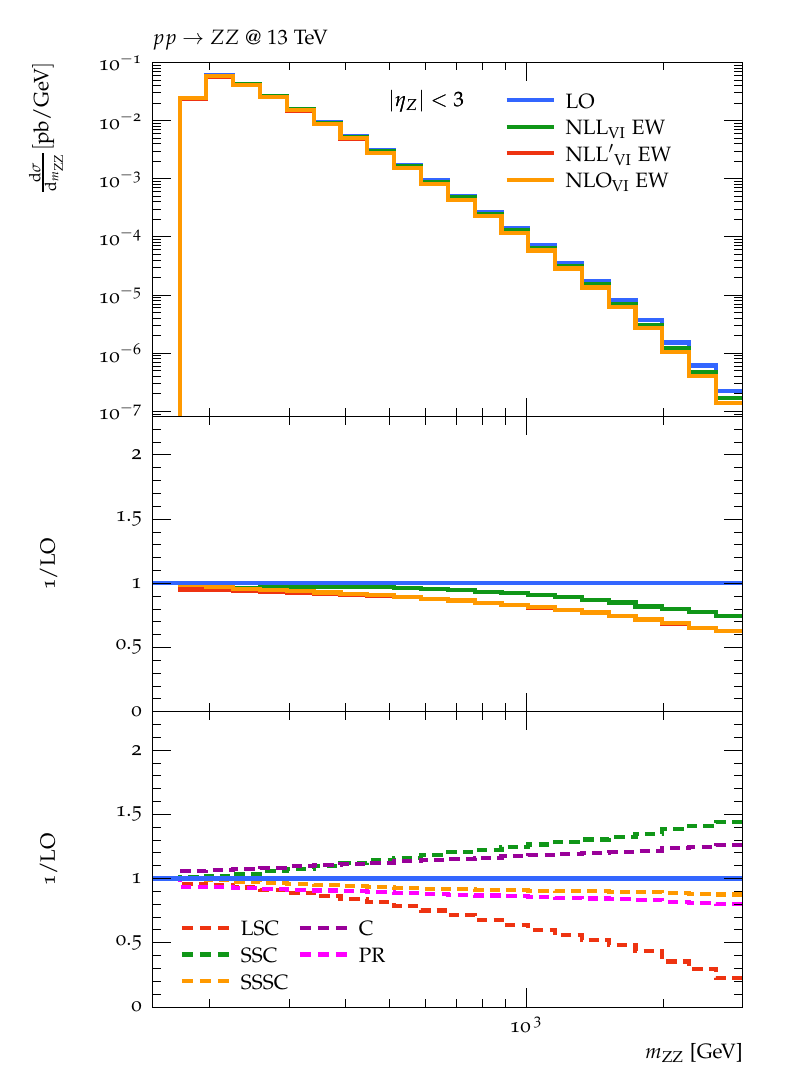}
\caption{Differential distribution in the  invariant mass  $m_{\rm{ZZ}}$  
in the inclusive phase-space (left) and with the requirement $|\eta_Z| < 3$ (right) in $pp \to ZZ$ at $\sqrt{s}=13 ~ \rm{TeV}$. Curves as in Fig.~\ref{fig:zj}.
 }
\label{fig:zz_mzz}
\end{figure*}

An imprint of the effect of large scale hierarchies in the forward region can also be observed in the  inclusive $m_{ZZ}$ distribution 
shown in~\reffi{fig:zz_mzz}~(left): the angular-dependent \SSC correction dominates over the \LSC correction, and the formally subleading \SSSC correction is of the same size as the \LSC one.  Overall, there is a very large cancellation between the different logarithmic effects yielding a total of $-10\%$ at \NLLp respectively $-15\%$ at \NVI for $m_{ZZ}=2$~TeV. The \NLL prediction is unreliable showing large and positive corrections for large $ZZ$ invariant masses. The observed behaviour originates from very forward configurations which introduce very large ratios of invariants, in violation of Eq.~\ref{la}. In any realistic analysis such very forward configurations will be removed via rapidity requirements on the decay products of the $Z$-bosons. In the right plot of~\reffi{fig:zz_mzz} we mimic such requirements by imposing $|\eta_Z|<3$ on the pseudorapidity of the $Z$-bosons.
This cut removes very forward configurations and introduces an indirect constraint on the minimal $m_{ZZ}$. In turn, the relative \SSC and \SSSC corrections in the high-$m_{ZZ}$ tail are largely reduced, while the \LSC correction remains unchanged, resulting in a remarkable sub-percent level agreement between \NLLp and \NVI in the entire considered invariant mass range. Still, the \SSSC correction and therefore the relative difference between \NLL and \NLLp reaches more than $10\%$ for large $m_{ZZ}$. In this case the difference between \NLL and \NLLp would largely overestimate the uncertainty of the \NLLp approximation.

\subsubsection*{$\mathbf{WW}$}

\begin{figure*}[tb]
\centering
\includegraphics[width=\setrelwidth\textwidth]{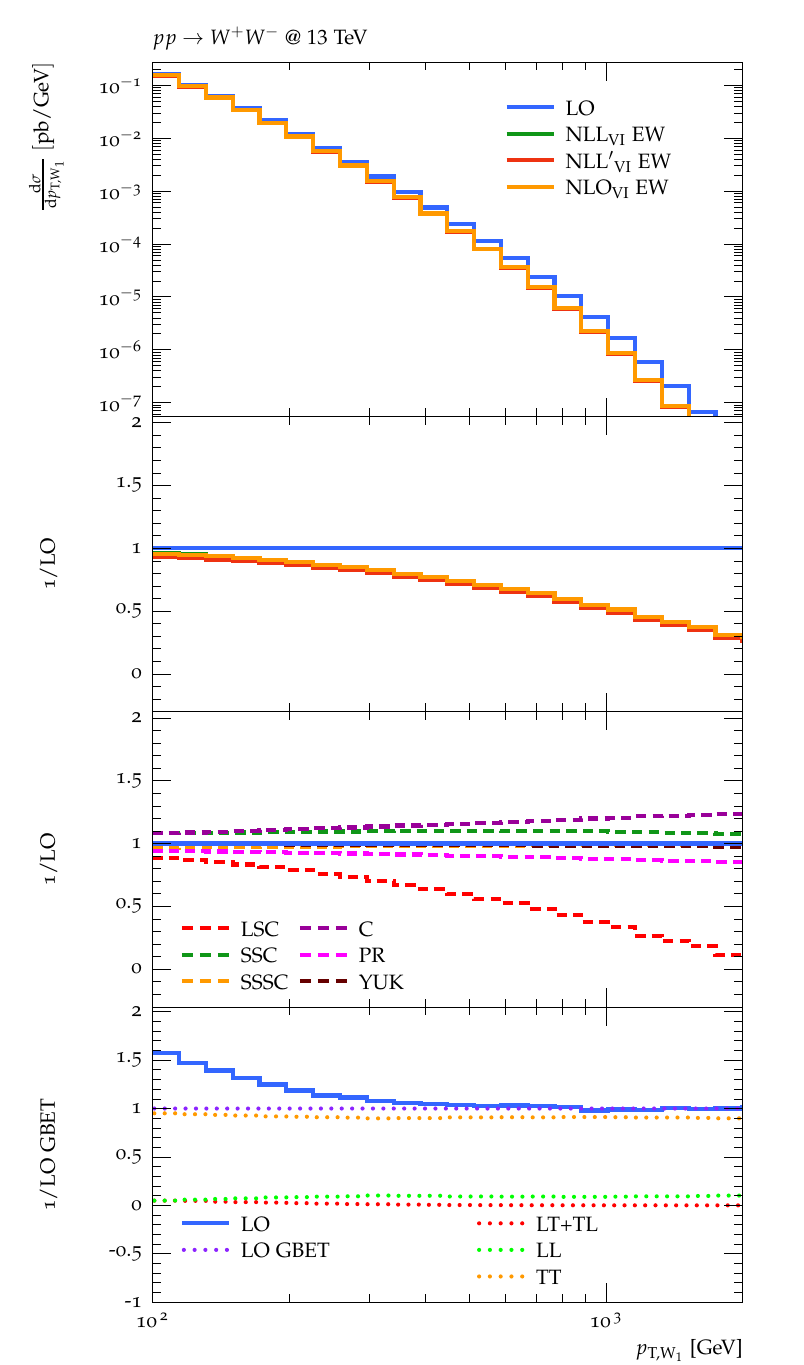}
	\includegraphics[width=\setrelwidth\textwidth]{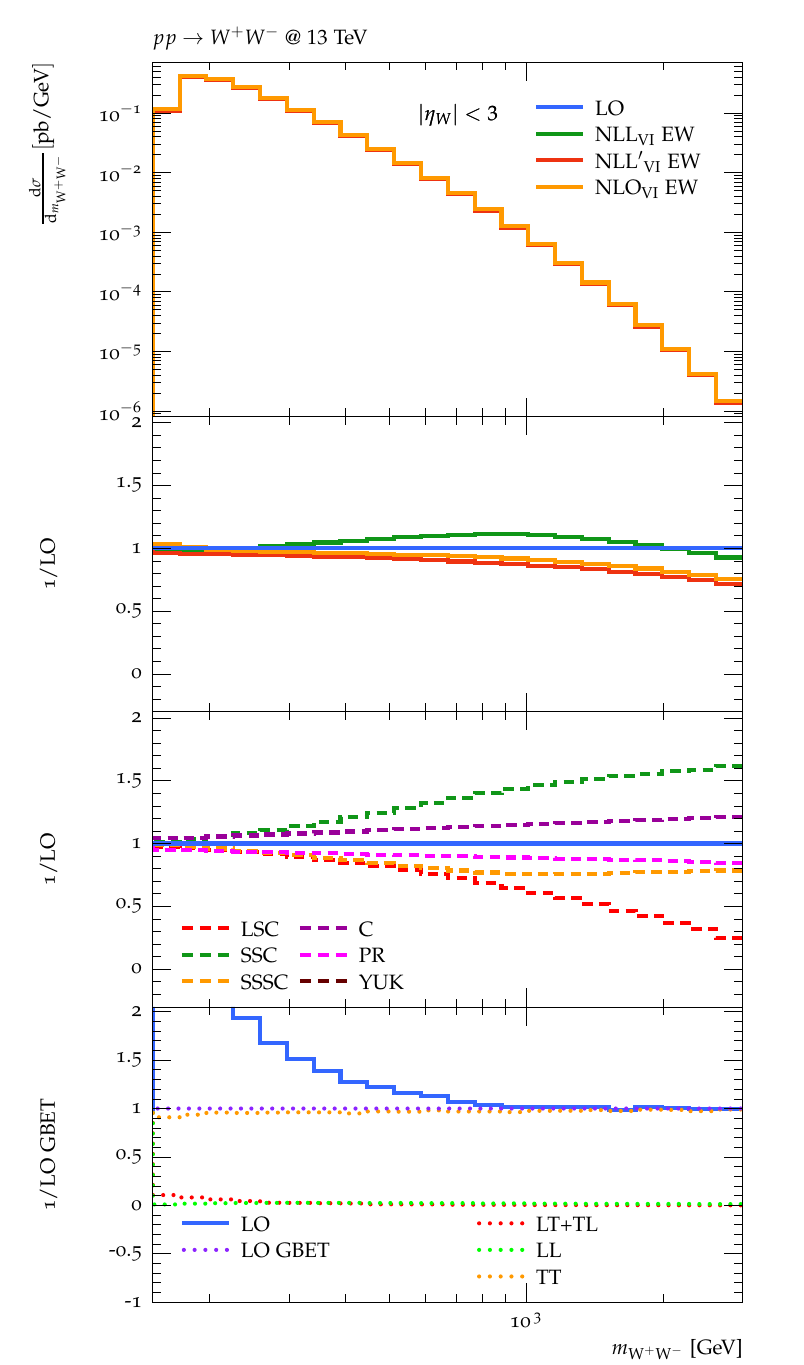}
\caption{Differential distribution in the inclusive transverse momentum $p_{\rm{T},W_1}$ (left) and of the invariant mass of the $WW$ system $m_{\rm{WW}}$ with $|\eta_W| < 3$  (right)
 in $pp \to W^+W^-$ at $\sqrt{s}=13 ~ \rm{TeV}$. Curves of the first three panels are as in Fig.~\ref{fig:zj}. In the third panel the different GBET polarisation fractions TT=$W^+_{\rm T}W^-_{\rm T}$ (dotted-orange), LL=$\phi^+\phi^-$ (dotted-green), LT+TL=$\phi^+W^-_{\rm T}+W_{\rm T}^+\phi^-$ (dotted-red) are shown with respect to the unpolarised LO GBET differential cross sections together with the physical LO (blue). 
 }
\label{fig:ww}
\end{figure*}

In~\reffi{fig:ww} we show the transverse momentum distribution of one of the back-to-back $W^{\pm}$-bosons (left), and the invariant mass distribution $m_{W^+W^-}$ with $|\eta_{W^{\pm}}|<3$ (right) for the process $pp \to W^+W^-$ with on-shell W-bosons.
At $p_{{\rm T},W_1}=1$~TeV the one-loop EW corrections yield $-50\%$ effects with a percent-level agreement among all the three considered one-loop predictions, qualitatively in-line to what we discussed above for  ~\reffi{fig:zz}; in the same way, condition \eqref{la} is clearly fulfilled and each single contribution behaves likewise.
In the  $m_{W^+W^-}$ invariant mass distribution we observe large angular-dependent \SSC corrections despite the pseudorapidity cut $|\eta_{W^{\pm}}|<3$. Nevertheless, this pseudorapidity cut removes pathological very forward phase-space regions and results in an agreement between \NLLp and \NVI at the one-percent level.  
Differences between \NLL and \NLLp reach up to $10-15\%$.
For both observables we show a third ratio plot to explicitly showcase the treatment of longitudinally polarised gauge bosons via the GBET. For on-shell $pp\to W^+ W^-$ the longitudinal-longitudinal (LL) and opposite transverse-transverse (TT) polarisations are not mass-suppressed.
For both observables in~\reffi{fig:ww} the TT configuration dominates, in particular in the invariant mass distribution where it saturates the unpolarised LO amplitude obtained via the GBET, whereas the LL configuration contributes about $10\%$ of the cross section for $p_{{\rm T},W_1}>300\,$GeV. 
However, a comparison against the physical LO highlights an underestimation of the LT, TL plus LL polarisation fractions in the low energy regime. The GBET is not expected to reliably approximate longitudinally polarised gauge bosons by means of the corresponding Goldstone partners in this regime. We checked that in this low-energy phase-space regime the reweighting introduced in Section \ref{sec:goldstoneTH} corrects the \NLLp prediction by up to $3\%$. At large energies where the one-loop EW corrections become  of significant size the GBET accurately reproduces the physical predictions and e.g. for $p_{{\rm T},W_1}>500$~GeV the reweighting factor Eq.~\eqref{eq:gbetreweight} agrees with unity at the permil level.


\subsubsection*{$\mathbf{ZZ+}$jet}

\begin{figure*}[tb]
\centering
       	\includegraphics[width=\setrelwidth\textwidth]{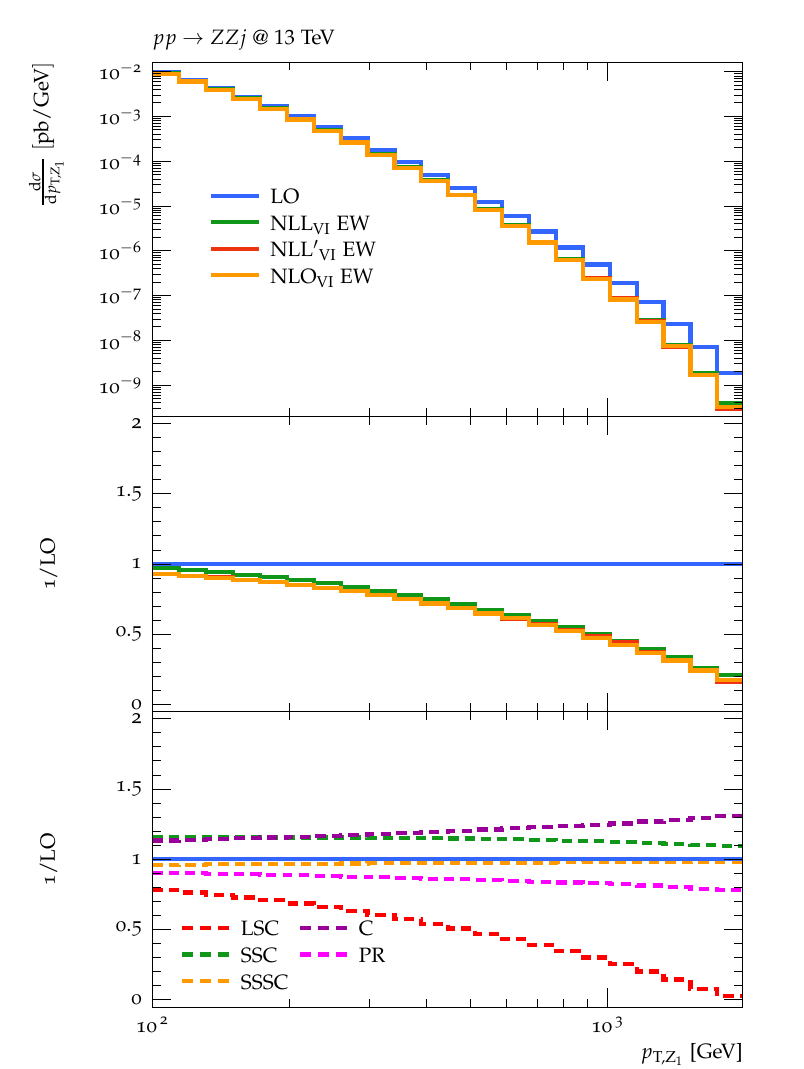}
	\includegraphics[width=\setrelwidth\textwidth]{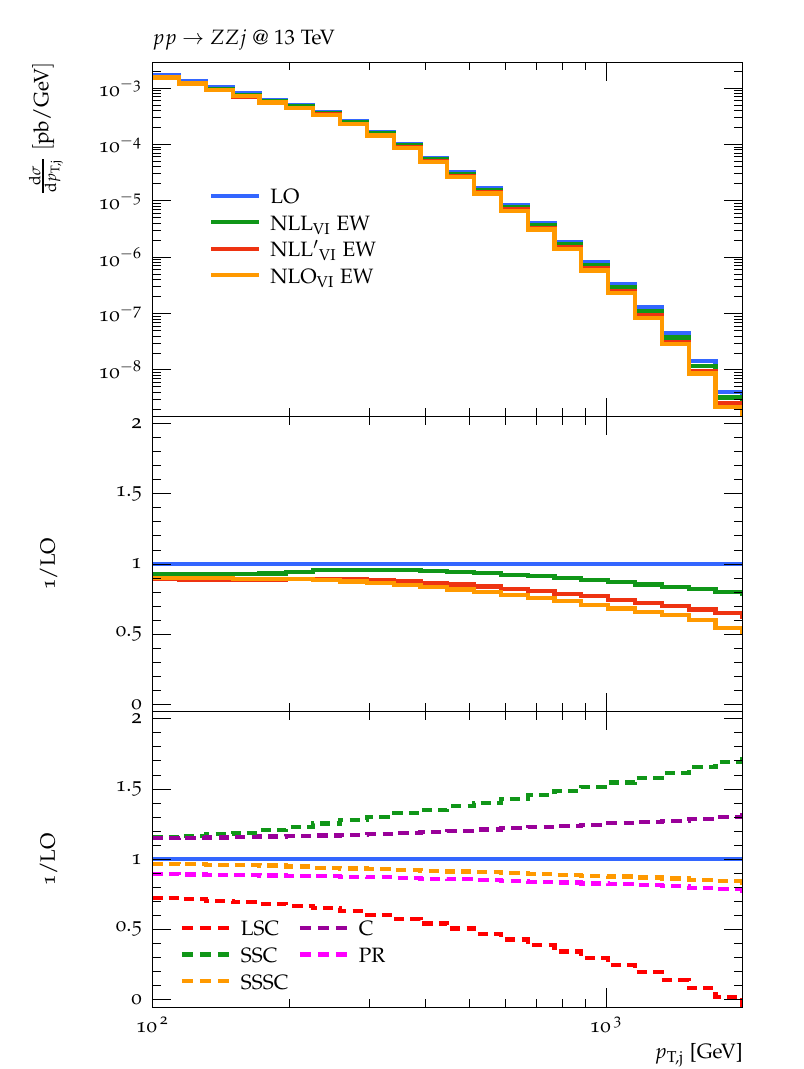}
\caption{Differential distribution in the transverse momentum of the hardest
$Z$ boson $p_{\rm{T},Z_1}$ (left) and of the jet $p_{\rm{T},j}$   (right)
 in $pp \to ZZj$ at $\sqrt{s}=13 ~ \rm{TeV}$. Curves as in Fig.~\ref{fig:zj}.}

\label{fig:zzj_ptZ}
\end{figure*}

We now consider hadronic $ZZ+$jet production and focus on the inclusive $p_{\rm{T},Z_1}$ (left) and $p_{\rm{T},j}$ (right) distributions in~\reffi{fig:zzj_ptZ}. In the case of the transverse momentum of the hardest $Z$-boson we observe sub-percent level agreement between the two predictions in LA and the \NVI one, with a typical virtual EW Sudakov behaviour that is dominated by large angular-independent \LSC effects. For this observable we observe the one-loop EW relative corrections to the $ZZj$ process to be identical to the relative corrections to the $ZZ$ process shown in \reffi{fig:zz} (left). This indicates a factorisation of QCD and EW effects motivating a multiplicative combination of higher-order QCD and EW corrections to the $pp \to ZZ$ process. 

For the transverse momentum distribution of the jet shown on the right of~\reffi{fig:zzj_ptZ} we observe an increasing impact of the \SSC correction towards higher $p_{\rm{T},j}$ leading to sizeable cancellations with the \LSC corrections. The total one-loop EW correction at $p_{\rm{T},j}=1$~TeV amounts to about $-35\%$ at \NVI respective $-30\%$ at \NLLp, and $-20\%$ at \NLL. The difference between \NLL and \NLLp can be interpreted as a conservative upper bound of the uncertainty of the \NLLp prediction due to logarithmic effects beyond the LA. 
 

\subsubsection[VVV]{$\mathbf{VVV}$}

In this subsection we consider tri-boson production, and as representative process $pp \to W^+W^-Z$. 
NLO EW corrections to tri-boson processes are known for various on-shell~\cite{Nhung:2013jta,Shen:2015cwj,Wang:2016fvj,Wang:2017wsy,Dittmaier:2017bnh,Frederix:2018nkq,Zhu:2020ous} and off-shell~\cite{Greiner:2017mft,Schonherr:2018jva,Dittmaier:2019twg,Cheng:2021gbx} combinations of $W,Z,\gamma, H$. 

\subsubsection*{$\mathbf{W^+W^-Z}$}

\begin{figure*}[tb]
\centering
	\includegraphics[width=\setrelwidth\textwidth]{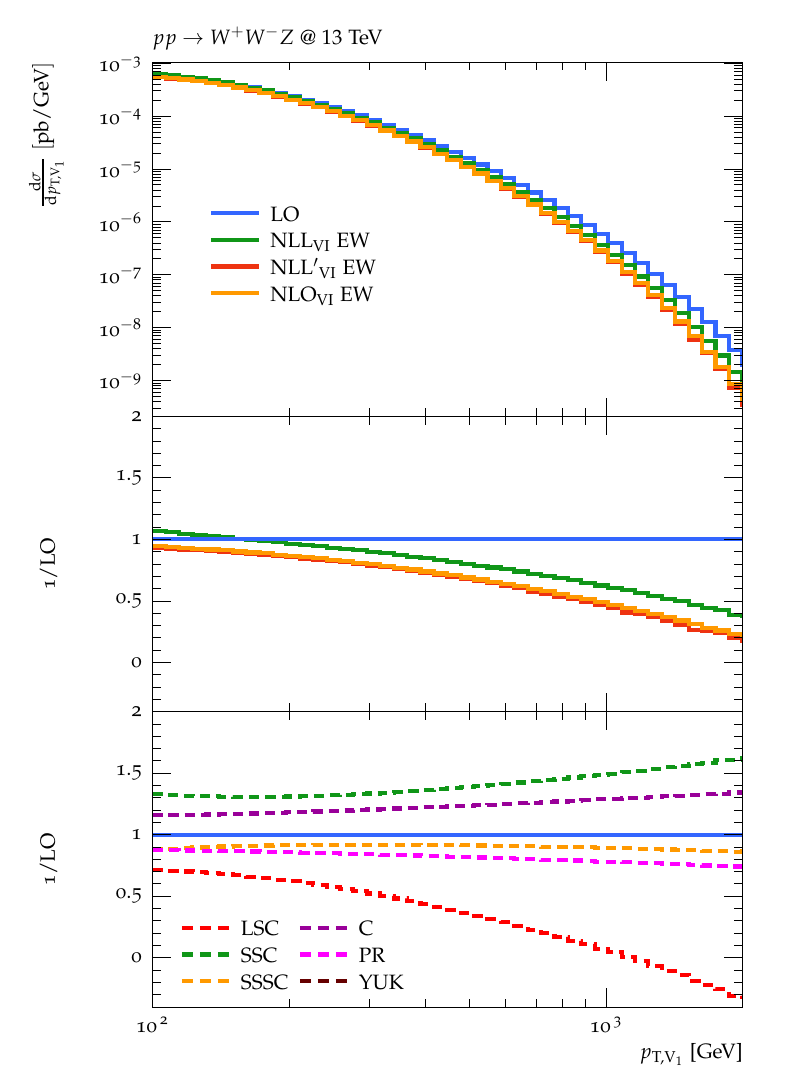}
	\includegraphics[width=\setrelwidth\textwidth]{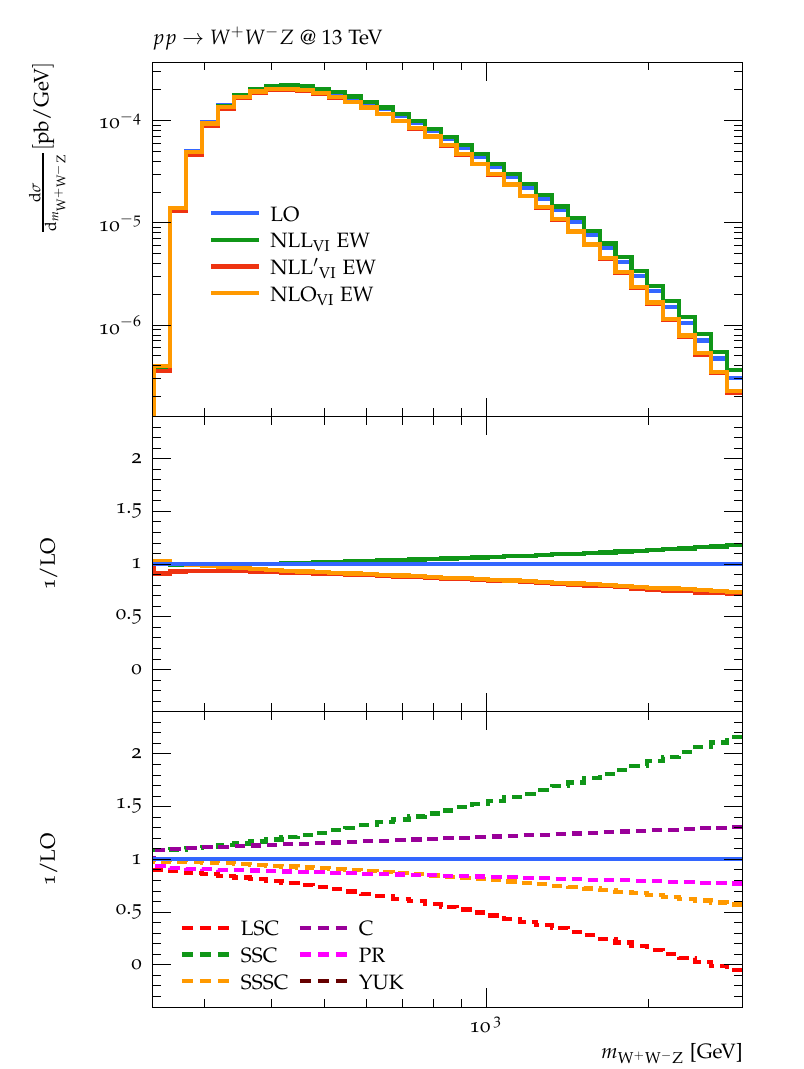}
\caption{Differential distribution in the transverse momentum of the hardest
vector boson $p_{\rm{T},V_1}$ (left) and of the invariant mass  $m_{\rm{WWZ}}$  (right)
 in $pp \to WWZ$ at $\sqrt{s}=13 ~ \rm{TeV}$. Curves as in Fig.~\ref{fig:zj}.}
\label{fig:wwz}
\end{figure*}

In~\reffi{fig:wwz} we show the differential transverse momentum distribution of the hardest gauge boson $V_1$ (left) and of the $W^+W^-Z$ invariant mass (right) for $pp \to W^+W^-Z$. Both distributions are inclusive without any phase-space restrictions.
For this process in the high-$p_{\rm T}$ tail EW corrections are very large. For $p_{\rm{T},V_1} > 1$ TeV the \LSC correction becomes larger than the entire LO contribution, resulting in overall one-loop EW corrections of $-60\%$ at \NVI and \NLLp, which agree at the percent-level, while the \SSSC terms induce about $10\%$ differences with the \NLL prediction. Considering the invariant mass distribution of the entire $W^+W^-Z$ system we observe very large cancellations between the \LSC and \SSC corrections, with increasing impact of the \SSSC terms towards high invariant masses, similar to the case of the $m_{ZZ}$ distribution discussed above. In the case of the $m_{WWZ}$ distribution shown here we observe excellent agreement at the percent-level between \NLLp and \NVI for $m_{WWZ}>400~$GeV up to very high invariant masses. The \NLL prediction on the other hand yields positive corrections with respect to LO at large invariant masses. 



\subsubsection[${t\bar t+X}$]{$\mathbf{t\bar t+X}$}

As a last example in the discussion of EW corrections to on-shell processes we consider $t \bar t W^+$ and $t \bar t W^++$jet production as a representative of $pp \to {t\bar t+X}$ processes, while in Appendix~\ref{appendixtth} we also show results for $t \bar t H$(+jet) production. These ${t\bar t+X}$ processes constitute important backgrounds in Higgs analyses and/or BSM searches, and offer a rich laboratory for tests of electroweak symmetry breaking. NLO EW corrections to $pp \to {t\bar t+X}$ processes have been presented in~\cite{Frixione:2015zaa,Frederix:2017wme,Frederix:2018nkq,Pagani:2023wgc}. To be precise in the following we consider the one-loop EW corrections to the leading QCD order of these processes, i.e. the LO $t \bar t W^+$ and $t \bar t W^++$jet processes are of $\ord(\alpha_S^2 \alpha)$ resp. $\ord(\alpha_S^3 \alpha)$. At this perturbative order the relative $\ord(\alpha)$ one-loop corrections are dominated by EW insertions into the LO matrix elements, while also QCD insertions into the subleading LO interference contribute. The \NVI predictions presented in the following contain both contributions and IR singularities are subtracted via a mixed QCD-QED $\mathbf{I}$-operator. In the case of the LA predictions \NLL and \NLLp we only consider the pure EW corrections with respect to the leading QCD amplitude, and neglect interference effects. Correspondingly, IR singularities are subtracted via a pure QED I-operator. 
Nevertheless, as shown in the following for the considered processes we find overall excellent agreement between the \NVI result and the LA approximation \NLLp~\footnote{As part of Ref.~\cite{Pagani:2021vyk} an extension of the \DPA has been presented which allows to formally include possible logarithmically enhanced effects in these interferences.}.

\subsubsection*{$\mathbf{t \bar tW^+}$}
\begin{figure*}[tb]
\centering
       \includegraphics[width=\setrelwidth\textwidth]{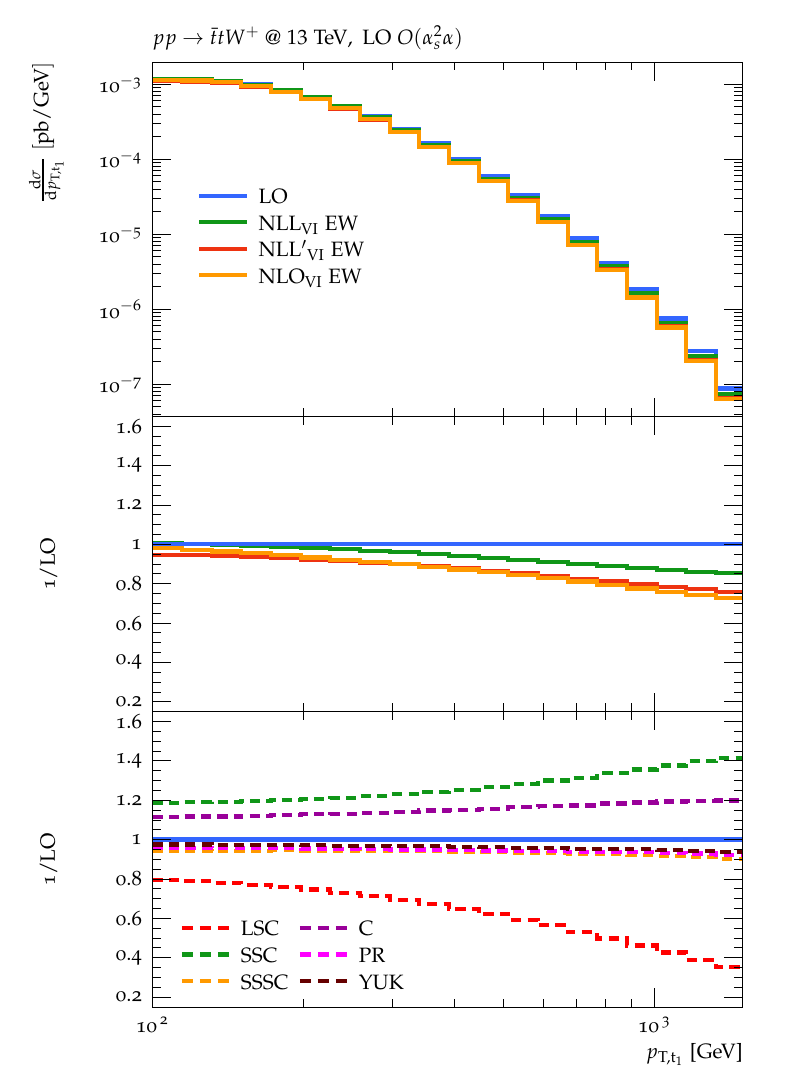}
	\includegraphics[width=\setrelwidth\textwidth]{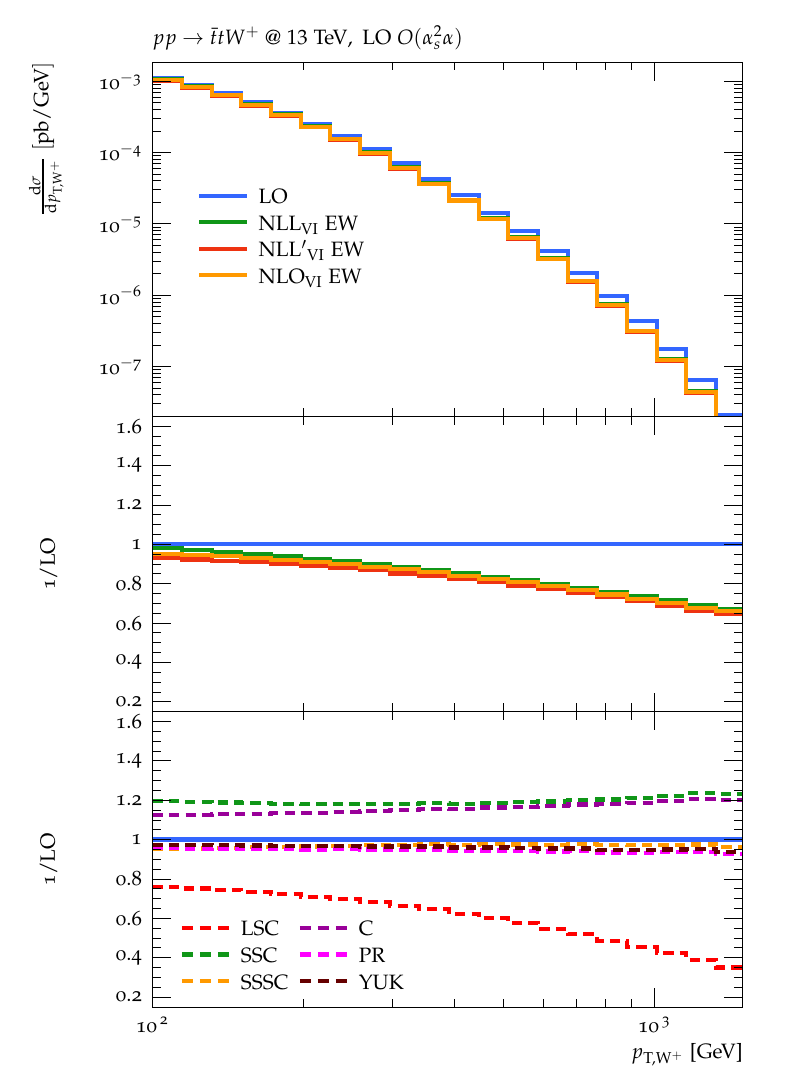}
\caption{Differential distribution in the transverse momentum of the hardest
top-quark $p_{\rm{T},t_1}$ (left) and of the $W$ boson $p_{\rm{T},W}$  (right)
 in $pp \to t\bar tW^+$ at $\sqrt{s}=13 \hspace{0.1 cm} \rm{TeV}$. Curves as in Fig.~\ref{fig:zj}, where additionally the \YUK term is shown (dashed brown).}
  \label{fig:ttw_ptw}
 \end{figure*}
 \begin{figure*}[tb]
     \includegraphics[width=\setrelwidth\textwidth]{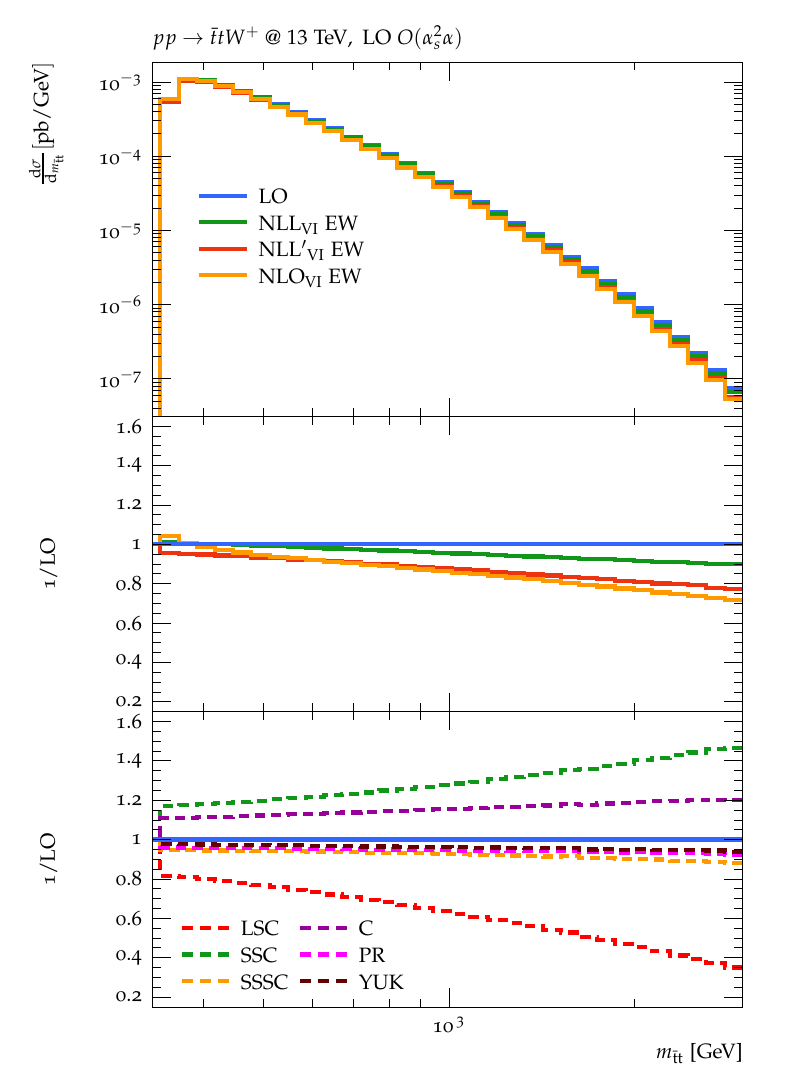}
	\includegraphics[width=\setrelwidth\textwidth]{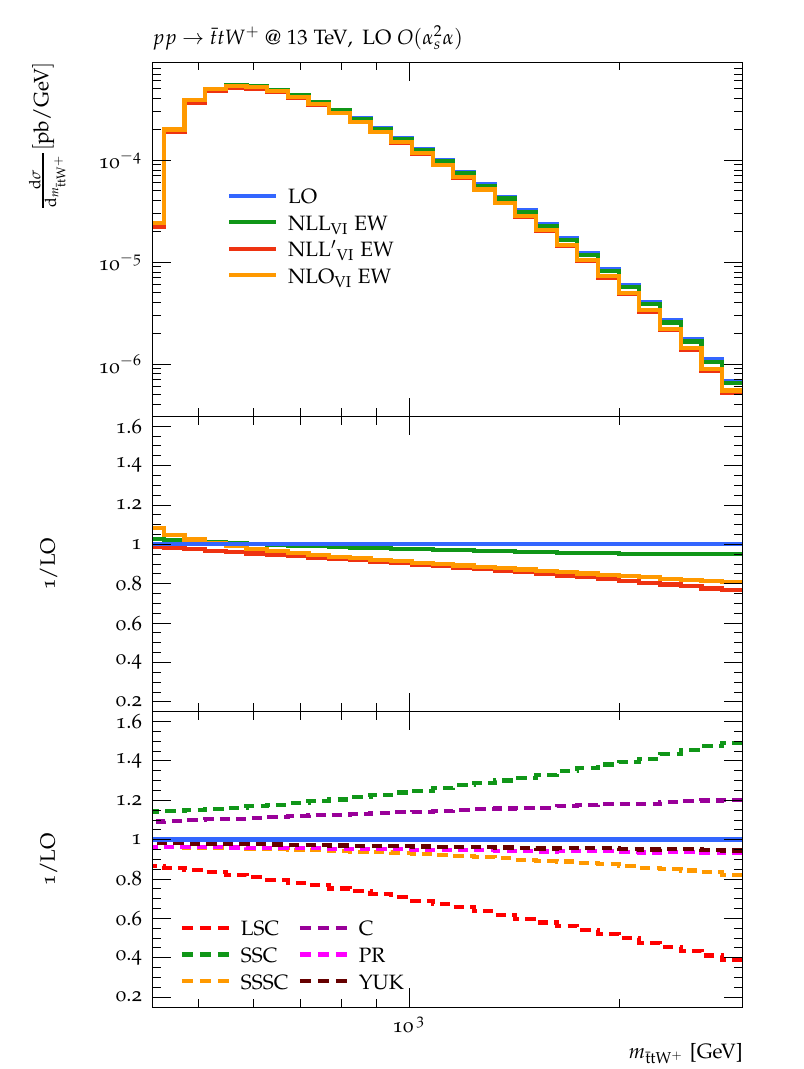}
\caption{Differential distribution in the invariant mass $m_{t \bar t}$ (left) and $m_{t \bar t W^+}$ (right) in $pp \to t\bar tW^+$ at $\sqrt{s}=13 \hspace{0.1 cm} \rm{TeV}$. Curves as in Fig.~\ref{fig:ttw_ptw} .}
 \label{fig:ttw_mass_ttw}
\end{figure*}

In~\reffi{fig:ttw_ptw} differential distributions in the transverse momentum of the hardest top-quark (left) and of the $W-$boson (right) are shown for $t\bar t W^+$ production. In contrast to all previous plots here we show explicitly the contribution  of the \YUK terms, as defined in Eq. \ref{yukawacorrections}. In both distributions we observe very good agreement between the \NVI prediction and the \NLLp prediction, which are both dominated by \LSC terms. However, in the case of the $p_{\rm{T},t_1}$ distribution the relative impact of the angular-dependent \SSC and \SSSC terms increases towards the tail of the distribution yielding sizeable cancellations with the \LSC correction. This results in an overall EW correction at the level of about $-25\%$ at $p_{\rm{T},t_1}=1$~TeV for \NVI and \NLLp, while the \NLL prediction yields about $-15\%$ at this level of transverse momentum. In the case of $p_{\rm{T},W^+}$ the angular-dependent terms show a significantly smaller enhancement towards the high-$p_T$ tail resulting in reduced cancellations between the different NLL terms, and overall one-loop EW corrections at the level of around $-30\%$ at $p_{\rm{T},W}=1$~TeV. Here, all three predictions agree at the percent level.

%
%

In Fig. \ref{fig:ttw_mass_ttw} we show the invariant mass distribution of the $tt$ system (left plot) and of the  $ttW$ system (right plot). In these distributions the LA condition, Eq.~\eqref{la}, is violated: the \SSC correction is as large as the \LSC one with opposite sign, resulting in sizeable cancellations amongst these. The \SSSC correction is as large as the \PR and \YUK corrections and reaches up to $-10 \%$ with respect to the LO in the tail of both distributions. The overall one-loop EW corrections reach about $-15 \%$ both at $m_{t\bar t}=1$~TeV and at $m_{t\bar t W}=1$~TeV with excellent agreement between \NVI and \NLLp at the percent level.

\subsubsection*{$\mathbf{t\bar tW^+}$jet}

\begin{figure*}[tb]
\centering
	\includegraphics[width=\setrelwidth\textwidth]{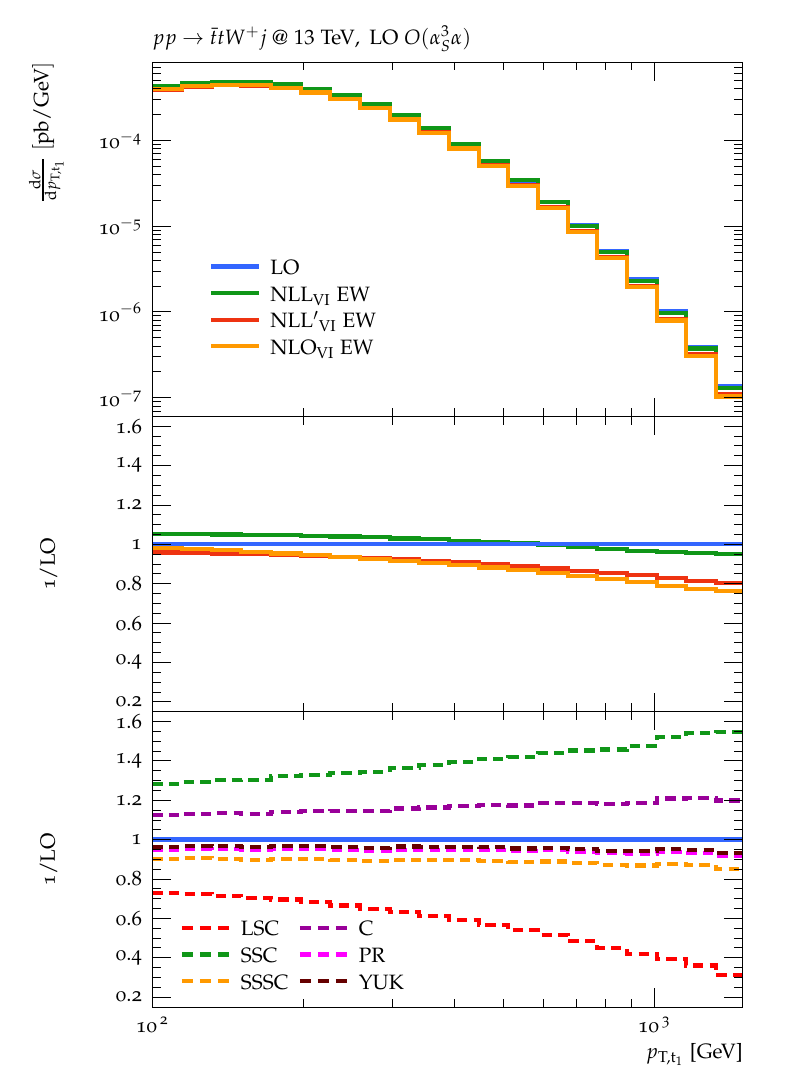}
	\includegraphics[width=\setrelwidth\textwidth]{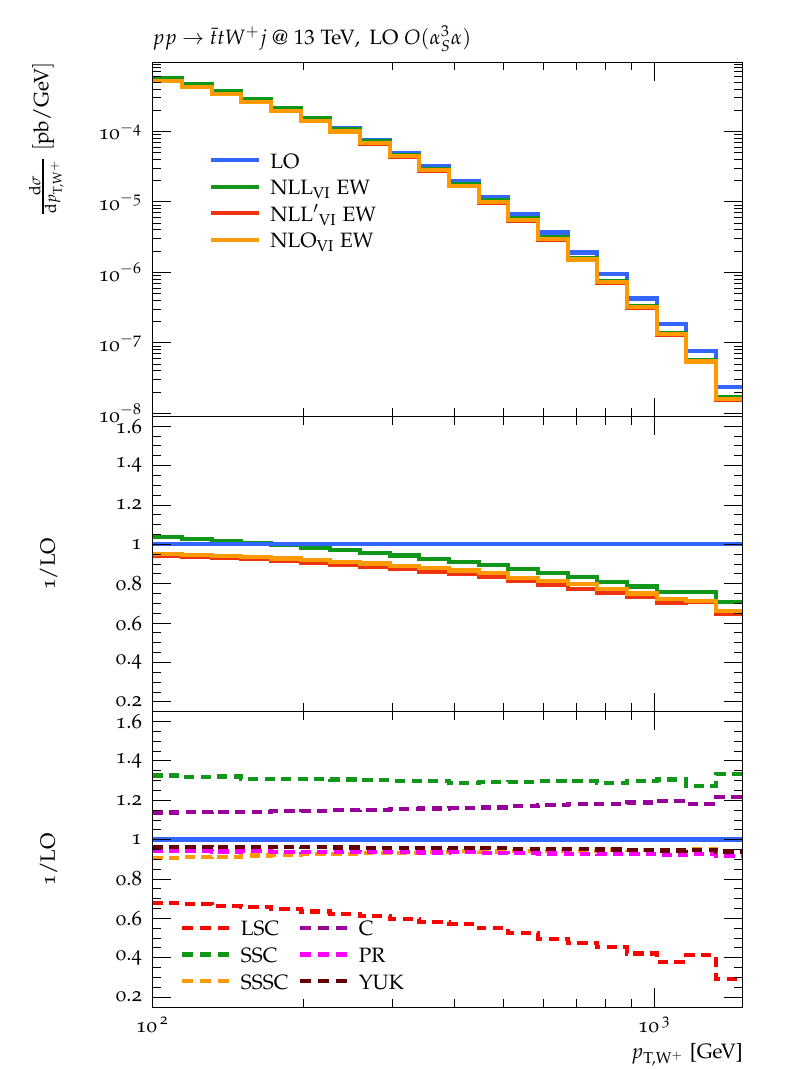}
\caption{Differential distribution in the transverse momentum of the hardest
top-quark $p_{\rm{T},t_1}$ (left) and of the $W$-boson $p_{\rm{T},W}$  (right)
 in $pp \to t\bar tW^+j$ at $\sqrt{s}=13 \hspace{0.1 cm} \rm{TeV}$. Curves as in Fig.~\ref{fig:ttw_ptw}.}
\label{fig:ttwj_ptt1}
	\includegraphics[width=\setrelwidth\textwidth]{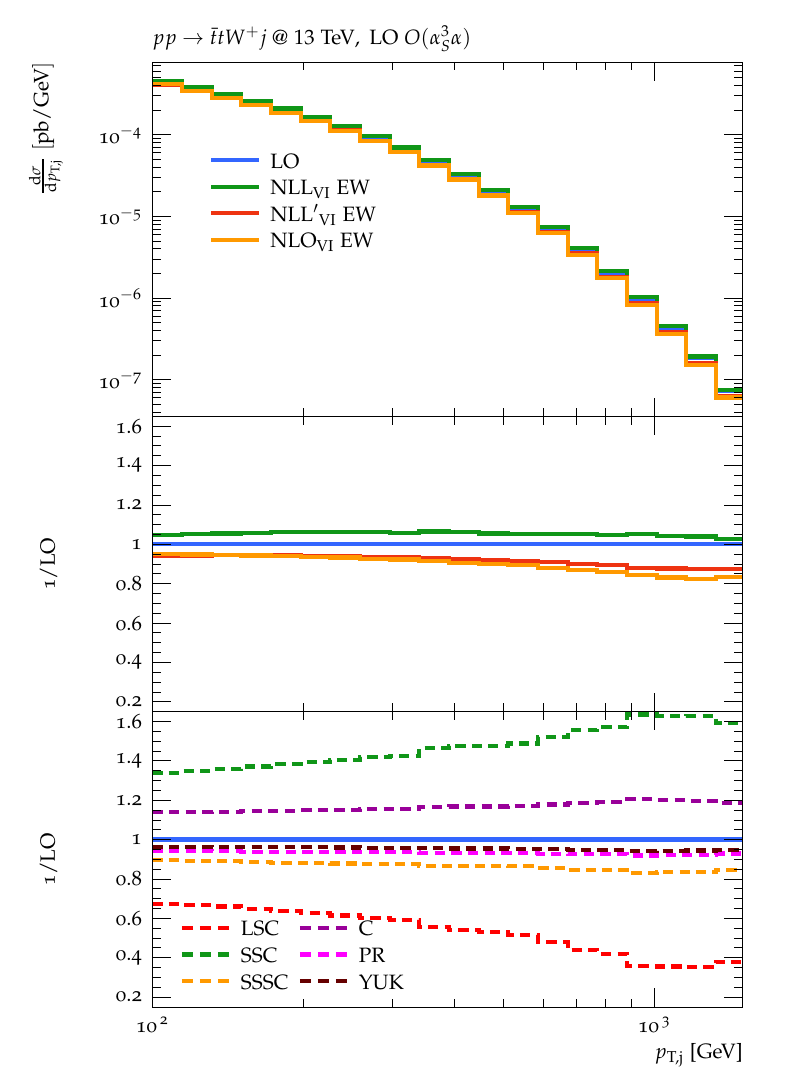}
       	\includegraphics[width=\setrelwidth\textwidth]{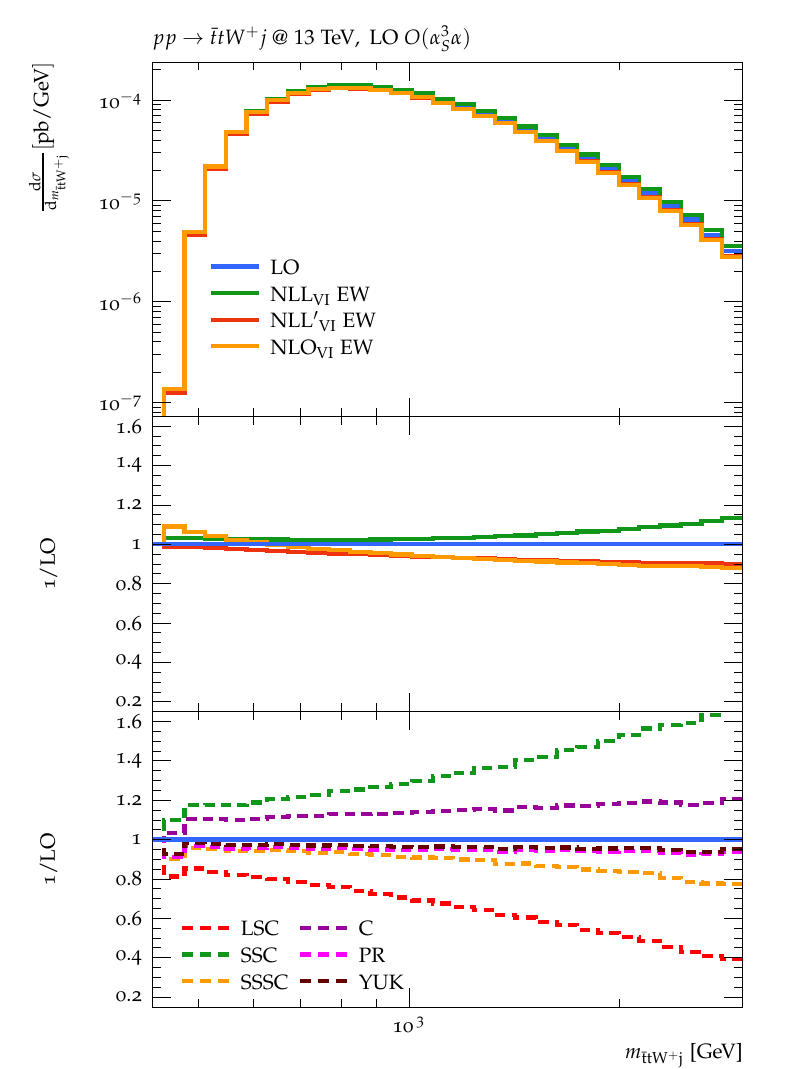}
\caption{Differential distribution in the  transverse momentum of the jet $p_{\rm{T},j_1}$  (left) and of the invariant mass $m_{\rm{t\bar tW^+j}}$ (right) in $pp \to t\bar tW^+j$ at $\sqrt{s}=13 \hspace{0.1 cm} \rm{TeV}$. Curves as in Fig.~\ref{fig:ttw_ptw}.} 
\label{fig:ttwj_ptj_mttwj}
\end{figure*}

Finally, we present one-loop EW predictions for $pp \to t\bar tW^++$jet, which to the best of our knowledge are presented here for the first time. In~\reffi{fig:ttwj_ptt1} we investigate the behaviour of the different higher-order EW corrections for the $p_{\rm{T},t_1}$ (left) and $p_{\rm{T},W}$ (right) 
distributions, and in~\reffi{fig:ttwj_ptj_mttwj} for the $p_{\rm{T},j}$ (left) and   $m_{\rm{t\bar tW^+j}}$ (right) distributions.
 As in the case of $pp \to t\bar tW^+$ shown above in \reffi{fig:ttw_ptw} for $p_{\rm{T},t_1}$ we observe large cancellations between the \LSC and \SSC corrections in the tail of the distribution. In fact, for  $t\bar tW^+$ production in association with a jet the relative impact of the angular-dependent \SSC term is even larger. However, also the impact of the negative \SSSC contribution is growing towards the tail of the distribution yielding an overall one-loop EW correction of about $-23\%$ at $p_{\rm{T},t_1}=1$~TeV for \NVI, and about $-20\%$ for \NLLp. At this level of transverse momentum the \NLL prediction only yields an overall correction of $-5\%$, significantly underestimating the size of the one-loop EW corrections. 
In contrast, the tail of the $p_{\rm{T},W}$ distribution (right) is clearly dominated by the \LSC correction,  while the \SSC correction is of the same size as the \COLL correction.  The \SSSC correction is again smaller and of the same size as the \PR and \YUK corrections reaching each up to about $\sim 5\%$. The overall one-loop EW correction at \NLLp agrees with the one at \NVI at below the percent level, while differing at the level of about $5\%$ with the \NLL prediction due to \SSSC effects.  

In the transverse momentum distribution of the jet (\reffi{fig:ttwj_ptj_mttwj}, left) we observe very large angular-dependent \SSC and \SSSC corrections, indicating a violation of the LA assumption, \eqref{la}. In turn, the two NLL predictions differ by up to $20\%$, while the \NLLp prediction agrees at the $1-5\%$ level with \NVI with overall corrections reaching about $-15\%$ at $p_{\rm{T},j}=1$~TeV, i.e. here the difference between the NLL predictions would largely overestimate the uncertainty associated to the \NLLp approximation.  A similar picture emerges for the invariant mass distribution of the entire $t\bar t W j$ system (right), which in-line with the results for other invariant mass distributions discussed above, receives large angular dependent DL corrections, with \SSC effects compensating the entire \LSC correction. The overall EW corrections remain below $-10\%$ in the entire considered invariant mass range, with excellent agreement between \NVI and \NLLp in the tail of the distribution.

%

\subsection{Off-shell processes} \label{Offshell}

In this section we investigate the scope of our implementation of EW one-loop corrections in LA for processes with internal resonances. For such processes the logarithmic approximation is not directly applicable due to the separation of scales between production and decay parts of a resonant process, and due to the interplay between different resonant and non-resonant topologies. Instead, the LA should be applied to the relevant hard scattering process, and the decay separately. In Section~\ref{sec:InternalinsTheory} we introduced such an algorithm based on a generalisation of our implementation of the logarithmic approximation to internal effective EW Sudakov vertices supplemented with probabilistic projectors based on the off-shell kinematics.
This algorithm allows for a consistent description of resonant processes including the effect of helicity correlations between production and decay parts of the process. 
In the following we showcase NLL EW results obtained with this resonance algorithm investigating $pp \to e^+e^-+$jet, i.e. off-shell $Z+$jet production, and $pp \to e^+ e^-  \mu^+ \mu^-$, i.e. off-shell $ZZ$ pair-production. We denote \NLLMR and \NLLpMR as the LA predictions based on this generalised version of the logarithmic EW approximation, where mass regularisation with $\lambda=m_W$ is used for the QED contributions.
By setting the photon mass regulator $\lambda$ to $\lambda=m_W$ we effectively integrate out  QED radiation up to the scale $m_W$ in the \NLLMR and \NLLpMR predictions. Instead in the \NVI prediction in DR QED radiation can be understood to be integrated up to all scales. Therefore, a comparison between these different predictions is not fully consistent.~\footnote{The current implementation of the resonance algorithm does not allow for a consistent use of dimensional regularisation. We leave a corresponding improvement to future work.}
  In the following we use for the width multiplier in the resonance projectors $w_{\text{rescale}}=20$. We checked that varying this multiplier by a factor of 2 does not change our predictions in a significant way. 
  
Additionally we consider results based on a naive application of the standard LA acting only on all external particles. We always include \SSSC terms in these predictions and denote them as "\NLLpMREXT".
Alternatively we also compare against results based on an on-shell modelling of only the relevant hard scattering process including one-loop EW corrections, with decays obtained via the hard decay handler (HDH) in \Sherpa. The latter preserves LO spin correlations, and the virtualities of the decaying particles are smeared according to a Breit-Wigner shape. Again, here we always include \SSSC terms.

\subsubsection[$e^+e^-+$jet]{$\mathbf{e^+e^-+}$jet} 

\begin{figure*}[tb]
\centering
	\includegraphics[width=\setrelwidth\textwidth]{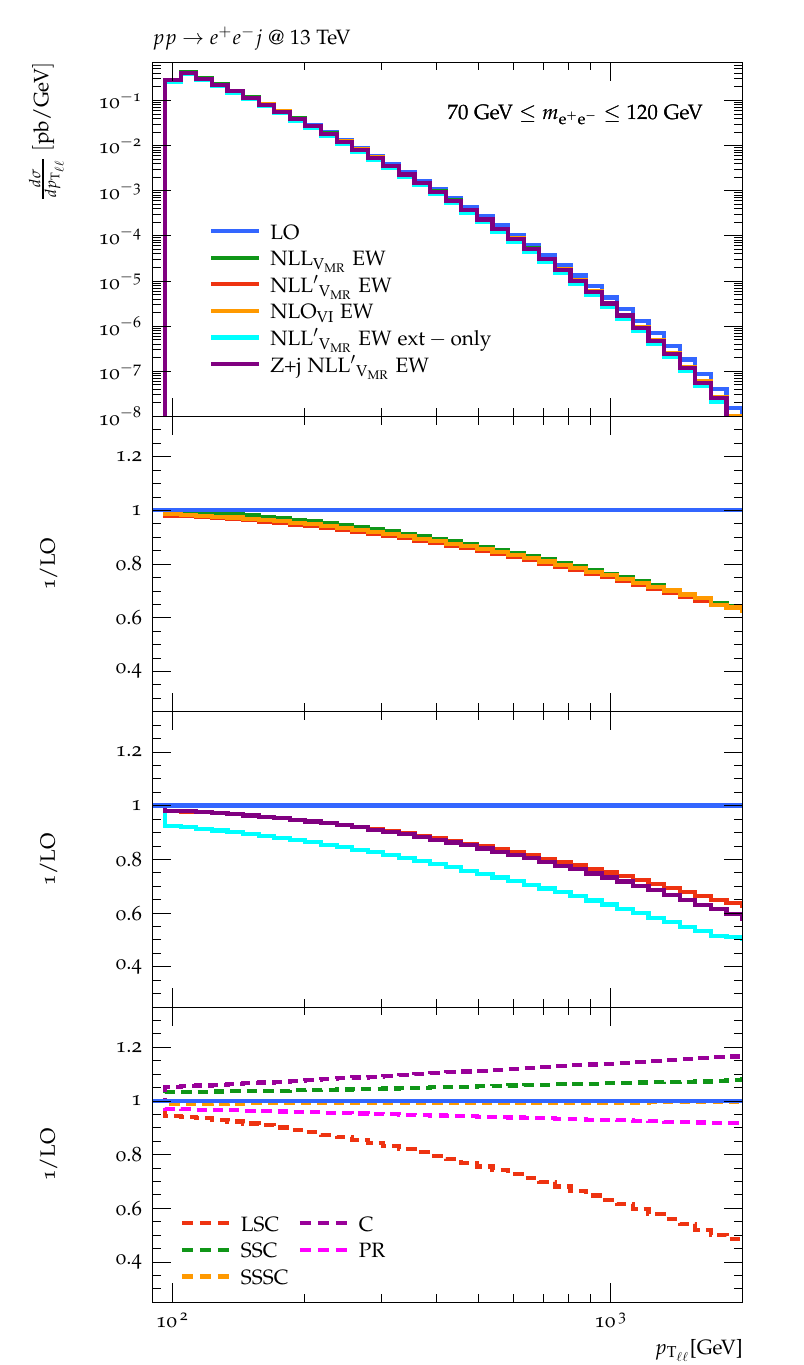}
	\includegraphics[width=\setrelwidth\textwidth]{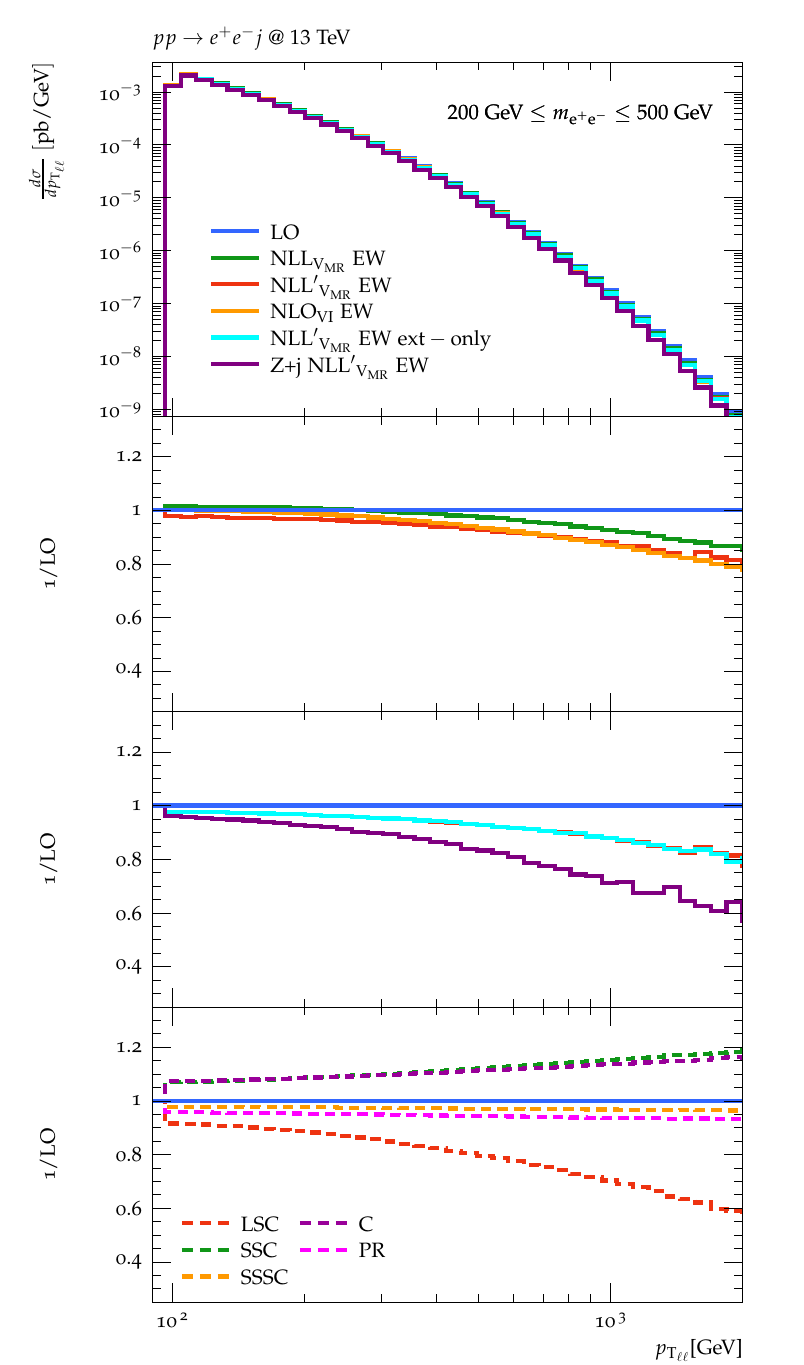}
\caption{Differential distribution in the transverse momentum of the dilepton system $p_{\textrm{T}, ll}$  in $pp \to e^+e^- j$ at $\sqrt{s}=13 \hspace{0.1 cm} \rm{TeV}$ for $70 \hspace{0.1 cm}  \text{GeV} \leq m_{ll} \leq 120 \hspace{0.1 cm}  \rm{GeV}$ (left) and $200 \hspace{0.1 cm}  \text{GeV} \leq m_{ll} \leq 500 \hspace{0.1 cm}  \text{GeV}$ (right). The first panel shows absolute predictions at LO~(blue), \NLLMR~(green), \NLLpMR~(red), \NVI~(orange), \NLLpMREXT~(cyan) and the \NLLp for corresponding on-shell $Z$+jet process with decays obtained via the HDH in \Sherpa~(purple).  Here the subscript "VI" indicates that the QED $I$-operator is added to the virtual amplitudes in DR. The second panel shows the relative corrections of the \NLLMR, \NLLpMR and \NVI predictions with respect to LO, while the third panel shows the same ratio for \NLLpMR, \NLLpMREXT, and for $Z+$jet \NLLpMR. The third panel shows the  various contributions at \NLLpMR in MR, as discussed in Section~\ref{sec2}, normalised to LO:
 \LSC (dashed red),  \SSC (dashed green),  \SSSC (dashed orange), \COLL (dashed purple), \PR (dashed magenta). }
\label{fig:pplljint_ptll}
\end{figure*}

For the process $pp \to e^+e^-+$jet, i.e. off-shell $Z/\gamma^*$+jet production, in \reffi{fig:pplljint_ptll} we show the differential distribution in the transverse momentum of the dilepton pair $p_{\textrm{T},ll}$ in two different invariant mass windows. On the left the invariant mass window is restricted to include the $Z$-boson resonance, i.e. $70 \hspace{0.1 cm}  \text{GeV} \leq m_{ll}  \leq 120 \hspace{0.1 cm} \rm{GeV}$, while on the right we restrict to high invariant mass configurations, $200 \hspace{0.1 cm}  \text{GeV} \leq m_{ll} \leq 500 \hspace{0.1 cm}  \text{GeV}$. 

In both invariant mass windows we observe percent-level agreement between the \NLLpMR prediction and the \NVI result over the entire considered transverse momentum range. In the window $70 \hspace{0.1 cm}  \text{GeV} \leq m_{ll} \leq 120 \hspace{0.1 cm} \rm{GeV}$ we are dominated by on-shell $Z$-boson production and decay. Correspondingly, we find excellent agreement between \NLL and \NLLpMR, as already shown in \reffi{fig:zj} (left). We also find very good agreement between the relative EW correction for the $Z$+jet simulation with HDH after-burner and our \NLLpMR prediction including resonance projectors. We verified, that the $1-5\%$ differences between these predictions at large $p_{\textrm{T},ll}$ originates from the $\gamma^*$ contribution together with the $Z-\gamma^*$ interference: removing any $\gamma^*$ contribution from \NLLpMR yields perfect agreement with the relative on-shell $Z+$jet prediction, while introducing $1-5\%$ differences with the \NVI prediction. In this mass window the naive application of the LA for only the external particles overestimates the relative EW corrections by $10-15\%$ as can be seen by comparing the \NLLpMREXT prediction with the \NLLpMR one in the second ratio panel of~\reffi{fig:pplljint_ptll}.    

We observe a very different situation for the   $200 \hspace{0.1 cm}  \text{GeV} \leq m_{ll} \leq 500 \hspace{0.1 cm}  \text{GeV}$ invariant mass window: here the relative correction obtained via an on-shell $Z$+jet simulation with a hard decay after-burner largely overestimates the \NLLpMR correction by up to $20\%$ which in turn perfectly agrees with the fully off-shell \NVI prediction in the complex mass scheme. In this off-shell phase-space region the naive application of the LA agrees at the sub-percent level with the \NLLpMR and \NVI predictions. In this regime any logarithmically enhanced EW corrections are indeed driven the off-shell di-lepton pair in the final-state.

\subsubsection[$e^+ e^-  \mu^+ \mu^-$]{$\mathbf{e^+ e^-  \mu^+ \mu^-}$}

\begin{figure*}[tb]
\centering
			\includegraphics[width=0.45\textwidth]{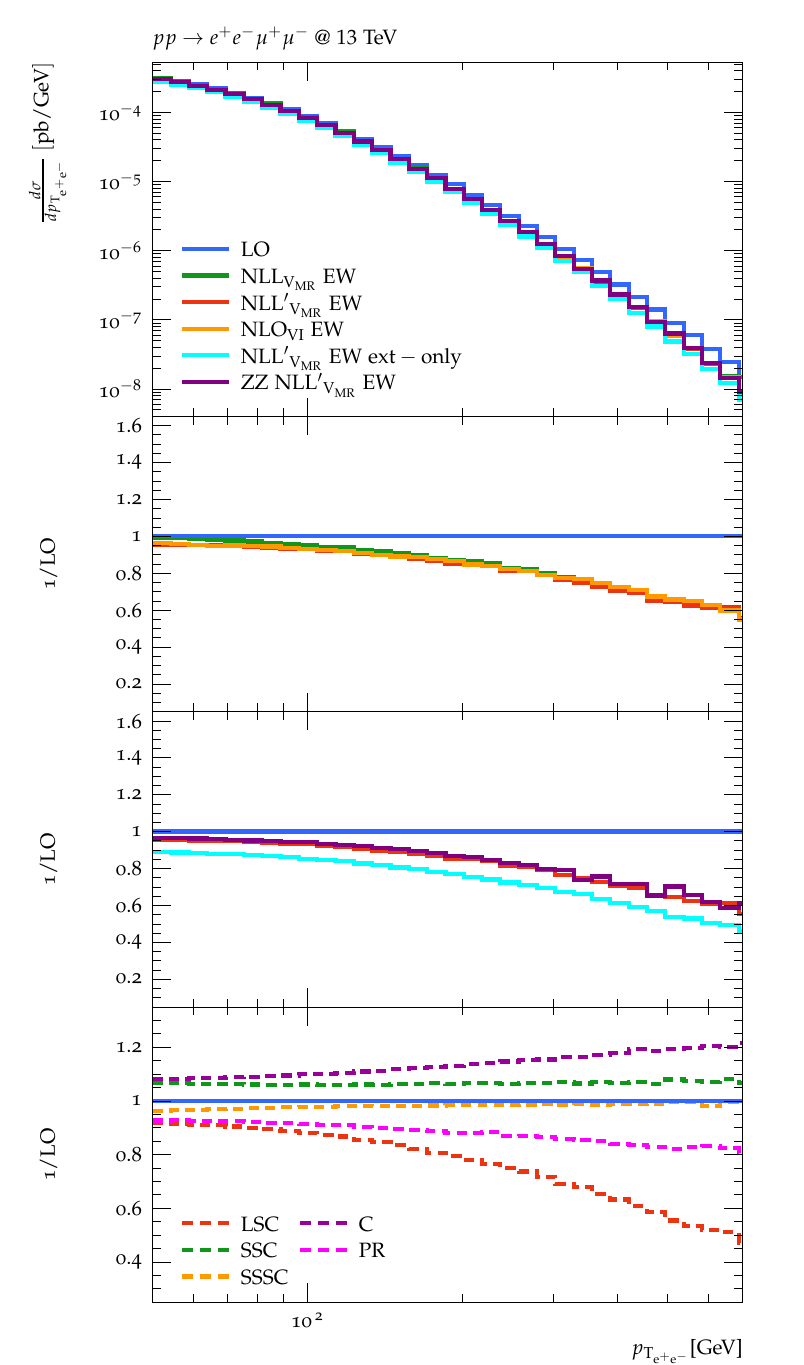}
	\includegraphics[width=0.45\textwidth]{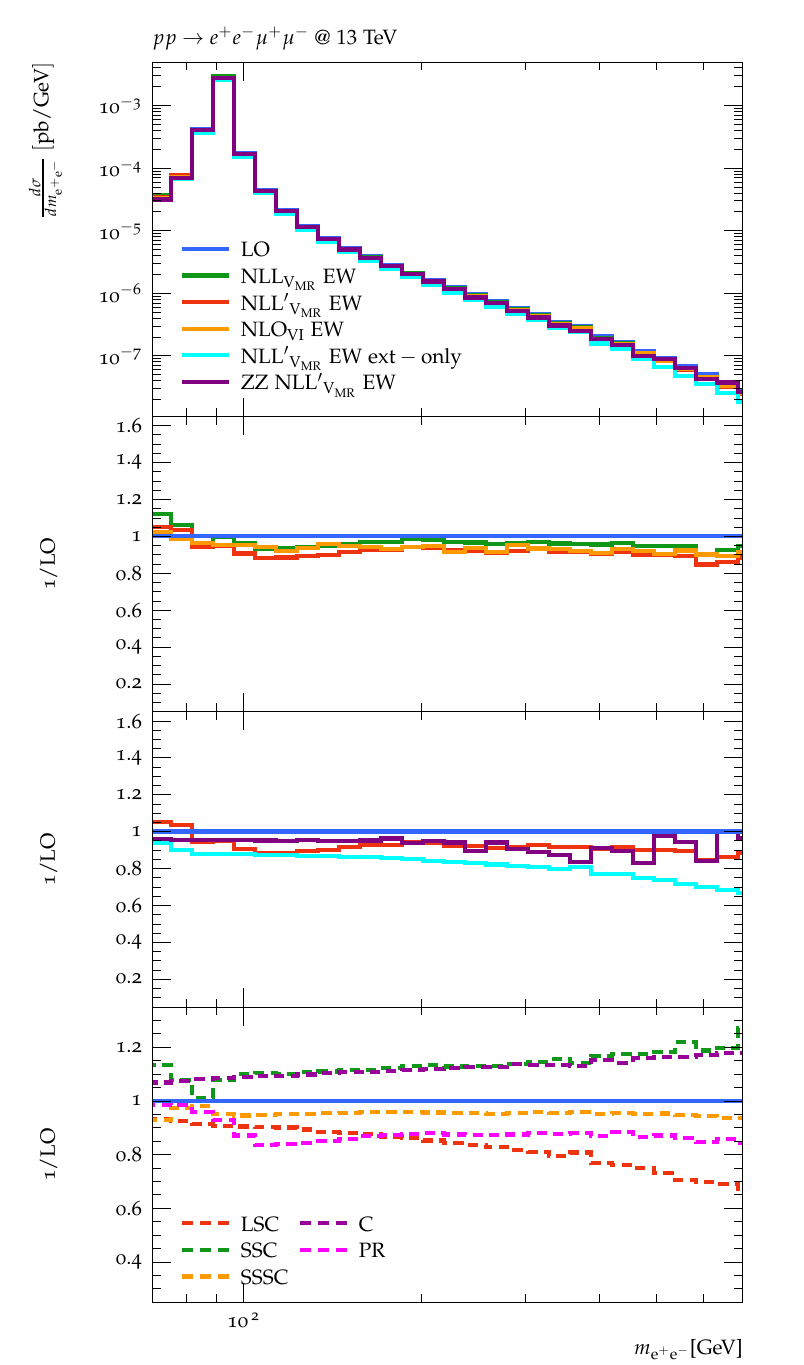}
\caption{Differential distribution in the transverse momentum $p_{\textrm{T}, ee}$ (left) and in the 
 invariant mass  $m_{ee}$ (right) of the di-electron system in $pp \to e^+e^- \mu^+ \mu^-$ at $\sqrt{s}=13 \hspace{0.1 cm} \rm{TeV}$. Curves as in Fig.~\ref{fig:pplljint_ptll}.}
\label{fig:ppllllint_massee}
\end{figure*}

We now consider the process $pp \to e^+ e^-  \mu^+ \mu^-$, i.e. off-shell $ZZ$ pair-production. In~\reffi{fig:ppllllint_massee} we show differential distributions in one of the reconstructed $Z$ bosons:  the transverse momentum $p_{\textrm{T}, e^+e^-}$ and the invariant mass $m_{e^+e^-}$. In both distribution we observe very good agreement between the \NVI prediction and the \linebreak \NLLpMR and \NLLMR predictions. In the case of the $p_{\textrm{T}, e^+e^-}$  all three predictions agree at the sub-percent level for $p_{\textrm{T}, e^+e^-} \geq 100\,$GeV, consistent with the observation in~\reffi{fig:zz} (left) for on-shell $ZZ$ production. Instead, for high $m_{e^+e^-}$ the \SSSC term introduces differences at the level of $5\%$ between \NLLpMR and \NLLMR, and here we also observe sizeable cancellations between the \LSC and \SSC corrections. For intermediate invariant masses above the $Z$ resonance, $100\, {\rm GeV} < m_{e^+e^-} < 150\,$GeV, the \NLLMR prediction agrees at the percent level with the \NVI prediction, while \NLLpMR yields about $5\%$ larger negative corrections.
In both distributions the on-shell $ZZ$ \NLLpMR description with hard-decay after-burner including Breit-Wigner smearing yields relative correction factors in agreement with the off-shell \NLLpMR description, while the naive application of the LA approximation for only the external particles overestimates the EW corrections by $10-15\%$.

\begin{figure*}[tb]
\centering
 	\includegraphics[width=0.45\textwidth]{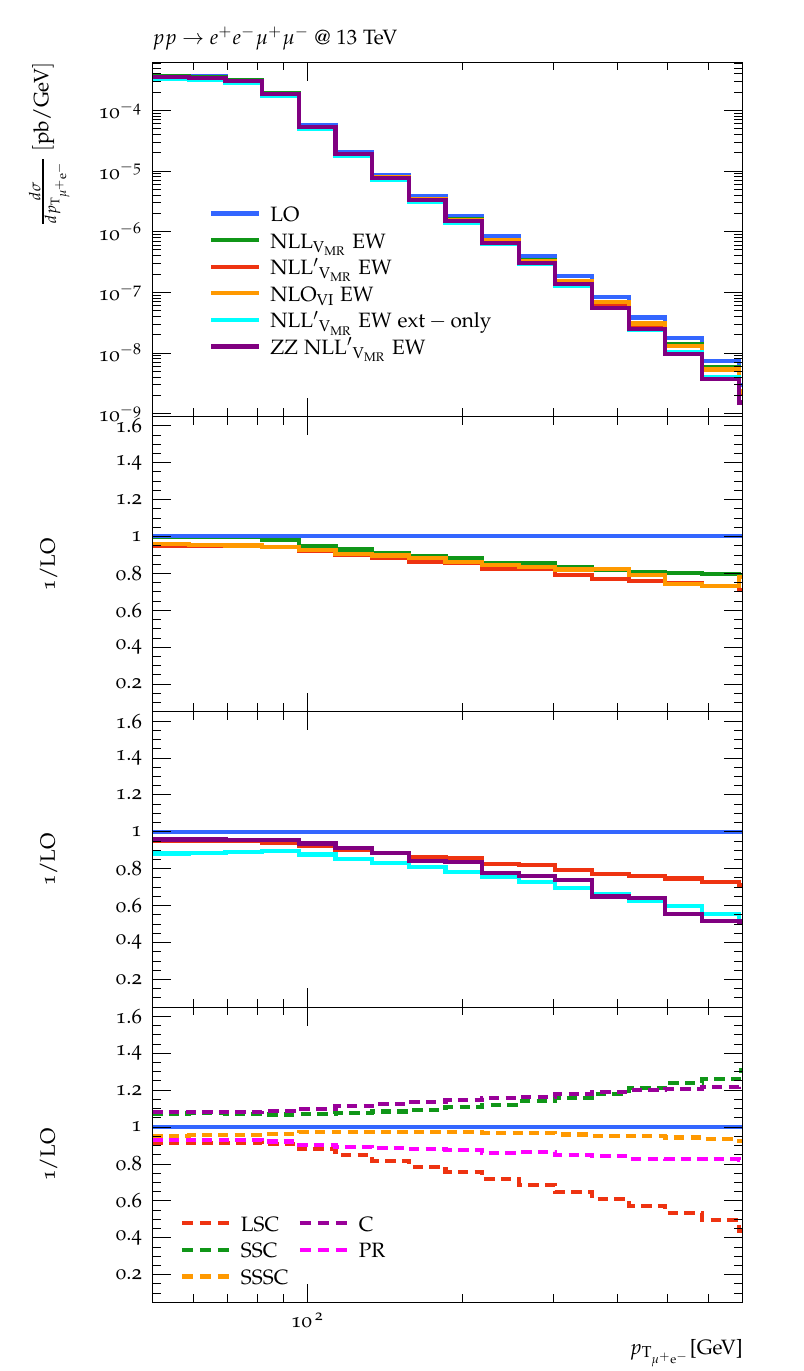}
		\includegraphics[width=0.45\textwidth]{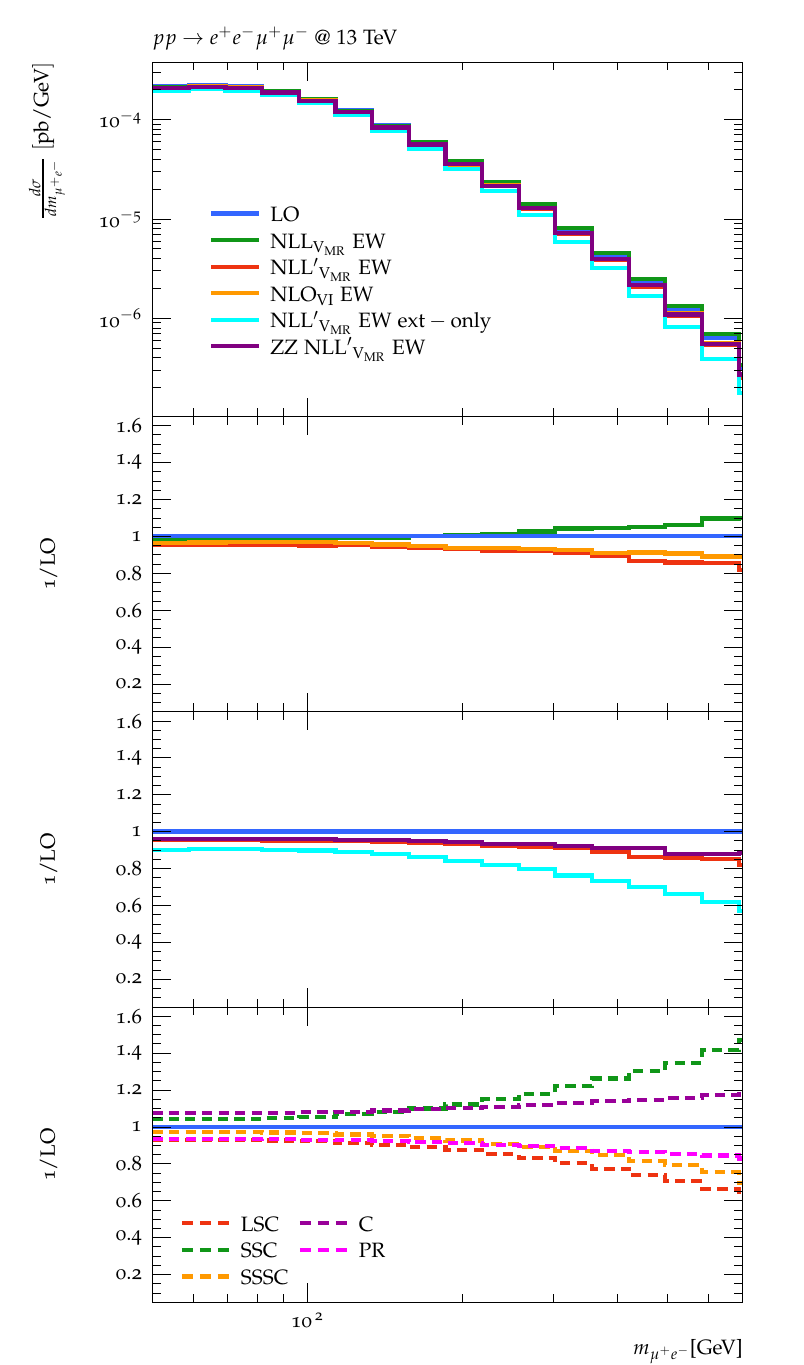}
\caption{Differential distribution in the in the transverse momentum $p_{\textrm{T}, \mu^+ e^-}$  (left)
and in the invariant mass  $m_{\mu^+ e^-}$ (right) in $pp \to e^+e^- \mu^+ \mu^-$ at $\sqrt{s}=13 \hspace{0.1 cm} \rm{TeV}$. Curves as in Fig.~\ref{fig:pplljint_ptll}.}
\label{fig:fig:ppllllint_mue}
\end{figure*}


In \reffi{fig:fig:ppllllint_mue} we look at the opposite-flavour di-lepton transverse momentum and invariant mass distributions. In the case of the transverse momentum distribution $p_{\textrm{T}, \mu^+ e^-}$  the on-shell description with subsequent decays overestimates the relative one-loop EW corrections by up to $20\%$, as does the naive application of the LA considering only the external particles, as a consequence of overlapping on- and off-shell configurations. However, these effects are correctly captured by our projector-based implementation of resonances providing a percent-level agreement of the \NLLpMR prediction with \NVI. Here, we observe up to $5\%$ differences between \NLLpMR and \linebreak \NLLMR at high $p_{\textrm{T}, \mu^+ e^-}$ induced by \SSSC effects.  
 In the case of the opposite-flavour invariant mass distribution the on-shell modelling yields a one-loop EW correction consistent with the  resonance-improved implementation yielding overall corrections of about $-15\%$ at   
 $m_{\mu^+ e^-}\approx 700~$GeV at \NLLpMR, which in turn agrees at the few percent level with the corresponding \NVI prediction. The \NLLMR prediction indicates a pathological behavior with positive overall EW corrections induces by very large \SSC effects, which are compensated by equally large \SSSC terms. Again, the naive implementation considering only external particles considerably overestimates the size of the one-loop EW corrections. 

The consistency between the on-shell predictions based on the HDH after-burner in Sherpa (purple lines) and the \NLLpMR predictions (red lines)  in the respective on-shell phase-spaces, together with the consistency of the \NLLpMR prediction and the naive implementation of the LA (\NLLpMREXT, light blue lines) can be seen as a validation of our implementation for resonance processes, where in all cases we observe very good agreement between \NLLpMR and \NVI. The agreement between \NLLpMR and the corresponding on-shell simulation in the on-shell phase-space also indicates small spin-correlation effects for the processes considered here. We leave a detailed study of such effects in processes subject to large spin correlation effects, including for off-shell top-quark production processes to  future work.

\section{Conclusions} \label{sec5}

We have presented an implementation of one-loop EW corrections in logarithmic  approximation (LA) in the amplitude generator \OpenLoops.
The implementation is based on a process-independent implementation of the algorithm of Refs.~\cite{Denner:2000jv,Denner:2001gw,Pozzorini:2001rs}  via effective EW Sudakov vertices for the evaluation of all soft-collinear double-logarithmic contributions and the collinear single-logarithmic contributions. Further single-logarithmic contributions originating from the UV are evaluated via a standard $\ord(\alpha)$ one-loop parameter renormalisation procedure restricted to the logarithmic level. This implementation allows to compute one-loop EW matrix elements for any SM process in LA consistently up to NLL EW resulting in a speed-up with respect to the full one-loop evaluation of up to two orders of magnitude.
We allow to include further sub-sub-leading angular-dependent logarithmic terms, which cannot entirely be controlled in a process-independent way, resulting in NLL' EW predictions. As suggested in Ref.~\cite{Pagani:2021vyk} the inclusion of such terms might nevertheless be mandatory for a reliable estimate of one-loop EW corrections in phase-space regions subject to large ratios of kinematic invariants, e.g. in tails of invariant mass distributions. Our implementation is fully general and can be applied to any tree-induced process in the Standard Model including for high-multiplicity and multi-jet final-states. Corrections for processes involving longitudinal gauge boson modes are evaluated via the Goldstone boson equivalence theorem, which is automatically applied. The implementation of parameter renormalisation contributions via the standard one-loop renormalisation procedure in \OpenLoops allows for a variety of parameter input schemes to be used.
The resulting NLL(') EW amplitudes are available via the standard interfaces of \OpenLoops and can thus easily be included in Monte Carlo event generators. QED contributions can be either treated in mass regularisation or in dimensional regularisation. The latter allows for a direct combination with NLO EW accurate real radiation contributions via infrared subtraction schemes.    

Our implementation is highly flexible and can be applied to models beyond the SM, and will allow for a streamlined extension to the two-loop NNLO EW level and/or for the inclusion of mass-suppressed effects. Furthermore, the effective vertex approach allowed us to generalise our implementation to the case of processes with internal resonances. To achieve this, depending on kinematic projectors we construct EW correction factors for any underlaying LO resonance topologies for processes involving the decay of massive particles, correctly mapping on the contributing helicity configurations. Different helicity configurations in general receive substantially different EW corrections, such that an on-shell approximation might not be sufficient.  
We leave the investigation of this implementation in regard of $\ord(\alpha)$ spin-correlation effects, to resonant top-quark processes, and t-channel topologies, e.g. in VBF/VBS processes, to future studies. 

Based on our implementation we have presented numerical predictions for a large set of representative phenomenologically relevant LHC processes comparing one-loop NLL(') EW results with full NLO EW one-loop corrections, highlighting the scope of the applied approximations. Overall we found very good agreement between the LA and the exact description, with a deterioration of the approximation towards phase-space regions subject to clear  multi-scale configurations such as for the tails of invariant mass distributions. 

The implementation will me made available in upcoming new versions of the \OpenLoops amplitude provider.

\section*{Acknowledgments}	
We thank Stefano Pozzorini for invaluable discussions and for comments on the manuscript. 
J.L. is supported by the Science and Technology Research Council (STFC) under the Consolidated Grants ST/T00102X/1 and ST/X000796/1, and the STFC Ernest Rutherford Fellowship ST/S005048/1. 
We acknowledge the use of the DiRAC Cumulus HPC facility under Grant No. PPSP226.

\begin{acknowledgement}

\end{acknowledgement}

\section{Appendix} \label{sec6}
\appendix
\section{Effective EW Sudakov CT vertices} \label{CTEffectiverules}
Here we report the full list of effective CT rules that we have implemented and which are derived from the EW Feynman rules~\cite{Denner:1991kt}, where all fields are understood to be incoming. In the following we use the notation $V_N=A,Z$ and $\Phi_N=H, \chi$ respectively for neutral gauge bosons and scalars. All effective two-point CT vertices are expressed in term of  the $\mathcal{I}$ coupling factors as introduced in Eq. \eqref{eq:effectiveCTrule_fermions} and defined by the relation $I_{\varphi_{i_{j}^{\prime}} \phiphantJ}^{V} = i e \mathcal{I}_{\varphi_{i_{j}^{\prime}} \phiphantJ}^{V}$; however, we use the short-hand notation $\mathcal{I}_{\phiphantJ}^V = \mathcal{I}_{\varphi_{i_{j}^{\prime}} \phiphantJ}^V$ when the field $\varphi_{i_{j}^{\prime}}$ is  unambiguously fixed by the external particle $\phiphantJ$ and the soft gauge boson $V$.

\paragraph{Fermion sector}

 \begin{equation}\nonumber
 \vcenter{\hbox{\begin{tikzpicture}
     \begin{feynman}
      \vertex (a) at ( 1, 0);
      \vertex  (e) at ( 0, 0);
      \vertex (c) at ( -2, 0);
      \vertex (u) at (-1, 0.75);
      \vertex (k) at ( -1, 0);
      
      \diagram* {
             (e) -- [plain, edge label=\(f^{\kappa}\)]
         (k),
          (k) -- [plain, edge label=\(f^{\kappa}\)]
         (c),
         (k) -- [photon, edge label=\(V_N\)]
         (u)
         };
    \end{feynman} $\hspace{0.5 cm} \longrightarrow \hspace{0.5 cm}$
  \end{tikzpicture}
 $\hspace{0.5 cm} \hspace{0.8 cm}$ 
 \begin{tikzpicture}
     \begin{feynman}
      \vertex (a) at ( 1, 0);
      \vertex  (e) at ( 0, 0);
      \vertex (c) at ( -2, 0);
      \vertex[dot] (k) at ( -1, 0) {\contour{black}{}};
       \vertex (u) at (-1, 0.3) {\(V_N\)};
       
      \diagram* {
             (e) -- [plain,  insertion={[size=2 pt, style=thick]0.5 }, edge label=\(f^{\kappa}\)]
         (k),
          (k) -- [plain, edge label=\(f^{\kappa}\)]
         (c),
         };
    \end{feynman}
  \end{tikzpicture}
  
  }}  = 
  ie \mathcal{I}_{f^{\kappa}}^{V_N}
\end{equation}

\begin{equation}\nonumber
 \vcenter{\hbox{\begin{tikzpicture}
     \begin{feynman}
      \vertex (a) at ( 1, 0);
      \vertex  (e) at ( 0, 0);
      \vertex (c) at ( -2, 0);
      \vertex (u) at (-1, 0.75);
      \vertex (k) at ( -1, 0);
      
      \diagram* {
             (e) -- [plain, edge label=\(f^{\prime \kappa}\)]
         (k),
          (k) -- [plain, edge label=\(f^{\kappa}\)]
         (c),
         (k) -- [photon, edge label=\(W^{\pm}\)]
         (u)
         };
    \end{feynman} $\hspace{0.5 cm} \longrightarrow \hspace{0.5 cm}$
  \end{tikzpicture}
 $\hspace{0.5 cm} \hspace{0.8 cm}$ 
 \begin{tikzpicture}
     \begin{feynman}
      \vertex (a) at ( 1, 0);
      \vertex  (e) at ( 0, 0);
      \vertex (c) at ( -2, 0);
       \vertex (u) at (-1, 0.3) {\(W^{\pm}\)};
      \vertex[dot] (k) at ( -1, 0) {\contour{black}{}};

      \diagram* {
             (e) -- [plain,  insertion={[size=2 pt, style=thick]0.5 }, edge label=\(f^{\prime \kappa}\)]
         (k),
          (k) -- [plain, edge label=\(f^{\kappa}\)]
         (c),
         };
    \end{feynman}
  \end{tikzpicture}
  
  }}  = 
  ie \mathcal{I}_{f^{\kappa}}^{W^{\pm}}
\end{equation}

\paragraph{Gauge Boson sector}

\begin{equation}\nonumber
 \vcenter{\hbox{\begin{tikzpicture}
     \begin{feynman}
      \vertex (a) at ( 1, 0);
      \vertex  (e) at ( 0, 0);
      \vertex (c) at ( -2, 0);
      \vertex (u) at (-1, 0.75);
      \vertex (k) at ( -1, 0);
      
      \diagram* {
             (e) -- [photon, edge label=\(W^{\mp}\)]
         (k),
          (k) -- [photon, edge label=\(W^{\pm}\)]
         (c),
         (k) -- [photon, edge label=\(V_N\)]
         (u)
         };
    \end{feynman} $\hspace{0.5 cm} \longrightarrow \hspace{0.5 cm}$
  \end{tikzpicture}
 $\hspace{0.5 cm} \hspace{0.8 cm}$ 
 \begin{tikzpicture}
     \begin{feynman}
      \vertex (a) at ( 1, 0);
      \vertex  (e) at ( 0, 0);
      \vertex (c) at ( -2, 0);
      \vertex[dot] (k) at ( -1, 0) {\contour{black}{}};      
      \vertex (u) at (-1, 0.3) {\(V_N\)};       
   
      \diagram* {

             (e) -- [photon,  insertion={[size=2 pt, style=thick]0.5}, edge label=\(W^{\mp}\)]
         (k),
          (k) -- [photon, edge label=\(W^{\pm}\)]
         (c),
         };
    \end{feynman}
  \end{tikzpicture}
  
  }}  = 
 ie \mathcal{I}_{W^{\pm}}^{V_N}
\end{equation}

\begin{equation}\nonumber
 \vcenter{\hbox{\begin{tikzpicture}
     \begin{feynman}
      \vertex (a) at ( 1, 0);
      \vertex  (e) at ( 0, 0);
      \vertex (c) at ( -2, 0);
      \vertex (u) at (-1, 0.75);
      \vertex (k) at ( -1, 0);
      
      \diagram* {
             (e) -- [photon, edge label=\(V_N\)]
         (k),
          (k) -- [photon, edge label=\(W^{\pm}\)]
         (c),
         (k) -- [photon, edge label=\(W^{\mp}\)]
         (u)
         };
    \end{feynman} $\hspace{0.5 cm} \longrightarrow \hspace{0.5 cm}$
  \end{tikzpicture}
 $\hspace{0.5 cm} \hspace{0.8 cm}$ 
 \begin{tikzpicture}
     \begin{feynman}
      \vertex (a) at ( 1, 0);
      \vertex  (e) at ( 0, 0);
      \vertex (c) at ( -2, 0);
      \vertex[dot] (k) at ( -1, 0) {\contour{black}{}};
      \vertex (u) at (-1, 0.3) {\(W^{\mp}\)};      \diagram* {

             (e) -- [photon,  insertion={[size=2 pt, style=thick]0.5}, edge label=\(V_N\)]
         (k),
          (k) -- [photon, edge label=\(W^{\pm}\)]
         (c),
         };
    \end{feynman}
  \end{tikzpicture}
  
  }}  = 
  ie \mathcal{I}_{W^{\pm}, V_N}^{W^{\mp}}
\end{equation}

\begin{equation}\nonumber
 \vcenter{\hbox{\begin{tikzpicture}
     \begin{feynman}
      \vertex (a) at ( 1, 0);
      \vertex  (e) at ( 0, 0);
      \vertex (c) at ( -2, 0);
      \vertex (u) at (-1, 0.75);
      \vertex (k) at ( -1, 0);
      
      \diagram* {
             (e) -- [photon, edge label=\(W^{\pm}\)]
         (k),
          (k) -- [photon, edge label=\(V_N\)]
         (c),
         (k) -- [photon, edge label=\(W^{\mp}\)]
         (u)
         };
    \end{feynman} $\hspace{0.5 cm} \longrightarrow \hspace{0.5 cm}$
  \end{tikzpicture}
 $\hspace{0.5 cm} \hspace{0.8 cm}$ 
 \begin{tikzpicture}
     \begin{feynman}
      \vertex (a) at ( 1, 0);
      \vertex  (e) at ( 0, 0);
      \vertex (c) at ( -2, 0);
      \vertex[dot] (k) at ( -1, 0) {\contour{black}{}};
      \vertex (u) at (-1, 0.3) {\(W^{\mp}\)};      \diagram* {

             (e) -- [photon,  insertion={[size=2 pt, style=thick]0.5}, edge label=\(W^{\pm}\)]
         (k),
          (k) -- [photon, edge label=\(V_N\)]
         (c),
         };
    \end{feynman}
  \end{tikzpicture}
  }}  = 
  ie \mathcal{I}_{V_N}^{W^{\mp}}
\end{equation}

\paragraph{Scalar sector}

 \begin{equation}\nonumber
 \vcenter{\hbox{\begin{tikzpicture}
     \begin{feynman}
      \vertex (a) at ( 1, 0);
      \vertex  (e) at ( 0, 0);
      \vertex (c) at ( -2, 0);
      \vertex (u) at (-1, 0.75);
      \vertex (k) at ( -1, 0);
      
      \diagram* {
             (e) -- [scalar, edge label=\(\chi\)]
         (k),
          (k) -- [scalar, edge label=\(H\)]
         (c),
         (k) -- [photon, edge label=\(Z\)]
         (u)
         };
    \end{feynman} $\hspace{0.5 cm} \longrightarrow \hspace{0.5 cm}$
  \end{tikzpicture}
 $\hspace{0.5 cm} \hspace{0.8 cm}$ 
 \begin{tikzpicture}
     \begin{feynman}
      \vertex (a) at ( 1, 0);
      \vertex  (e) at ( 0, 0);
      \vertex (c) at ( -2, 0);
      \vertex[dot] (k) at ( -1, 0) {\contour{black}{}};
      \vertex (u) at (-1, 0.3) {\(Z\)};      \diagram* {

             (e) -- [scalar, insertion={[size=2 pt, style=thick]0.5}, edge label=\(\chi\)]
         (k),
          (k) -- [scalar, edge label=\(H\)]
         (c),
         };
    \end{feynman}
  \end{tikzpicture}
  
  }}  = 
  ie \mathcal{I}_{H}^{Z}
\end{equation}

  \begin{equation}\nonumber
 \vcenter{\hbox{\begin{tikzpicture}
     \begin{feynman}
      \vertex (a) at ( 1, 0);
      \vertex  (e) at ( 0, 0);
      \vertex (c) at ( -2, 0);
      \vertex (u) at (-1, 0.75);
      \vertex (k) at ( -1, 0);
      
      \diagram* {
             (e) -- [scalar, edge label=\(H\)]
         (k),
          (k) -- [scalar, edge label=\(\chi\)]
         (c),
         (k) -- [photon, edge label=\(Z\)]
         (u)
         };
    \end{feynman} $\hspace{0.5 cm} \longrightarrow \hspace{0.5 cm}$
  \end{tikzpicture}
 $\hspace{0.5 cm} \hspace{0.8 cm}$ 
 \begin{tikzpicture}
     \begin{feynman}
      \vertex (a) at ( 1, 0);
      \vertex  (e) at ( 0, 0);
      \vertex (c) at ( -2, 0);
      \vertex[dot] (k) at ( -1, 0) {\contour{black}{}};
      \vertex (u) at (-1, 0.3) {\(Z\)};      \diagram* {

             (e) -- [scalar, insertion={[size=2 pt, style=thick]0.5}, edge label=\(H\)]
         (k),
          (k) -- [scalar, edge label=\(\chi\)]
         (c),
         };
    \end{feynman}
  \end{tikzpicture}
  
  }}  = 
  ie \mathcal{I}_{\chi}^{Z}
\end{equation}

 \begin{equation}\nonumber
 \vcenter{\hbox{\begin{tikzpicture}
     \begin{feynman}
      \vertex (a) at ( 1, 0);
      \vertex  (e) at ( 0, 0);
      \vertex (c) at ( -2, 0);
      \vertex (u) at (-1, 0.75);
      \vertex (k) at ( -1, 0);
      
      \diagram* {
             (e) -- [scalar, edge label=\(\phi^{\mp}\)]
         (k),
          (k) -- [scalar, edge label=\(\phi^{\pm}\)]
         (c),
         (k) -- [photon, edge label=\(V_N\)]
         (u)
         };
    \end{feynman} $\hspace{0.5 cm} \longrightarrow \hspace{0.5 cm}$
  \end{tikzpicture}
 $\hspace{0.5 cm} \hspace{0.8 cm}$ 
 \begin{tikzpicture}
     \begin{feynman}
      \vertex (a) at ( 1, 0);
      \vertex  (e) at ( 0, 0);
      \vertex (c) at ( -2, 0);
      \vertex[dot] (k) at ( -1, 0) {\contour{black}{}};
      \vertex (u) at (-1, 0.3) {\(V_N\)};      \diagram* {

             (e) -- [scalar, insertion={[size=2 pt, style=thick]0.5}, edge label=\(\phi^{\mp}\)]
         (k),
          (k) -- [scalar, edge label=\(\phi^{\pm}\)]
         (c),
         };
    \end{feynman}
  \end{tikzpicture}
  
  }}  = 
  ie \mathcal{I}_{\phi^{\pm} }^{V_N}
\end{equation}
 
\begin{equation}\nonumber
 \vcenter{\hbox{\begin{tikzpicture}
     \begin{feynman}
      \vertex (a) at ( 1, 0);
      \vertex  (e) at ( 0, 0);
      \vertex (c) at ( -2, 0);
      \vertex (u) at (-1, 0.75);
      \vertex (k) at ( -1, 0);
      
      \diagram* {
             (e) -- [scalar, edge label=\(\Phi_N\)]
         (k),
          (k) -- [scalar, edge label=\(\phi^{\pm}\)]
         (c),
         (k) -- [photon, edge label=\(W^{\mp}\)]
         (u)
         };
    \end{feynman} $\hspace{0.5 cm} \longrightarrow \hspace{0.5 cm}$
  \end{tikzpicture}
 $\hspace{0.5 cm} \hspace{0.8 cm}$ 
 \begin{tikzpicture}
     \begin{feynman}
      \vertex (a) at ( 1, 0);
      \vertex  (e) at ( 0, 0);
      \vertex (c) at ( -2, 0);
      \vertex[dot] (k) at ( -1, 0) {\contour{black}{}};
      \vertex (u) at (-1, 0.3) {\(W^{\mp}\)};      \diagram* {

             (e) -- [scalar, insertion={[size=2 pt, style=thick]0.5}, edge label=\(\Phi_N\)]
         (k),
          (k) -- [scalar, edge label=\(\phi^{\pm}\)]
         (c),
         };
    \end{feynman}
  \end{tikzpicture}
  
  }}  = 
  ie \mathcal{I}_{\phi^{\pm}, \Phi_N}^{W^{\mp}}
\end{equation}

\begin{equation}\nonumber
 \vcenter{\hbox{\begin{tikzpicture}
     \begin{feynman}
      \vertex (a) at ( 1, 0);
      \vertex  (e) at ( 0, 0);
      \vertex (c) at ( -2, 0);
      \vertex (u) at (-1, 0.75);
      \vertex (k) at ( -1, 0);
      
      \diagram* {
             (e) -- [scalar, edge label=\(\phi^{\pm}\)]
         (k),
          (k) -- [scalar, edge label=\(\Phi_N\)]
         (c),
         (k) -- [photon, edge label=\(W^{\mp}\)]
         (u)
         };
    \end{feynman} $\hspace{0.5 cm} \longrightarrow \hspace{0.5 cm}$
  \end{tikzpicture}
 $\hspace{0.5 cm} \hspace{0.8 cm}$ 
 \begin{tikzpicture}
     \begin{feynman}
      \vertex (a) at ( 1, 0);
      \vertex  (e) at ( 0, 0);
      \vertex (c) at ( -2, 0);
      \vertex[dot] (k) at ( -1, 0) {\contour{black}{}};
      \vertex (u) at (-1, 0.3) {\(W^{\mp}\)};      \diagram* {

             (e) -- [scalar, insertion={[size=2 pt, style=thick]0.5}, edge label=\(\phi^{\pm}\)]
         (k),
          (k) -- [scalar, edge label=\(\Phi_N\)]
         (c),
         };
    \end{feynman}
  \end{tikzpicture}
  
  }}  = 
  ie \mathcal{I}_{\Phi_N}^{W^{\mp}}
\end{equation}
with $\mathcal{I}_{W^{\pm}}^{V_N} =\mathcal{I}_{V_N}^{W^{\mp}} = \mathcal{I}_{W^{\pm}, V_N}^{W^{\mp}}$ as well as $\mathcal{I}_{\phi^{\pm}}^{V_N} =\mathcal{I}_{\Phi_N}^{W^{\mp}} = \mathcal{I}_{\phi^{\pm}, \Phi_N}^{W^{\mp}}$ and $\mathcal{I}_{H}^{Z} =\mathcal{I}_{\chi}^{Z}$.

\section{Logarithmic approximation of self-energy integrals}\label{appendixLA}

The generic two point functions
 \begin{equation} \label{DL_diagrams}
     \vcenter{\hbox{\begin{tikzpicture}
    \begin{feynman}
      \vertex (a) at ( -1.5, 0);
      \vertex (d) at ( 1.5, 0);
    \vertex (b) at ( -0.75, 0) ;
        \vertex (c) at ( 0.75, 0);
      \diagram* {
    (a)--[plain, edge label=\(p\)] (b),
       (b) -- [plain, half left, looseness=1.5,  edge label'=\(m_0\)] (c),
              (c) -- [plain, half left, looseness=1.5,  edge label'=\(m_1\)] (b),
       (c) -- [edge label=\(p\)]  (d)
         };
\end{feynman}
  \end{tikzpicture}}}
\end{equation}
with explicit integral representations in $D=4-2\epsilon$
\begin{equation}
\begin{aligned}
&\frac{\mathrm{i}}{(4 \pi)^2} B_{\{0, \mu, \mureg \nu\}}\left(p, m_0, m_1\right):=\mu^{4-D} \\ 
&\int \frac{\mathrm{d}^D q}{(2 \pi)^D} \frac{\left\{1, q_\mu, q_\mu q_\nu\right\}}{\left(q^2-m_0^2+\mathrm{i} \varepsilon\right)\left[(q+p)^2-m_1^2+\mathrm{i} \varepsilon\right]}
\end{aligned}
\end{equation}
depend on the mass scales $p^2, m_0^2, m_1^2, \mureg^2$. For the evaluation of the parameter renormalisation contributions in logarithmic approximation we restrict to scale hierarchies
\begin{equation}\label{hierarchyLA}
\mureg^2 = s \gg p^2, m_0^2, m_1^2\,,
\end{equation}
and we only consider large logarithms of the ratios of these scales to the finite part, neglecting all mass suppressed contributions.
In this approximation the one-loop tadpole function $A_0$ reads 
\begin{equation}\label{eq:azero}
A_0\left(m_0\right) \stackrel{\text { LA }}{=} m_0^2 \log \frac{\mureg^2}{m_0^2}.
\end{equation}
We can derive explicit LA results for the scalar two-point function and its derivative from Ref.~\cite{Denner:1991kt},
\begin{equation}\label{inputLA}
\begin{aligned}
B_0\left(p, m_0, m_1\right) & \stackrel{\mathrm{LA}}{=} \log \frac{\mureg^2}{m_0^2}+\frac{1}{2}\left(1-\frac{m_0^2-m_1^2}{p^2}\right) \log \frac{m_0^2}{m_1^2}\\ 
&+ \frac{1}{2} \frac{m_0 m_1}{p^2}\left(r-\frac{1}{r}\right) \log r^2 , \\
p^2 B_0^{\prime}\left(p, m_0, m_1\right) & \stackrel{\mathrm{LA}}{=} \frac{1}{2}\left\{\frac{m_1^2-m_0^2}{p^2} \log \frac{m_1^2}{m_0^2} \right. \\
&\left. -\left[\frac{m_0 m_1}{p^2}\left(r-\frac{1}{r}\right)+\frac{r^2+1}{r^2-1}\right] \log r^2\right\}\,,
\end{aligned}
\end{equation}
with 
\begin{equation}
r+\frac{1}{r}=\frac{m_0^2+m_1^2-p^2-\mathrm{i} \varepsilon}{m_0 m_1}\,.
\end{equation}
Expressions for $B_1,B_{00}, B_{11}$ and their derivatives $B_1', B_{00}'$ can be derived from \ref{inputLA}. In general these results depend on the hierarchy of scales $p^2, m_0^2,m_1^2$, distinguishing four cases
\begin{equation}\label{4casesLA}
\begin{aligned} 
& (a) \quad  m_i^2 \ll p^2  \text{ and } p^2-m_{1-i}^2 \ll p^2 \quad \text {for }i=0 \text { or }\quad i=1, \\
& (b) \quad \hspace{-0.1 cm} \operatorname{not} \quad (a) \text { and } m_i^2 \ngtr p^2 \text { for } i=0,1 , \\
& (c) \quad  m_0^2=m_1^2 \gg p^2 \\
& (d) \quad  m_i^2 \gg p^2 \nless m_{1-i}^2 \text{ for }i=0 \text{ or } i=1\,,
\end{aligned}
\end{equation}
where case $(a)$ is only relevant for diagrams involving a photon with $m_i^2=\lambda ^2$, such that
\begin{equation}
(a') \quad m_i^2=\lambda^2 \ll p^2=m_{1-i}^2 \text { for } i=0 \text { or } i=1
\end{equation}
is the actual situation which can occur.
Apart from the derivatives $B_0'$ and $B_1'$ which give rise to additional IR divergences in the case $(a')$, all the considered 2-point functions have the same logarithmic approximation for each one of the cases in \eqref{4casesLA}:
\begin{equation}
\begin{aligned}
B_{0}\left(p, m_{0}, m_{1}\right) & \stackrel{\text { LA }}{=} \log \frac{\mureg^{2}}{M^{2}}\,, \\
B_{1}\left(p^{2}, m_{0}, m_{1}\right) & \stackrel{\text { LA }}{=}-\frac{1}{2} \log \frac{\mureg^{2}}{M^{2}}\,,\\
\frac{1}{p^2} B_{00}\left(p^2, m_0, m_1\right)& \stackrel{\text { LA }}{=} \frac{3 m_0^2+3 m_1^2-p^2}{12 p^2} \log \frac{\mureg^2}{M^2}\,, \\
\frac{1}{p^2} g^{\mu \nu} B_{\mu \nu}\left(p^2, m_0, m_1\right)& \stackrel{\text { LA }}{=} \frac{m_0^2+m_1^2}{p^2} \log \frac{\mureg^2}{M^2}\,,
\end{aligned}
\end{equation}
where $M^2:=\max \left(p^2, m_0^2, m_1^2\right)$ sets the scale of the logarithms. From last equation one can derive $B_{11}$ via
\begin{align}\nonumber
g^{\mu \nu} B_{\mu \nu}&\left(p, m_0, m_1\right)=\\
D B_{00}&\left(p^2, m_0, m_1\right)+p^2 B_{11}\left(p^2, m_0, m_1\right).
\end{align} 
Derivatives of two-point functions are vanishing for the cases $(b),\text{}(c) \text{ and } (d)$, i.e.
\begin{equation}
p^2 B_0^{\prime}\left(p, m_0, m_1\right)=p^2 B_1^{\prime}\left(p^2, m_0, m_1\right) \stackrel{\mathrm{LA}}{=} 0,
\end{equation}
while for $(a')$
\begin{equation}
\begin{aligned}
p^2 B_0^{\prime}\left(p, m_0, m_1\right)  \stackrel{\text { LA }}{=} \frac{1}{2} \log \frac{m_{1-i}^2}{m_i^2}&=\frac{1}{2} \log \frac{p^2}{\lambda^2}\,, \\
p^2 B_1^{\prime}\left(p^2, m_0, m_1\right)+\frac{1}{2} p^2 B_0^{\prime}\left(p, m_0, m_1\right) & \stackrel{\text { LA }}{=}-\frac{1}{4} \log \frac{m_0^2}{m_1^2}\,.
\end{aligned}
\end{equation}


\section{List of relevant \OpenLoops parameters}\label{app:parameters}

In Tab. \ref{tab:par} we list allinput parameters and switches available in \OpenLoops
that are relevant for the selection and evaluation of the one-loop EW corrections in logarithmic approximation. These parameters can be set via the \texttt{set\_parameter} routine in \OpenLoops, see Appendix A.2 in Ref.~\cite{Buccioni:2019sur}.

\begin{table*}[h!]
\centering
\begin{tabular}{ccl}
\hline
\hline
parameter & type/ default value   &description \\
\hline
\hline
 \multirow{2}{*}{\texttt{nllew}\textunderscore \texttt{on}}  & \multirow{2}{*}{int, default=0} &0: NLO EW\\
 & &1: NLL(') EW\\ 
\hline
\multirow{13}{*}{\texttt{nllew\textunderscore contrib}}  & \multirow{13}{*}{int, default=1} & 0: none\\
 & &1: NLL' EW = LSC + SSC + S-SSC + C \\
 &.& \hspace{2.5cm}+ YUK + PR \\
 & &2: NLL EW = LSC + SSC + C + YUK + PR \\
 & &10:  LSC\\
 & &11:  SSC\\
 & &12:  S-SSC\\
 & &13:  Coll\\
 & &14:  WFR\\
 & &15:  C=Coll+WFR\\
 & &17:  YUK\\
 & &18:  PR\\
 & &19:  PR+YUK+WFR via standard UV CTs\\
 & &20:  DL = \LSC + \SSC + \SSSC \\
\hline
\multirow{3}{*}{\texttt{nllew}\textunderscore \texttt{softv}}  & \multirow{3}{*}{int, default=1} &  1: $\gamma+Z+W^{\pm}$ contribution (SM) \\
 & &2: $Z+W^{\pm}$ contribution (weak) \\
 & &3: only $\gamma$ contribution (QED)\\
\hline
\multirow{2}{*}{\texttt{nllew}\textunderscore \texttt{phase}}  & \multirow{2}{*}{int, default=1} &0: complex phase in DL off\\
 & &1: complex phase in DL on\\ 
\hline
 \multirow{2}{*}{\texttt{nllew}\textunderscore \texttt{irreg }}  & \multirow{2}{*}{int, default=1} & 0: mass regularisation\\
 & &1: dimensional regularisation\\
 \hline
 \texttt{nllew}\textunderscore \texttt{mreg } & double, default=80.399 & 
 photon mass regularisation scale $\lambda$ \\
  \hline
 \texttt{nllew}\textunderscore \texttt{mregf } & double, default=1. & 
 fermion mass regularisation scale $M_{\varphi}$ \\
 \hline
\multirow{3}{*}{\texttt{nllew\textunderscore resswitch}}  & \multirow{3}{*}{int, default=0} &0: corrections only on external states\\
 & &1: corrections only on resonance processes \\
 & &2: corrections on resonance processes \\
 & &\hspace{0.5cm}and external states  \\
 \hline
\texttt{nllew\textunderscore rescaling} & double, default=10.& $w_{\text{rescale}}$: width rescaling factor for \\
 & &   resonance projectors\\
\hline\hline
\end{tabular}
\caption{Available input parameters and switches that control the contributions of the one-loop EW corrections in logarithmic approximation in \OpenLoops.}
\label{tab:par}
\end{table*}


\section{Amplitude-level validation}\label{appendixEnergyScans}

Here we investigate amplitude-squared level numerical results for 
individual polarisation configurations for several partonic processes, 
including some of the processes studied at cross section level in
Section \ref{sec4}. We consider scans in energy and in angles
varying $\sqrt{s}$ in the range $[0.5, 100]$ TeV for a fixed scattering
angle $ \theta=120 ^{\circ}$ in $2 \to 2 $ processes, or varying the
latter in the domain $[0^{\circ},180^{\circ}] $ while keeping the energy
fixed at $\sqrt{s}=13$ TeV. 
Polarisations $\lambda$ in the subscripts are labelled as follows: 
(L/R) stand for left- and right-handed fermions while $\pm 1$ and $0$ are, 
in this order, the transverse and longitudinal polarisations of vector bosons.
We compare predictions in logarithmic approximation against complete one-loop
results employing dimensional regularisation. Resulting infrared poles are removed
by consistently adding the Catani-Seymour's QED $\mathbf{I}_{\rm QED}$-operator, as
discussed at the beginning of Section~\ref{sec4}. Additionally we 
show predictions in mass regularisation with $\lambda=m_W$. We
consider NLL EW and NLL' EW predictions, where the latter includes \SSSC terms,
which cannot completely be controlled in LA.  By comparing amplitude-level predictions
at NLL(') EW against corresponding predictions at full one-loop level we check
for any residual logarithmically enhanced effects beyond the LA.
Beyond the processes shown here we checked various other 
$2 \to 2/3/4$ processes at amplitude-squared level and
did not find any large remaining residual logarithmic effects for phase-space regions
fulfilling Eq.~\ref{la}.

\begin{figure*}[tb]
\centering
 \includegraphics[width=\setrelwidth\textwidth]{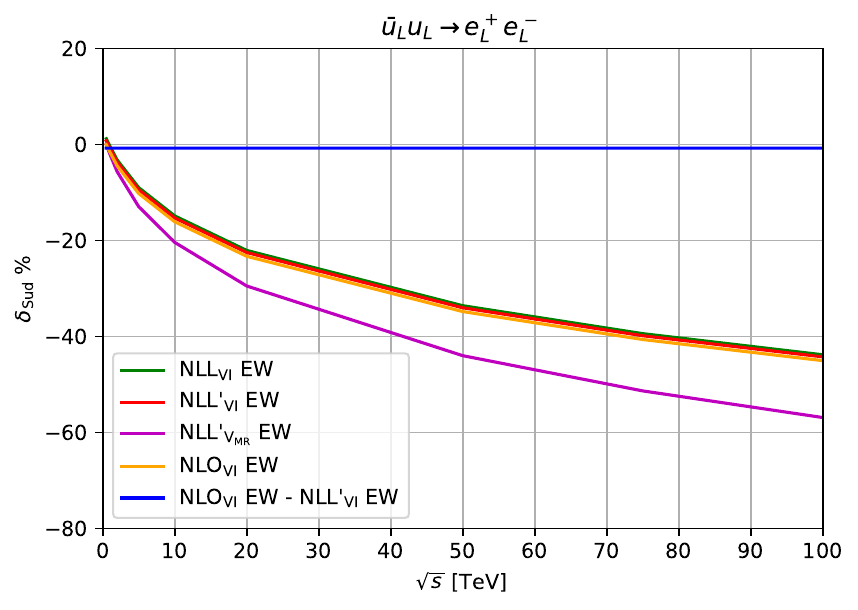}
 \includegraphics[width=\setrelwidth\textwidth]{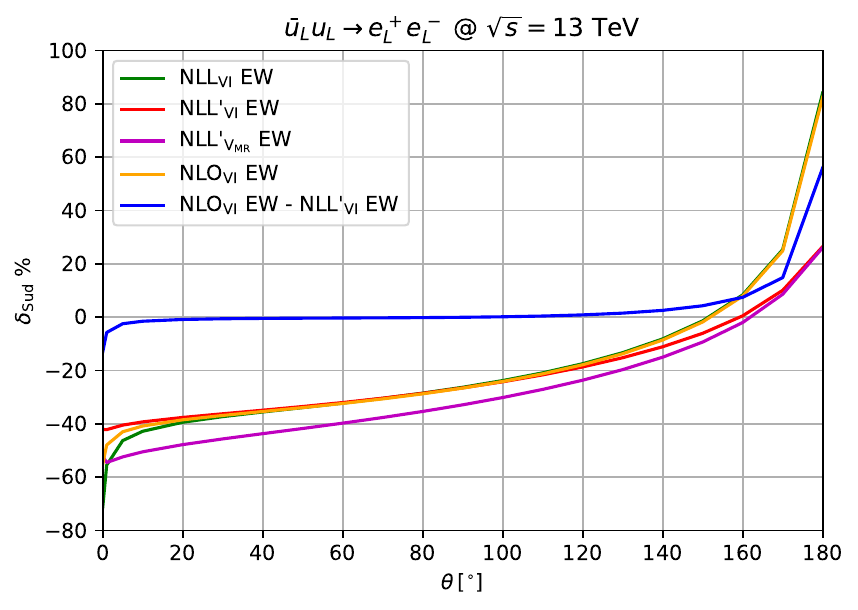}
\caption{Scans in energy (left) and scattering angle (right) for the partonic process $\bar{u}_L u_L \to e^+_L e^-_L$. The plot shows the relative corrections of the different one-loop EW predictions with respect to LO: \NLL~(solid green), \NLLp~(solid red), \NLLpMR~(solid magenta), \NVI~(solid orange) and the relative difference between \NVI and \NLLp (solid blue), where the subscript "VI" indicates that the QED $\mathbf{I}_{\rm QED}$-operator is added to the virtual amplitudes in DR.}
\label{fig:ES_eeuu}
\end{figure*}

In~\reffi{fig:ES_eeuu} we show scans in energy (left) and angle (right) for the partonic channel $\bar{u}_L u_L \to e^+_L e^-_L$. Being a not mass-suppressed configuration, the relative difference between \NVI and \NLLp results is constant over the entire considered energy range, and in the angle scan only spoiled for very backward/forward configurations ($\theta < 10^{\circ}$ or $\theta > 130^{\circ}$). 
Furthermore, there is almost perfect agreement between \NLL and \NLLp. Indeed the LA condition \eqref{la} 
is only fulfilled for kinematic regions that are not subject to large ratios of scales, which is the case for very backward/forward configurations.

\begin{figure*}[tb]
\centering
 \includegraphics[width=\setrelwidth\textwidth]{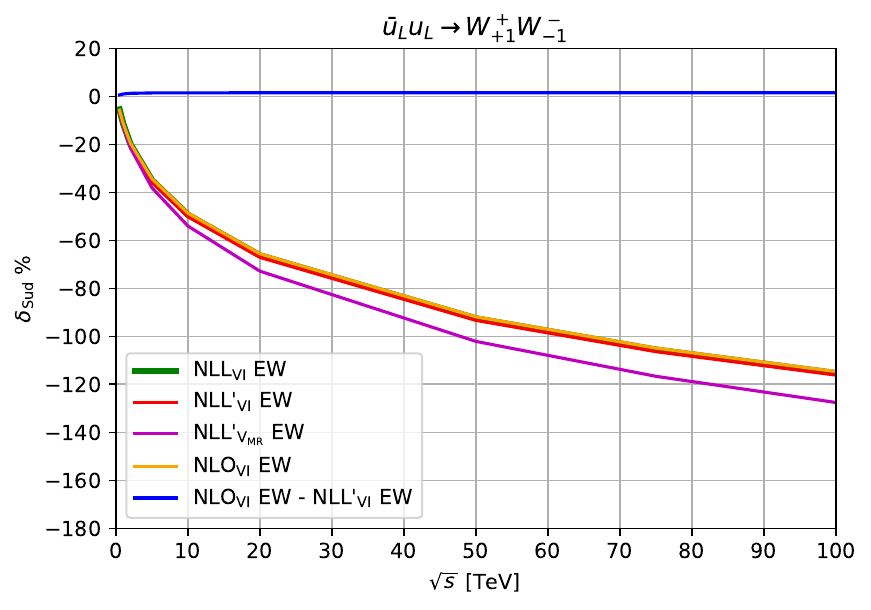}
  \includegraphics[width=\setrelwidth\textwidth]{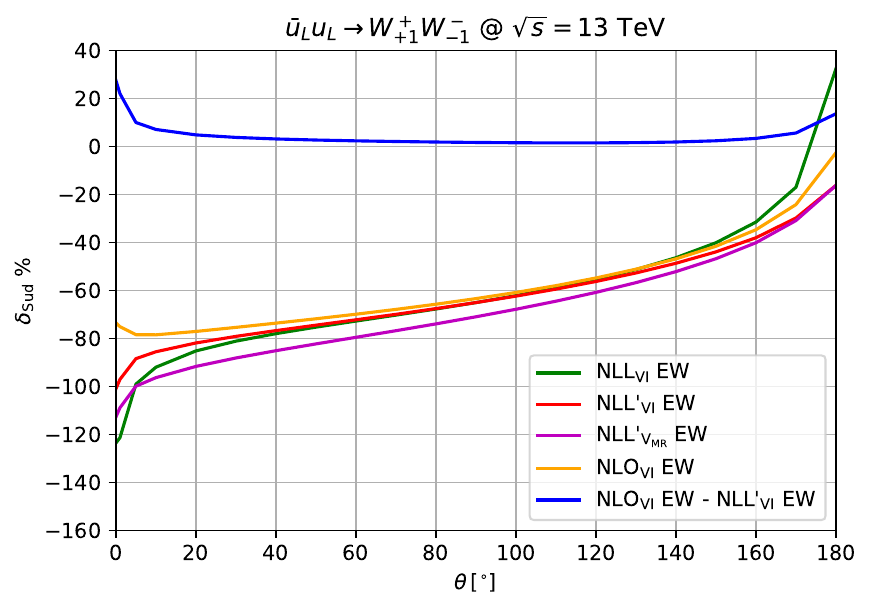}
\caption{Scans in energy (left) and scattering angle (right) for the partonic process $\bar{u}_L u_L \to W^+_{+1}W^-_{-1}$. Curves as in Fig.~\ref{fig:ES_eeuu}.}
\label{fig:ES_ww}
\end{figure*}

We continue our amplitude-level validation by considering the partonic channel $\bar{u} u \to W^+_{+1} W^-_{-1}$, i.e. considering transverse-polarised vector bosons. In Fig. \ref{fig:ES_ww} we show a scan in the energy (left) and in the scattering angle $\theta$ (right). 
The energy scan shows a sub-percent level agreement between \NVI and the LA approximation, as well as a constant relative difference in the entire examined range as expected for a not mass-suppressed configuration. Again we also find perfect overlap between \NLL and \NLLp. The scan in the scattering angle reveals a constant sub-percent level agreement between \NVI and \NLLp in the central regime, which is spoiled only for very backward/forward configurations.

\begin{figure*}[tb]
\centering
 \includegraphics[width=\setrelwidth\textwidth]{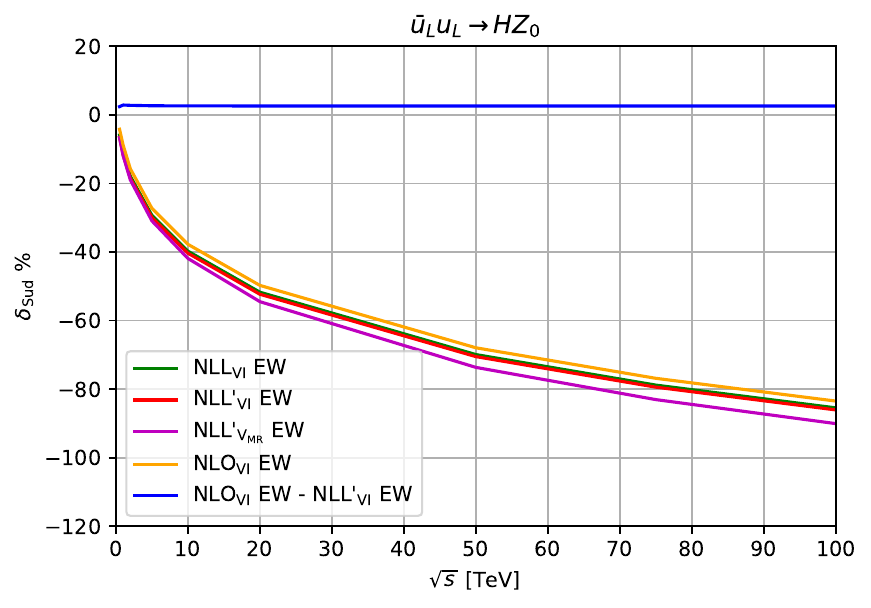}
  \includegraphics[width=\setrelwidth\textwidth]{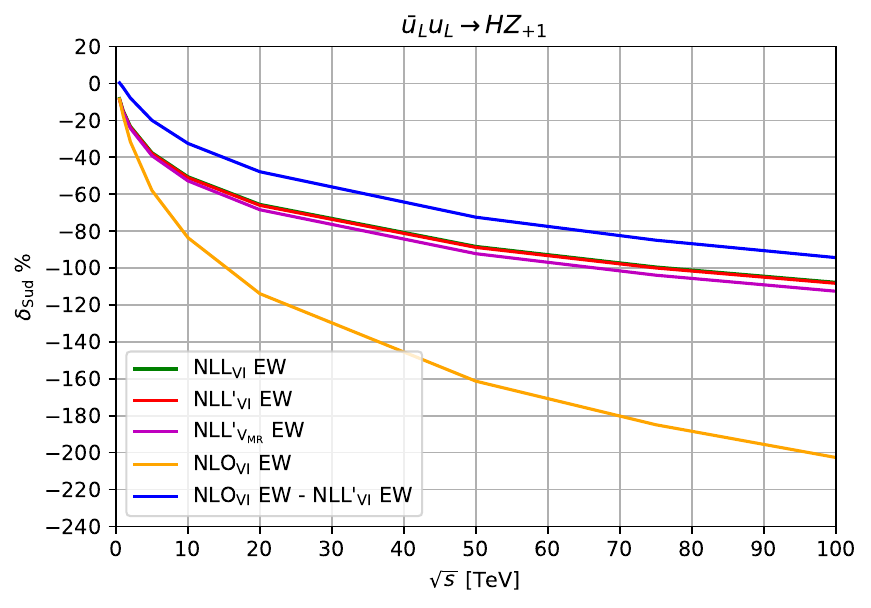}
\caption{Energy scans for the partonic processes $\bar{u}_L u_L \to Z_{+1}H$ and $\bar{u}_L u_L \to Z_{0}H$. Curves as in Fig.~\ref{fig:ES_eeuu}.}
\label{fig:ES_zh}
\end{figure*}

Next, we analyse the partonic process $\bar{u} u \to ZH$. In Fig. \ref{fig:ES_zh} we show energy scans for two polarisation configurations of the $Z$-boson: either longitudinal (left) or transverse (right), whereas quarks are left-handed in both cases.
For a longitudinally polarised $Z$-boson the amplitude is not-mass suppressed and, the difference between \NVI and \NLLp results in a constant in the entire energy range. Moreover, the agreement between these two predictions is at the percent level, as is the difference with respect to \NLL.
On the contrary, for a transversely polarised $Z$-boson the amplitude is mass-suppressed.
Indeed, the LA approximation clearly fails to reproduce the \NVI prediction: there is a clear residual
logarithmic behaviour not captured by the LA.

\begin{figure*}[tb]
\centering
 \includegraphics[width=\setrelwidth\textwidth]{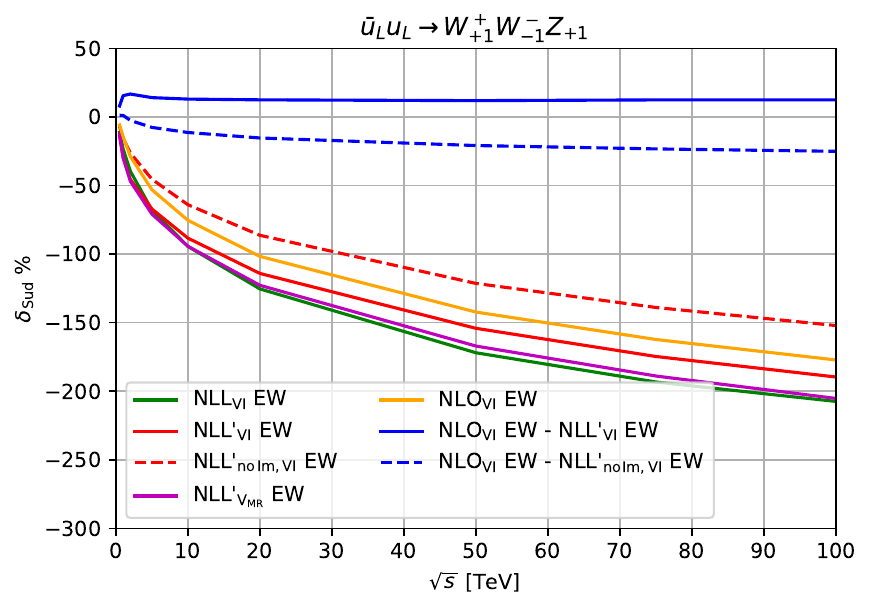}
  \includegraphics[width=\setrelwidth\textwidth]{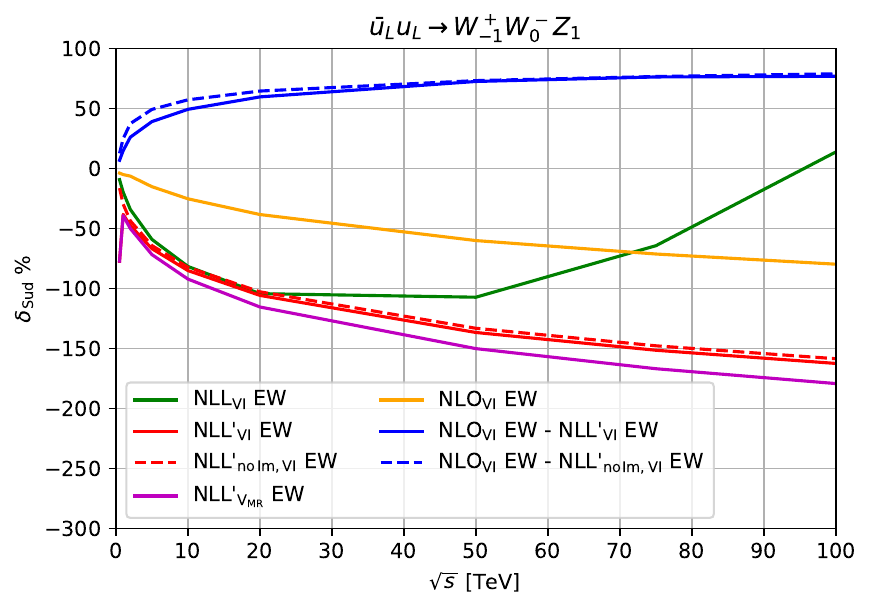}
\caption{Energy scan for the partonic processes $\bar{u}_L u_L \to W^+_{+1}W^-_{-1}Z_{1}$ and $\bar{u}_L u_L \to W^+_{-1}W^-_{0}Z_{1}$. Curves as in Fig.~\ref{fig:ES_eeuu} with the addition of \NLLpzero (dashed red) and its relative difference with \NVI  (dashed blue).}
\label{fig:ES_wwz_LLpmp}
\end{figure*}

Finally we consider the partonic process $\bar{u} u \to W^+ W^- Z$ in Fig. \ref{fig:ES_wwz_LLpmp}. We show results for the polarisation configurations with left-handed quarks and all transverse (left) or two transverse and one longitudinal (right) vector bosons; in addition to the curves already shown in the previous plots, we include the \NLLpzero  (dashed red) prediction and its relative difference with \NVI (dashed blue). In this prediction the imaginary part of $C_0|_{\text{LA}}$, Eq.~\ref{C0imaginary}, is not included. The configuration where all vector bosons are transversely polarised is not mass-suppressed and the LA including the imaginary part of $C_0|_{\text{LA}}$ provides percent-level constant agreement 
with \NVI. Excluding the complex phase in \NLLpzero  (dashed red) reveals a clear residual logarithmic difference with the \NVI prediction. 
For configurations with one of the three gauge bosons being longitudinal, as shown on the right, 
the amplitude is mass-suppressed. Correspondingly the LA completely fails to account for the one-loop EW corrections.

\section{Further numerical results}\label{appendixtth}

\subsubsection*{$\mathbf{t\bar tH}$}

\begin{figure*}[h]
\centering
	\includegraphics[width=\setrelwidth\textwidth]{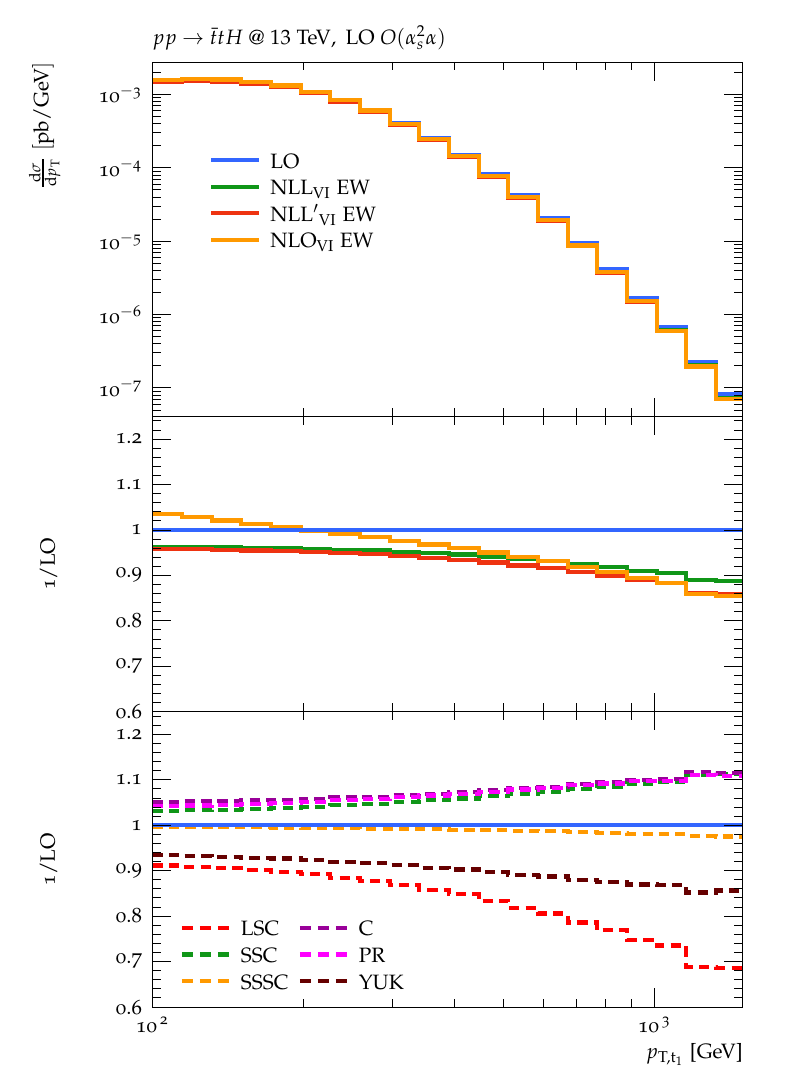}
       	\includegraphics[width=\setrelwidth\textwidth]{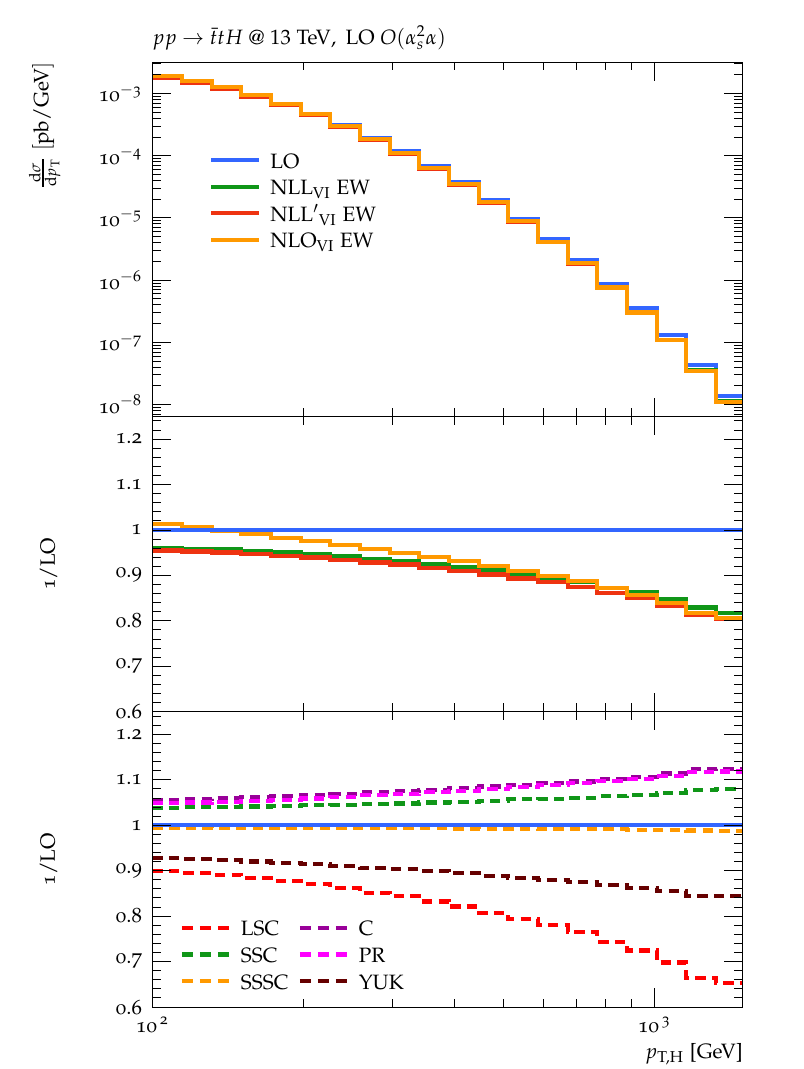}
\caption{Differential distribution in the transverse momentum of the hardest
$t$ quark $p_{\rm{T},t_1}$ (left) and of the $H$ boson $p_{\rm{T},H}$  (right)
 in $pp \to ttH$ at $\sqrt{s}=13 \hspace{0.1 cm} \rm{TeV}$.  Curves as in Fig.~\ref{fig:zj}. }
\label{fig:tth_pt}
	\includegraphics[width=\setrelwidth\textwidth]{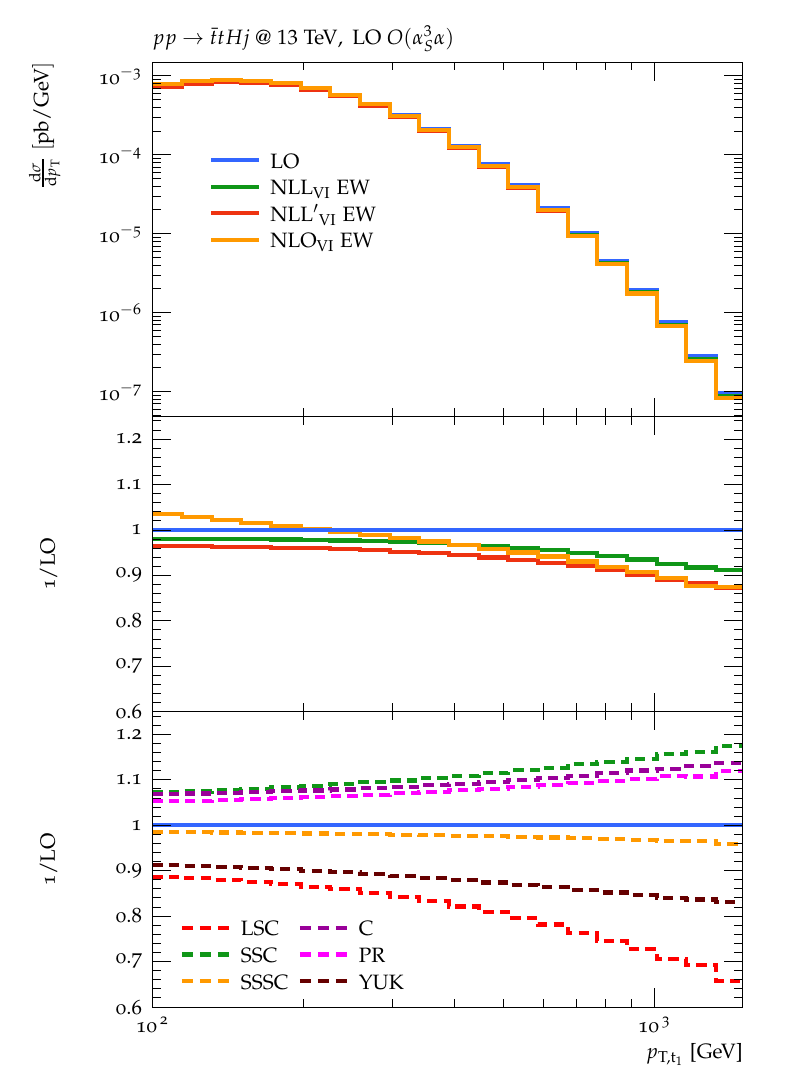}
       	\includegraphics[width=\setrelwidth\textwidth]{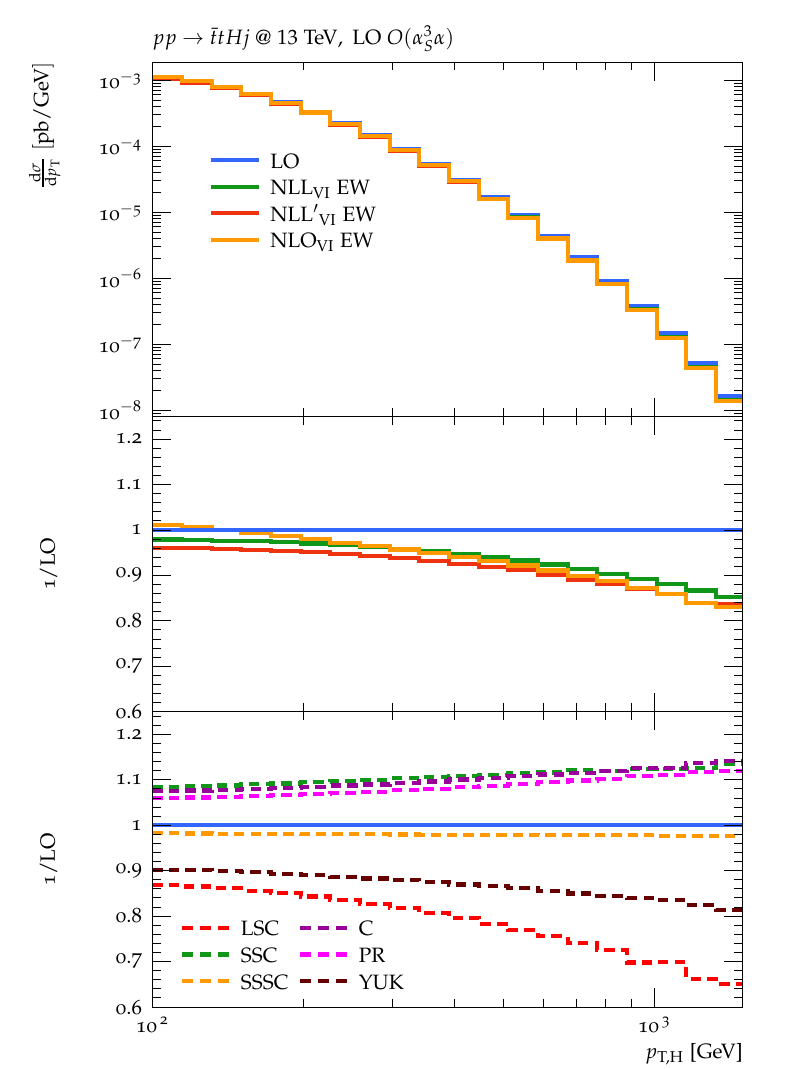}
\caption{Differential distribution in the transverse momentum of the hardest
$t$ quark $p_{\rm{T},t_1}$ (left) and of the $H$ boson $p_{\rm{T},H}$  (right)
 in $pp \to ttHj$ at $\sqrt{s}=13 \hspace{0.1 cm} \rm{TeV}$.  Curves as in Fig.~\ref{fig:zj}.}
\label{fig:tthj_pt}
\end{figure*}

In \reffi{fig:tth_pt} and \reffi{fig:tthj_pt} we show additional numerical results for  $t\bar t H$ and  $t\bar t H +$jet production respectively. In both cases on the left we show the $p_{\rm{T},t_1}$ distribution, and on the right we show the $p_{\rm{T},H}$ distribution.

%


\newpage
\clearpage

\bibliographystyle{JHEP}
\bibliography{ewsud}

\end{document}